\documentclass[aps,prx,reprint,superscriptaddress,longbibliography,nofootinbib,floatfix]{revtex4-2}
\usepackage[T1]{fontenc}
\usepackage[utf8]{inputenc}
\usepackage{xcolor}
\usepackage{mathtools}
\usepackage{amsmath}
\usepackage{amsthm}
\usepackage{amssymb}
\usepackage[bookmarks=false,
 breaklinks=false,pdfborder={0 0 1},backref=false,colorlinks=true]
 {hyperref}
\hypersetup{
 linkcolor=blue,citecolor=blue,urlcolor=blue,
 pdftitle={Mixed State Deep Thermalization under Generic Quantum Errors},
 pdfauthor={Xie-Hang Yu, Pavel Kos},
 pdfsubject={Quantum Physics},
 pdfkeywords={deep thermalization, projected ensemble, mixed states, quantum channels, Choi state, dual-unitary circuits, random quantum circuits, random density matrices}}

\makeatletter

\usepackage{bm}
\usepackage{braket}
\usepackage{graphicx}
\usepackage{tikz}
\usetikzlibrary{decorations.pathreplacing}

\newtheorem{proposition}{Proposition}
\newtheorem{lemma}[proposition]{Lemma}
\newcommand{\tr}{\operatorname{Tr}}
\newcommand{\id}{I}
\newcommand{\MSPE}{\mathcal{E}_{\mathrm{MSPE}}}
\newcommand{\Ncal}{\mathcal{N}}

\definecolor{mygreen}{RGB}{34,139,34}
\newcommand{\Wgategreen}[2]{%
\draw[very thick] (#1-0.5,#2+0.5)--(#1+0.5,#2-0.5);
\draw[very thick] (#1-0.5,#2-0.5)--(#1+0.5,#2+0.5);
\draw[thick,fill=mygreen!70,rounded corners=2pt]
  (#1-0.35,#2+0.35) rectangle (#1+0.35,#2-0.35);
\node at (#1,#2) {\scriptsize$k$};
}
\newcommand{\Measurement}[2]{%
\draw[very thick] (#1-0.5,#2-0.5)--(#1,#2)--(#1+0.5,#2-0.5);
\draw[fill=orange!30,draw=black,very thick] (#1,#2) circle (0.35);
}

\definecolor{additiongreen}{RGB}{0,125,55}

\definecolor{newversionblue}{RGB}{0,70,190}

\makeatother

\begin{document}
\title{Mixed State Deep Thermalization under Generic Quantum Errors}
\author{Xie-Hang Yu}
\affiliation{Department of Computing and Mathematical Sciences,
California Institute of Technology, Pasadena, CA 91125, USA}
\author{Pavel Kos}
\affiliation{Department of Physics, Faculty of Mathematics and Physics, University of Ljubljana, Jadranska 19, SI-1000 Ljubljana, Slovenia}
\date{\today}
\begin{abstract}
Pure-state deep thermalization predicts that, after sufficient scrambling, projected ensembles approach universal distributions that contain information beyond the reduced density matrix. In realistic quantum systems, however, noise is unavoidable and the projected states are generally mixed. Whether mixed-state deep thermalization admits an analogous universal limiting distribution has remained unclear.
Here we answer this question for quantum dynamics containing local quantum channels, i.e., errors. We show that the late-time mixed-state projected ensemble is completely determined by the normalized Choi state of the error channel and is universal, independent of the microscopic details of the underlying unitary dynamics. We establish this result analytically for dual-unitary circuits, global Haar-random dynamics, and local Haar-random circuits at large local dimension, and verify it numerically in qubit circuits and kicked Ising dynamics. 
In a solvable setting, we further find exponentially fast convergence to the limiting ensemble. Our results extend deep thermalization to noisy quantum dynamics and establish a direct correspondence between quantum errors and universal random mixed-state ensembles.
\end{abstract}
\maketitle

\section{Introduction}

Which properties of a local quantum error remain observable
after chaotic evolution has spread it through a many-body
system? We address this question through states conditioned on
measurements of the surroundings.  For a retained region $A$, measuring
its complement $B$ gives conditional states $\rho_{N_A}(\alpha)$ with Born
probabilities $P_\alpha$. Their average is the reduced density matrix
used in conventional thermalization~\cite{DAlessioEtAl2016ETH,VidmarRigol2016GGE},
whereas higher moments of the ensemble resolve fluctuations among
outcomes.  We ask which data of the error channel determine this distribution when
local noise acts during the evolution and every final measurement
outcome on $B$ is retained.

Modern quantum simulators can read out most of a many-body system
in each experimental run and record the full set of measurement
outcomes. For a pure global state and rank-one projective measurements
on $B$, the conditional states are pure, and the projected ensemble
$\mathcal{E}=\{P_{\alpha},\ket{\psi(\alpha)}_{A}\}$ probes the
distribution of wave functions in the Hilbert space and
contains information about quantum information scrambling,
including whether quantum correlations between subsystems are visible
to measurements~\cite{SherryRoy2024MeasurementInvisible}.
For generic chaotic dynamics, the
projected ensemble approaches a limiting distribution which depends
only on coarse-grained properties such as symmetries and conserved
charges. This form of quantum equilibration is known as deep
thermalization~\cite{HoChoi2022Exact,ClaeysLamacraft2022Biunitarity,CotlerEtAl2023Designs,IppolitiHo2023Purification,MarkEtAl2024MaximumEntropy,MarkEndresIppolitiHoChoi2026Review}.
Haar statistics emerge under suitable scrambling and measurement
conditions~\cite{WilmingRoth2022HighTemperature}, whereas conservation
laws, symmetries, and kinetic constraints select charge-resolved,
Scrooge-type, or otherwise restricted
ensembles~\cite{ChangShrotriyaHoIppoliti2025Charge,MokHaugHoPreskill2026NatureStingy,VarikutiBandyopadhyay2024Symmetry,MannaRoySreejith2025ConservedCharges,BhoreDesaulesPapic2023Constrained},
with Gaussian analogues for free fermions and continuous-variable
bosons~\cite{LucasEtAl2023FreeFermions,LiuHuangHo2024Gaussian}.
Constraints on quantum resources such as coherence can even prevent
deep thermalization while the reduced state remains
thermal~\cite{LiuIppolitiHo2026CoherenceTransition,FengLiuChengHoIppoliti2026ResourceLocalizability}.

Projected ensembles have been measured in Rydberg and
superconducting simulators~\cite{ChoiEtAl2023PreparingRandom,YanEtAl2026Superconducting},
and emergent randomness has been used for benchmarking, random-state
generation, and noise
learning~\cite{MarkEtAl2023BenchmarkingErgodic,ZhangEtAl2025Holographic,ShawEtAl2025NoiseLearning,MarkEndresIppolitiHoChoi2026Review}.
These applications, as well as shadow tomography from emergent
state designs~\cite{McGinleyFava2023ShadowTomography} and the
conversion of classical into quantum randomness by chaotic
dynamics~\cite{MokEtAl2025ClassicalRandomness}, rely on knowing which
ensemble the dynamics produces.
However, experimental systems are inevitably subject to incoherent
errors, which render the states in the projected ensemble mixed and
raise the question of whether deep thermalization persists in noisy
dynamics. In our previous work~\cite{YuHoKos2025Mixed}, we introduced
the mixed-state projected ensemble (MSPE) by considering incomplete
measurements on $B$, which can equivalently be interpreted as completely depolarizing
channels acting during or after the dynamics.  We found that the MSPE approaches
a generalized Hilbert--Schmidt ensemble characterized by a single
parameter that encodes the number of errors.
Initially mixed states evolving unitarily provide a different
setting, whose limiting ensemble retains the initial density-matrix
spectrum~\cite{SherryRoy2025Mixed}.
Partial and observable-projected ensembles also generate mixed
conditional states through incomplete bath
measurements~\cite{MandalClaeysRoy2025PartialProjected,SherryMandalRoy2025InformationPhases,MilekhinMurciano2025ObservableProjected, HuangMokHaugHo2026ProjectedKraus}.

These results leave open the case of a generic quantum error.
For example, partial depolarization, dephasing, and amplitude damping
retain different information about the input state and cannot in general
be described only by a single parameter, like the purity of the output.
In the weak-noise limit, the bit-string
statistics~\cite{ChristopoulosChanDeLuca2025UniversalOverlaps} of noisy random
circuits become nearly independent of the specific noise
channel~\cite{SauliereLamiBoyerDeNardisDeLuca2026NoisyAnticoncentration}.
It is therefore not clear which random-density-matrix ensemble should
emerge after scrambling, or how much microscopic information about the
error remains at late times.

In this work, we study mixed-state deep thermalization by inserting dilute local quantum channels into the bulk dynamics while retaining all measurement outcomes on $B$. Unlike protocols based on incomplete measurements, mixedness is therefore generated entirely by the physical error channel. We find that, after sufficient scrambling, the information that determines the MSPE is encoded only in the normalized Choi state of the channel. Because the errors are local, this information remains finite and serves as macroscopic data characterizing the late-time ensemble.
The resulting limiting distribution is a Born-weighted normalized correlated Wishart ensemble, which we refer to as the \emph{channel-induced ensemble (CIE)}. It is universal in the sense that, for a fixed error channel, it is independent of the microscopic details of the unitary dynamics and of the initial pure state. Different error channels can nevertheless select different limiting ensembles, with the Choi states of the error channels serving as the relevant local data retained after deep thermalization.

A unitary error recovers the pure-state Haar ensemble, while
complete depolarization reproduces our earlier generalized
Hilbert--Schmidt result~\cite{YuHoKos2025Mixed}.
Partial depolarization interpolates between these limits; general CPTP
maps select other Choi spectra.
The approach to the limiting law contains additional information:
in the solvable setting, channels with equal Choi spectra have the
same limit, but their transient corrections can depend on the Choi
state's input--output structure. Under the replica-continuation
assumptions stated in the derivation, we obtain exponential convergence.
Its moment-dependent prefactors distinguish a relaxation rate from
the time needed to attain a given accuracy, consistent with the
broader distinction between design times and asymptotic
rates~\cite{IppolitiHo2022DistinctDesignTimes,IppolitiHo2023Purification,ShrotriyaHo2025Nonlocality,MandalSherryRoy2026Locality,HerrmannFritzschBacker2026Timescales}.

We analytically derive the convergence to CIE  in dual-unitary
circuits~\cite{BertiniKosProsen2019Correlations,BertiniClaeysProsen2026SpaceTimeReview}. Dual unitarity has also been extended to noisy circuits in which every
gate is replaced by a channel, at the cost of additional space--time
unitality constraints~\cite{KosStyliaris2023SpaceTimeChannels}; here,
by contrast, the errors are dilute and otherwise generic. The same
ensemble follows for global Haar dynamics and for local Haar-random circuits at large local
dimension.
Numerical simulations of qubit circuits and kicked Ising
dynamics away from dual unitarity further support the conjecture that this enable emerges in general.
Section~\ref{sec:setup} defines the ensemble and the main result;
Secs.~\ref{sec:dual-unitary}--\ref{sec:local-haar} present the analytical
derivations, Sec.~\ref{sec:numerics} gives the numerical tests, and
Sec.~\ref{sec:discussion} discusses extensions.

\section{The mixed-state projected ensemble and the main result}

\label{sec:setup}

\subsection{Mixed-state projected ensemble}

We consider a many-body system of qudits with local Hilbert-space
dimension $d$. At the final time, the system is divided into a retained
subsystem $A$, containing $N_{A}$ qudits and having dimension $D_{A}=d^{N_{A}}$,
and a measured complement $B$ containing $N_B$ qudits and having dimension $D_B=d^{N_B}$. We assume that the measurement of $B$ is complete and that every outcome is retained.

The system is initialized in a pure state and undergoes unitary evolution interspersed with a dilute set of local error channels, $\{\mathcal{N}_1,\mathcal{N}_2,\ldots\}$. Consequently, the final state is generally mixed and is denoted by $\rho(t)$.
Let $\{\ket{\alpha}_{B}\}$ be a complete orthonormal product basis
on $B$ (a product over sites, or over pairs of sites in Sec.~\ref{sec:dual-unitary}). 
The unnormalized mixed state on $A$ conditioned on outcome $\alpha$
is 
\begin{equation}
\widetilde{\rho}_{N_{A}}(\alpha)={}_{B}\!\braket{\alpha|\rho(t)|\alpha}_{B}.\label{eq:unnormalized-conditional-state}
\end{equation}
Its Born probability and normalized conditional state are 
\begin{equation}
P_{\alpha}=\tr\widetilde{\rho}_{N_{A}}(\alpha),\qquad\rho_{N_{A}}(\alpha)=\frac{\widetilde{\rho}_{N_{A}}(\alpha)}{P_{\alpha}}.\label{eq:probability-and-state}
\end{equation}
The MSPE is therefore given by
\begin{equation}
\MSPE=\{P_{\alpha},\rho_{N_{A}}(\alpha)\}_{\alpha},\label{eq:mspe-definition}
\end{equation}
and its $k$th moment is 
\begin{equation}
\rho^{(k)}_{N_{A}}=\sum_{\alpha}P_{\alpha}\rho^{\otimes k}_{N_{A}}(\alpha)=\sum_{\alpha}P^{1-k}_{\alpha}\widetilde{\rho}^{\otimes k}_{N_{A}}(\alpha).\label{eq:mspe-moment}
\end{equation}
The first moment is the reduced density matrix on $A$. The higher
moments distinguish ensembles that have the same average state but
different fluctuations and eigenvalue statistics.

For a candidate reference ensemble with $k$th moment operator
$\rho_{\mathrm{ref}}^{(k)}$, we quantify the deviation of the MSPE
from the reference ensemble using the normalized Schatten distance
\begin{equation}
\Delta_{\xi}^{(k)}
=
\frac{
\left\|
\rho_{N_A}^{(k)}-\rho_{\mathrm{ref}}^{(k)}
\right\|_{\xi}
}{
\left\|
\rho_{\mathrm{ref}}^{(k)}
\right\|_{\xi}
}.
\label{eq:moment-distance}
\end{equation}
Here $\|O\|_\xi=(\tr[(O^\dagger O)^{\xi/2}])^{1/\xi}$ is the Schatten norm.
We say that the MSPE deeply
thermalizes to the reference ensemble if, for every fixed $k$,
$\Delta_{\xi}^{(k)}$ vanishes in the late-time limit. The decay of $\Delta_\xi^{(k)}$ with time indicates the rate of approach to the reference ensemble.
For $\xi=1$, $\Delta_1^{(k)}/2$ is the trace distance between the two moment states and serves as a natural measure of how well they can be distinguished by measurements.
The Hilbert--Schmidt distance, $\xi=2$, is also widely used because it can be calculated directly from moment overlaps, as in the local Haar analysis of Sec.~\ref{sec:local-haar}.

\subsection{The channel-induced Choi--Wishart reference ensemble}

Here we introduce the relevant reference ensemble for the MSPE.
Let 
\begin{equation}
\ket{\phi_{0}}=\frac{1}{\sqrt{d}}\sum^{d}_{i=1}\ket{i}\ket{i}\label{eq:maximally-entangled-state}
\end{equation}
and define the normalized Choi state of the local channel by~\cite{Choi1975CompletelyPositive}
\begin{equation}
M_{\Ncal_i}=(\operatorname{id}\otimes\Ncal_i)\left(\ket{\phi_{0}}\!\bra{\phi_{0}}\right).\label{eq:normalized-choi-state}
\end{equation}
For a CPTP map, $M_{\Ncal_i}\geq0$ and $\tr M_{\Ncal_i}=1$. It acts
on an auxiliary space $C_i=C_{i}^1\otimes C_{i}^2$ of dimension $d^{2}$,
corresponding to the input and output indices of the channel.
For several error channels, we introduce the joint Choi state
\begin{equation}
    M=\bigotimes_i M_{\Ncal_i}.
\end{equation}
Pictorially, this can be expressed as
\[
\begingroup
\definecolor{choiInk}{HTML}{222222}
\definecolor{choiFrame}{HTML}{777777}
\definecolor{choiChannel}{HTML}{E7E7E7}
\definecolor{choiKet}{HTML}{9B565B}
\definecolor{choiBra}{HTML}{62527F}
\newcommand{\channelChoiCopy}[2]{%
  \begin{scope}[shift={(#1,0)}]
    \draw[choiFrame,line width=0.45pt,dash pattern=on 2pt off 1.6pt]
      (0,0) rectangle (118,148);
    \node[anchor=south east,font=\normalsize] at (118,157)
      {$M_{\mathcal N_{#2}}$};
    \draw[fill=choiChannel,line width=0.75pt]
      (48,38)--(48,51)--(78,51)--(78,97)
      --(48,97)--(48,110)--(104,110)--(104,38)--cycle;
    \node[font=\normalsize] at (91,74) {$\mathcal N_{#2}$};
    \draw[choiKet] (67,-10)--(67,38);
    \draw[choiKet] (67,51)--(67,61)
      ..controls (67,77) and (19,75)..(19,49)--(19,-10);
    \draw[choiBra] (67,110)--(67,158);
    \draw[choiBra] (67,97)--(67,87)
      ..controls (67,71) and (19,73)..(19,99)--(19,158);
    \node[anchor=south] at (19,175) {$C_{#2}^{1}$};
    \node[anchor=south] at (67,175) {$C_{#2}^{2}$};
  \end{scope}%
}
\begin{tikzpicture}[x=0.55pt,y=0.55pt,draw=choiInk,text=choiInk,
  line width=0.75pt,line cap=round,line join=round,
  every node/.style={inner sep=0pt,font=\small}]
  \node[font=\normalsize] at (-35,74) {$M=$};
  \channelChoiCopy{0}{1}
  \node[font=\normalsize] at (142,74) {$\otimes$};
  \channelChoiCopy{168}{2}
  \node[font=\normalsize] at (313,74) {$\otimes$};
  \node[font=\normalsize] at (342,74) {$\cdots$};
\end{tikzpicture}%
\endgroup
,
\]
which acts on the space $C=\bigotimes_i C_i$ of dimension $D_C=d^{2n}$ for $n$ channels. 

Let $\ket{\psi_X}$ be Haar distributed on $A\otimes C$, and arrange its
coefficients into a rectangular $D_A\times D_C$ matrix $X$:

\begin{equation}
\ket{\psi_X}
=
\frac{
\sum_{a=1}^{D_A}
\sum_{\sigma=1}^{D_C}
X_{a\sigma}\ket{a}_A\ket{\sigma}_C
}{
\sqrt{\operatorname{Tr}(XX^\dagger)}
}.
\label{eq:coefficient-matrix}
\end{equation}
Here, $X$ is drawn from a scaled complex Ginibre ensemble, whose entries
are independent and identically distributed centered circular complex
Gaussian random variables satisfying
\begin{equation}
\mathbb{E}[X_{a\sigma}]=0,
\qquad
\mathbb{E}\!\left[
X_{a\sigma}X_{b\tau}^{*}
\right]
=
\frac{1}{D_A}\delta_{ab}\delta_{\sigma\tau}.
\label{eq:ginibre-covariance}
\end{equation}
In particular, each column of $X$ has unit norm on average, or,
equivalently,
\begin{equation}
\mathbb{E}\!\left[X^\dagger X\right]=I_C.
\end{equation}
We denote the corresponding normalized probability measure by
$d\mu_G(X)$.

For a fixed positive-semidefinite, unit-trace matrix $M$ on
$C$, we define
\begin{equation}
\begin{aligned}
\widetilde{\rho}_{M}(X)
&=
X M X^\dagger,\\
P_{M}(X)
&=
\operatorname{Tr}\widetilde{\rho}_{M}(X),\\
\rho_{M}(X)
&=
\frac{\widetilde{\rho}_{M}(X)}{P_{M}(X)}.
\end{aligned}
\label{eq:correlated-wishart-state}
\end{equation}
Pictorially, we have
\begin{equation}
\scalebox{0.85}{\begingroup
\definecolor{wishartInk}{HTML}{222222}
\definecolor{wishartKet}{HTML}{E9B0B2}
\definecolor{wishartBra}{HTML}{D5CBED}
\definecolor{wishartKetLeg}{HTML}{9B565B}
\definecolor{wishartBraLeg}{HTML}{62527F}
\definecolor{wishartMatrix}{HTML}{E7E7E7}
\begin{tikzpicture}[x=0.82pt,y=0.9pt,draw=wishartInk,text=wishartInk,
  baseline=(current bounding box.center),line width=0.75pt,
  line cap=round,line join=round,
  every node/.style={inner sep=0pt,font=\normalsize}]
  \node[anchor=east] at (-15,60) {$\widetilde{\rho}_M(X)=$};
  \foreach \x in {15,33,51,119,137,155}{
    \draw[wishartKetLeg] (\x,106)--(\x,118);
    \draw[wishartBraLeg] (\x,14)--(\x,2);
  }
  \node at (85,112) {$\cdots$};
  \node at (85,8) {$\cdots$};
  \foreach \x in {72,98}{
    \draw[wishartKetLeg] (\x,84)--(\x,72);
    \draw[wishartBraLeg] (\x,48)--(\x,36);
  }
  \draw[fill=wishartKet] (0,84) rectangle (170,106);
  \node at (85,95) {$X$};
  \draw[fill=wishartMatrix] (58,48) rectangle (112,72);
  \node at (85,60) {$M$};
  \draw[fill=wishartBra] (0,14) rectangle (170,36);
  \node at (85,25) {$X^\dagger$};
\end{tikzpicture}%
\endgroup
}.
\end{equation}
The corresponding Born-weighted probability measure is
\begin{equation}
d\nu_{M}(X)
=
P_{M}(X)d\mu_G(X),
\label{eq:born-weighted-measure}
\end{equation}
which is normalized.
We denote the resulting ensemble by 
\begin{equation}
\begin{aligned}
\mathcal{E}_{M}
&=
\left\{
d\nu_{M}(X),\rho_{M}(X)
\right\}\\
&=
\left\{
P_{M}(X)d\mu_G(X),\rho_{M}(X)
\right\}.
\end{aligned}
\label{eq:choi-ensemble-definition}
\end{equation}
Its $k$th moment operator is equivalently given by
\begin{equation}
\begin{aligned}
\rho_{M}^{(k)}
&=
\int d\nu_{M}(X)\,
\rho_{M}(X)^{\otimes k}\\
&=
\int d\mu_G(X)\,
P_{M}(X)^{1-k}
\left(
\widetilde{\rho}_{M}(X)
\right)^{\otimes k}.
\end{aligned}
\label{eq:choi-ensemble-moment}
\end{equation}
We refer to $\mathcal{E}_{M}$ as the CIE.
Algebraically, it is a Born-weighted normalized correlated Wishart
ensemble. When $P_{M}(X)$ is constant or sharply concentrated around
its mean value, the Born weighting becomes immaterial, and the
conventional normalized correlated Wishart ensemble is recovered.

\begin{proposition}[Mixed-State Deep Thermalization (informal)] After sufficient scrambling, dilute local quantum errors do not destroy deep thermalization; instead, they select the limiting ensemble. More precisely, the late-time MSPE deeply thermalizes to the CIE $\mathcal{E}_{M}$, which is determined solely by the normalized Choi state $M$ of the error process and is independent of the microscopic details of the underlying unitary dynamics. \label{prop:central} \end{proposition}

Proposition~\ref{prop:central} states that, for fixed $M$, the
limiting distribution is independent of both the microscopic realization
of the unitary dynamics and the precise space--time location of the error
channel. In this sense, the limiting distribution is universal: the
error process enters only through its normalized Choi state. Moreover,
one can show that the limiting distribution depends on $M$ only
through its spectrum.
Indeed, writing $M=V\Lambda V^\dagger$ with $V$ unitary and $\Lambda$ diagonal, invariance of the Ginibre measure under $X\mapsto XV$ shows that $XMX^\dagger$, including its trace weight, has the same distribution as $X\Lambda X^\dagger$.

The physical picture is simple. The unitary evolution first produces a single many-body state. An ensemble arises after measuring the complementary subsystem $B$: for sufficiently scrambling dynamics, the conditional states on $A$ sample the Haar ensemble, as in pure-state deep thermalization.
A local quantum channel deforms this Haar ensemble in a specific way. Scrambling makes the random amplitudes universal, represented by a Ginibre matrix $X$, while the information about the channel survives through its normalized Choi state $M$. The resulting unnormalized conditional state is therefore
$
\widetilde{\rho}_M(X)=XMX^\dagger,
$
with the corresponding Born weight
$
P_M(X)=\operatorname{Tr}(XMX^\dagger).
$
Thus measurements provide the universal Haar randomness, while the channel determines how these random components are mixed through its Choi state. 

This result connects familiar pure- and mixed-state
ensembles. If $\Ncal$ is a unitary channel, its normalized Choi state
$M$ is pure and therefore has a single nonzero eigenvalue equal
to unity. In this case, the correlated Wishart matrix has rank one, and
$\mathcal{E}_{M}$ reduces to the pure-state Haar ensemble. In
the opposite limit, if $\Ncal$ is the completely depolarizing channel,
\begin{equation}
\Ncal(\rho)
=
\frac{\tr\rho}{d}I_d,
\qquad
M
=
\frac{I_{d^{2}}}{d^{2}},
\label{eq:complete-depolarization}
\end{equation}
then Eq.~\eqref{eq:correlated-wishart-state} reduces to an ordinary
induced random density matrix~\cite{ZyczkowskiSommers2001Induced,SommersZyczkowski2004Statistical,ZyczkowskiEtAl2011Generating} and reproduces the generalized
Hilbert--Schmidt ensemble introduced in Ref.~\cite{YuHoKos2025Mixed}.

In the remainder of this paper, we establish Proposition~\ref{prop:central} analytically in several solvable models and test it numerically in generic chaotic dynamics.
We first consider a single local channel and its projected ensemble.
We then consider multiple channels separated by long scrambling regions.
In this dilute-error limit, their permutation weights (defined in Sec.~\ref{sec:dual-unitary}) factorize, so the effective Choi state is the tensor product of the individual Choi states,
\begin{equation}
M
=
\bigotimes_i M_{\Ncal_i}.
\end{equation}

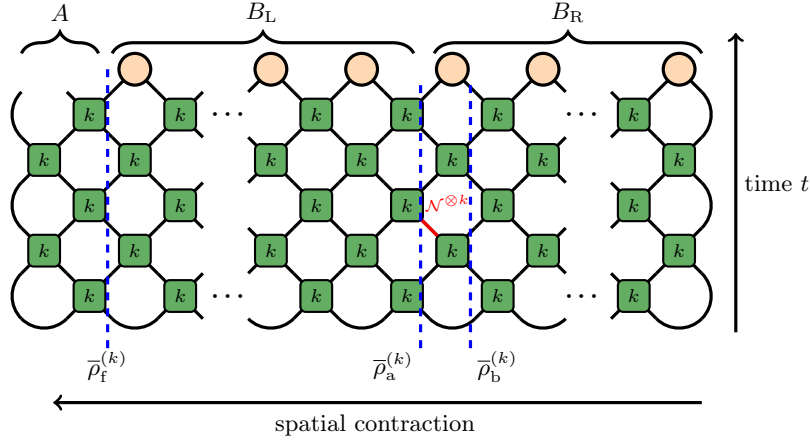
\begin{figure*}
\centering
\begin{tikzpicture}[scale=0.60]
\foreach \x in {0,2,5,7,9,12}{
  \foreach \y in {0,2,4}{\Wgategreen{\x}{\y}}
}
\foreach \x in {-1,1,4,6,8,10,13}{
  \foreach \y in {1,3}{\Wgategreen{\x}{\y}}
}
\foreach \x in {1,4,6,8,10,13}{\Measurement{\x}{5}}
\foreach \x in {-2,0,3,5,7,9,12}{
  \draw[very thick] (\x+0.5,-0.5) to[out=-45,in=225] (\x+1.5,-0.5);
}
\foreach \y in {-1,1,3}{
  \draw[very thick] (13.5,\y+0.5) to[out=45,in=315] (13.5,\y+1.5);
  \draw[very thick] (-1.5,\y+0.5) to[out=135,in=225] (-1.5,\y+1.5);
}
\node at (3.1,0) {$\bm{\cdots}$};
\node at (3.1,4) {$\bm{\cdots}$};
\node at (10.9,0) {$\bm{\cdots}$};
\node at (10.9,4) {$\bm{\cdots}$};
\draw[decorate,decoration={brace,amplitude=8pt},very thick]
  (0.5,5.35)--(7.15,5.35);
\draw[decorate,decoration={brace,amplitude=8pt},very thick]
  (7.6,5.35)--(13.5,5.35);
\draw[decorate,decoration={brace,amplitude=8pt},very thick]
(-1.5,5.35)--(0.2,5.35);
\node at (-0.65,6.25) {$A$};
\node at (3.8,6.25) {$B_{\mathrm L}$};
\node at (10.55,6.25) {$B_{\mathrm R}$};
\draw[very thick,red] (7.8,1.2)--(7.3,1.7);
\foreach \x/\y in {7/2,8/1}{
  \draw[thick,fill=mygreen!70,rounded corners=2pt]
    (\x-0.35,\y+0.35) rectangle (\x+0.35,\y-0.35);
  \node at (\x,\y) {\scriptsize$k$};
}
\draw[very thick,dashed,blue] (0.4,-1.15)--(0.4,5.0);
\node at (0.4,-1.55) {$\overline{\rho}_{\mathrm f}^{(k)}$};
\draw[very thick,dashed,blue] (7.3,-1.15)--(7.3,4.8);
\node at (6.7,-1.55) {$\overline{\rho}_{\mathrm a}^{(k)}$};
\node[red,font=\fontsize{6}{7}\selectfont] at (7.90,2.05) {$\Ncal^{\otimes k}$};
\draw[very thick,dashed,blue] (8.4,-1.15)--(8.4,4.8);
\node at (9.0,-1.55) {$\overline{\rho}_{\mathrm b}^{(k)}$};
\draw[->,very thick] (14.25,-0.8)--(14.25,5.8);
\node[right] at (14.25,2.5) {time $t$};
\draw[->,very thick] (13.5,-2.35)--(-0.8,-2.35);
\node at (6.3,-2.85) {spatial contraction};
\end{tikzpicture}
\caption{Spatial contraction of the folded $k$-replica dual-unitary
network, which gives rise to the ensemble raw moments $\overline{\rho}_{N_A}^{(k)}$. Green blocks represent $(U\otimes U^*)^{\otimes k}$; orange
circles denote replicated maximally entangled-basis measurement tensors,
with the retained outcome shared across replicas and summed over; and the
red segment denotes $\Ncal^{\otimes k}$. Contraction proceeds from right
to left through $B_{\mathrm R}$ and $B_{\mathrm L}$. Blue dashed lines
mark the raw tensors $\overline{\rho}_{\mathrm b}^{(k)}$,
$\overline{\rho}_{\mathrm a}^{(k)}$, and
$\overline{\rho}_{\mathrm f}^{(k)}$ before the
channel, after the channel, and after the second measured region,
respectively. The vertical arrow shows the original circuit-time
direction.}
\label{fig:du-spatial-contraction}
\end{figure*}

\section{Exactly solvable dual-unitary circuits}
\label{sec:dual-unitary}

We now establish Proposition~\ref{prop:central} in an exactly solvable
one-dimensional circuit. The calculation isolates the two ingredients
that determine the limiting ensemble: scrambling projects the replicated
dynamics onto permutation sectors, while the local error assigns a
Choi-dependent weight to each sector. We first consider one local channel
and return to well-separated channels at the end of the section.

\subsection{Brick-wall circuits and dual-unitary gates}

We consider a small modification of the exactly solvable protocol introduced in
Ref.~\cite{YuHoKos2025Mixed}, as illustrated in
Fig.~\ref{fig:du-spatial-contraction}. 
A chain of qudits is initialized
in a product of maximally entangled pairs and evolved under a
depth-$t$ brick-wall circuit. The complementary subsystem $B$ is
measured pairwise in a maximally entangled basis, represented by the
orange circles in the figure, and all measurement outcomes are
retained. A single-site CPTP map $\Ncal$, represented by the red leg
in Fig.~\ref{fig:du-spatial-contraction}, is inserted between the two
measured regions $B_{\mathrm L}$ and $B_{\mathrm R}$. We assume the
absence of conservation laws and focus on the regime
$t+1\geq N_A$.

For a two-qudit gate $U$, we define
$U_{ij}^{kl}=\bra{k,l}U\ket{i,j}$ and introduce the reshuffled gate
$\widetilde U$ through
$(\widetilde U)_{ik}^{jl}=U_{ij}^{kl}$. The gate $U$ is dual unitary
if both $U$ and $\widetilde U$ are
unitary~\cite{BertiniKosProsen2019Correlations,BertiniClaeysProsen2026SpaceTimeReview}. Since the projected ensemble
is characterized by its higher moments, as defined in
Eqs.~\eqref{eq:mspe-moment} and \eqref{eq:choi-ensemble-moment}, we
consider $k$ replicas of the evolution. In the folded $k$-replica
tensor network, the forward and backward evolutions are paired
together. Each green gate $\begin{tikzpicture}[scale=0.60]
    \Wgategreen{0}{0}
\end{tikzpicture}$ in
Fig.~\ref{fig:du-spatial-contraction} therefore represents
$(U\otimes U^*)^{\otimes k}$ in the temporal direction, which is equivalent to $(\widetilde U\otimes\widetilde U^*)^{\otimes k}$
in the spatial direction.

We draw the gates from an open set of dual-unitary gates whose
brick-wall products are dense in the unitary group~\cite{SuzukiMitaraiFujii2022Computational,IppolitiHo2023Purification,YuHoKos2025Mixed}.
First, at fixed $t$ and positive integer moment order, we take the measured regions $B_{\mathrm L}$ and $B_{\mathrm R}$ to be infinitely long.
This projects the spatial transfer matrix onto the permutation subspace defined below.
We then take $t\to\infty$ at fixed $d$ and $N_A$.
The vertical time-like cut contains $t+1$ qudits and has dimension
$d^{t+1}$. The normalized
overlap of distinct permutation operators is at most
$d^{-(t+1)}$ (cf. Eq.~\eqref{eq:du-permutation-gram}). Thus the second limit, $t\to\infty$, removes the
corrections due to their nonzero overlaps.
For finite measured regions, the exact projection is replaced
by a finite power of the spatial transfer matrix. The finite-$B$
corrections are therefore
controlled by how quickly this power approaches the permutation
projector. For solvable noise-free dual-unitary circuits,
Ref.~\cite{ArvindHunterJones2026FiniteBath} derives finite-bath
design bounds under quantitative local-mixing, backflow, and
boundary-contraction assumptions, which apply here.
We also test convergence of the MSPE moments to the CIE moments
numerically at finite $B$ in the qubit circuits of
Sec.~\ref{sec:numerics}.

\subsection{Replica construction and Born weights}
\label{sec:du-replica-construction}

\begin{figure}[t]
\centering
\resizebox{\columnwidth}{!}{\begingroup
\definecolor{replicaKet}{HTML}{E9B0B2}
\definecolor{replicaBra}{HTML}{D5CBED}
\definecolor{replicaKetLeg}{HTML}{9B565B}
\definecolor{replicaBraLeg}{HTML}{62527F}
\definecolor{replicaInk}{HTML}{1C1C1C}

\newcommand{\replicaStateCopy}[3]{%
  \begin{scope}[shift={(#1,#2)}]
    \draw[fill=replicaBra] (20,-20) rectangle (164,60);
    \foreach \leg in {28,72,116}{
      \draw[replicaBraLeg] (\leg+20,-20)--(\leg+20,-52);
    }
    \draw[fill=replicaKet] (0,0) rectangle (144,80);
    \foreach \leg in {28,72,116}{
      \draw[white,line width=4.8pt,line cap=butt]
        (\leg,0)--(\leg,-32);
      \draw[replicaKetLeg] (\leg,0)--(\leg,-32);
    }
    \ifnum#3=1\relax
      \foreach \leg in {28,72,116}{
        \draw[white,line width=4.8pt,line cap=butt]
          (\leg,-32)..controls (\leg,-73) and (\leg+20,-86)
          ..(\leg+20,-52);
        \draw (\leg,-32)..controls (\leg,-73) and (\leg+20,-86)
          ..(\leg+20,-52);
      }
    \fi
  \end{scope}%
}

\begin{tikzpicture}[x=1pt,y=-1pt,draw=replicaInk,text=replicaInk,
  line width=1.5pt,line cap=round,line join=round,
  every node/.style={inner sep=0pt,font=\normalsize}]
  \path[use as bounding box] (50,0) rectangle (672,472);

  \foreach \x/\y in {330/197.333,636/95.333}{
    \foreach \offset in {0,10,20}{
      \fill[replicaInk] (\x+\offset,\y-\offset/3)
        circle[radius=1.35pt];
    }
  }

  \replicaStateCopy{458}{120}{1}
  \replicaStateCopy{368}{150}{1}
  \replicaStateCopy{158}{220}{0}
  \replicaStateCopy{68}{250}{0}

  \draw[decorate,decoration={brace,mirror,amplitude=8pt},line width=1.4pt]
    (61,395)--(319,309);
  \node[scale=2.3] at (197,376) {$k$};
  \node[scale=1.7] at (203,397) {open};
  \draw[decorate,decoration={brace,mirror,amplitude=8pt},line width=1.4pt]
    (361,295)--(649,199);
  \node[scale=2.3] at (512,271) {$q$};
  \node[scale=1.7] at (518,292) {traced};

  \draw[->,>=stealth,line width=1.7pt] (88,463)--(664,271);
  \node[rotate=18.435,scale=2.2]
    at (388,403) {$r=k+q$ replicas};

  \draw[fill=replicaKet,line width=1.1pt] (68,65) rectangle (92,78);
  \node[anchor=base west,scale=1.8] at (103,77) {ket};
  \draw[fill=replicaBra,line width=1.1pt] (146,65) rectangle (170,78);
  \node[anchor=base west,scale=1.8] at (181,77) {bra};
\end{tikzpicture}%
\endgroup}
\caption{Replica construction with $r=k+q$ copies, shown in order from the lower left to the upper right, where $q\geq0$ is an integer. Each replica represents one copy of the conditional states $\widetilde\rho_{N_A}(\alpha)$, with ket and bra indices shown separately in pale red and pale purple. The first $k$ copies keep their retained-region indices open. In the other $q$ copies, loops connect corresponding ket and bra indices to take the trace in Eq.~\eqref{eq:appendix-raw-spectator-trace}. All copies share the same outcome $\alpha$.}
\label{fig:replica-construction}
\end{figure}
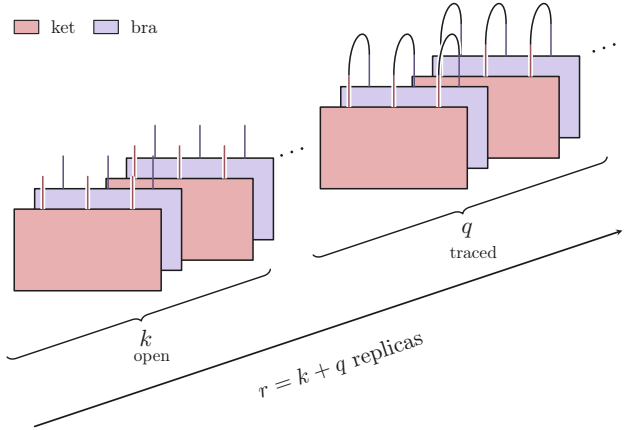

For $k>1$, the physical moment in Eq.~\eqref{eq:mspe-moment} contains the negative power $P_\alpha^{1-k}$, which cannot be represented by a contraction of $k$ copies alone. We therefore first contract a positive integer number of copies and recover the Born weight by replica continuation, see Fig.~\ref{fig:replica-construction} for an illustration. We take $r=k+q$ copies with an integer $q\geq0$, where all the copies share the same outcome $\alpha$. The first $k$ copies remain uncontracted, while the other $q$ copies are traced out
~\cite{IppolitiHo2023Purification,ChanDeLuca2024Projected,YuHoKos2025Mixed}.

At integer order $r$, we calculate the following trace-normalized raw moments of the circuit and reference ensembles, defined as
\begin{equation}
\begin{aligned}
\overline\rho_{N_A}^{(r)}
&=\frac{\displaystyle\sum_\alpha
\widetilde\rho_{N_A}(\alpha)^{\otimes r}}
{\displaystyle\sum_\alpha P_\alpha^r},\\
\overline\rho_{M_{\Ncal}}^{(r)}
&=\frac{\displaystyle\int d\mu_G(X)\,
\widetilde\rho_{M_{\Ncal}}(X)^{\otimes r}}
{\displaystyle\int d\mu_G(X)\,P_{M_{\Ncal}}(X)^r}.
\end{aligned}
\label{eq:du-raw-replica-tensors}
\end{equation}
Both operators are positive and have unit trace, but they are generally
different from the physical Born-weighted moments.

Tracing the last $q$ copies gives
\begin{equation}
\begin{aligned}
\tr_{k+1,\ldots,r}\overline\rho_{N_A}^{(r)}
&=\frac{\displaystyle\sum_\alpha P_\alpha^q
\widetilde\rho_{N_A}(\alpha)^{\otimes k}}
{\displaystyle\sum_\alpha P_\alpha^{k+q}},\\
\tr_{k+1,\ldots,r}\overline\rho_{M_{\Ncal}}^{(r)}
&=\frac{\displaystyle\int d\mu_G(X)P_{M_{\Ncal}}(X)^q
\widetilde\rho_{M_{\Ncal}}(X)^{\otimes k}}
{\displaystyle\int d\mu_G(X)P_{M_{\Ncal}}(X)^{k+q}}.
\end{aligned}
\label{eq:appendix-raw-spectator-trace}
\end{equation}
Each traced copy gives one factor of $P_\alpha$ in the first identity, or $P_{M_{\Ncal}}(X)$ in the second.
Continuing to $q=1-k$ makes both denominators equal to one and gives the physical moments in Eqs.~\eqref{eq:mspe-moment} and \eqref{eq:choi-ensemble-moment}.
This gives the following direct consequence under the replica continuation used here.

\begin{lemma}[Replica continuation from raw moments]
\label{lem:raw-to-physical-moments}
Consider two ensembles of density matrices, each with unit total Born weight and finite, nonzero raw normalization. By analytic continuation, equality of their trace-normalized raw moments at all positive integer orders implies equality of their physical Born-weighted moments at any fixed order.
\end{lemma}

It is therefore sufficient to match Eq.~\eqref{eq:du-raw-replica-tensors} for every positive integer order.

\subsection{Permutation-space reduction}

We next evaluate the raw moments at positive integer replica number $k$. Here $k$ denotes a generic integer replica number; when continuing to the physical moments, it is replaced by $r=k+q$. The dual-unitary structure greatly simplifies the replicated dynamics. After contraction through a sufficiently long measured region, all information except the permutation structure of the $k$ replicas is lost. 
The replicated state can therefore be described by only $k!$ coefficients, one for each permutation $g\in S_k$. The role of the error channel will be to modify these coefficients, while the subsequent evolution and measurements preserve them.

Let $S_k$ be the symmetric group on the $k$ replicas. For $g\in S_k$,
define the permutation operator on $k$ replicas, each containing $L$
qudits, by
\begin{equation}
\rho_{L,k}(g)=
\sum_{i_1,\ldots,i_k=1}^{d^L}
\ket{i_{g(1)},\ldots,i_{g(k)}}\bra{i_1,\ldots,i_k}.
\label{eq:du-permutation-operator}
\end{equation}
Let $\mathcal C(g)$ denote the set of disjoint cycles of $g$, and let $|c|$ be the length of a cycle $c\in\mathcal C(g)$.
The quantity $|g|:=k-|\mathcal C(g)|$ is the minimum number of transpositions whose product is $g$.
The normalized Gram matrix is
\begin{equation}
G_L^{(k)}(g,g')
:=d^{-Lk}\tr\!\left[\rho_{L,k}(g)^\dagger\rho_{L,k}(g')\right]
=d^{-L|g^{-1}g'|}.
\label{eq:du-permutation-gram}
\end{equation}
Thus the permutation operators are nonorthogonal at finite $L$, but
their off-diagonal overlaps vanish exponentially at fixed $k$ as
$L\to\infty$.

Throughout this section, unbarred moments denote the physical Born-weighted moments, whereas an overbar denotes the trace-normalized raw moments introduced in Eq.~\eqref{eq:du-raw-replica-tensors}. When only relative permutation coefficients are relevant, we suppress their common normalization. The permutation operators $\rho_{L,k}(g)$ from Eq.~\eqref{eq:du-permutation-operator} should not be confused with either of these moment tensors.

\begin{lemma}[Permutation-space reduction~\cite{YuHoKos2025Mixed}]
\label{lem:du-permutation-reduction}
Consider the above dual-unitary protocol with complete maximally
entangled-basis measurements and gates drawn from the open set specified
above. At fixed positive integer $k$, contraction through an infinitely
long measured region projects the replicated time-like tensor onto
$\operatorname{span}\{\rho_{t+1,k}(g):g\in S_k\}$. When $t+1\geq N_A$, the retained circuit in region $A$ induces a
map $W$ from the time-like cut to the output Hilbert space of $A$,
satisfying $WW^\dagger=\id_A$. Moreover, for every $g\in S_k$,
\begin{equation}
W^{\otimes k}\rho_{t+1,k}(g)(W^\dagger)^{\otimes k}
=\rho_{N_A,k}(g).
\label{eq:du-retained-map}
\end{equation}
Thus the retained region preserves every
permutation coefficient without mixing distinct sectors.
\end{lemma}

The picture is that the permutation-symmetric right boundary assigns equal coefficients to all permutation sectors in the raw moments before the error channel $\mathcal{N}$. The channel then modifies these sector weights, while the second measured region, $B_{\mathrm L}$, projects the resulting time-like raw tensor $\overline{\rho}_{\mathrm a}^{(k)}$ back onto the permutation subspace, giving $\overline{\rho}_{\mathrm f}^{(k)}$. Finally, the retained circuit in region $A$ maps $\overline{\rho}_{\mathrm f}^{(k)}$ to the physical output on $A$, which is the desired raw moment $\overline{\rho}_{N_A}^{(k)}$.

\subsection{Generic local channel and permutation weights}

We now figure out how the initially equal weights of different permutations are affected by a local quantum channel, making the weights $g$ dependent. 
Since the right boundary in Fig.~\ref{fig:du-spatial-contraction} factorizes across the replica copies, its overlap with $\rho_{t+1,k}(g)$ is independent of $g\in S_k$. Therefore, the time-like raw state before applying the channel can be written as
\begin{equation}
\overline{\rho}_{\mathrm b}^{(k)}
=\sum_{g\in S_k}\rho_{t+1,k}(g),
\end{equation}
up to a normalization constant.
We now evaluate how the channel changes these raw permutation
coefficients. Using
$\rho_{t+1,k}(g)=\rho_{t,k}(g)\otimes\rho_{1,k}(g)$, the time-like state
immediately after the channel is
\begin{equation}
\begin{aligned}
\overline{\rho}_{\mathrm a}^{(k)}
=
\sum_{g\in S_k}
\rho_{t,k}(g)\otimes
\Ncal^{\otimes k}\!\left(\rho_{1,k}(g)\right).
\end{aligned}
\label{eq:du-channel-cuts}
\end{equation}
After contracting through the second extended measured region
$B_{\mathrm L}$, Lemma~\ref{lem:du-permutation-reduction} implies that
the final time-like state takes the form
\begin{equation}
\overline{\rho}_{\mathrm f}^{(k)}
=
\sum_{g\in S_k}
\alpha_{\mathrm f}(g)\rho_{t+1,k}(g).
\label{eq:du-final-timelike-state}
\end{equation}
The coefficients $\alpha_{\mathrm f}(g)$ therefore determine the raw
integer-replica tensor. They are
fixed by the overlaps with the permutation sectors. Indeed, each
$\rho_{t+1,k}(g)$ is a fixed point of the spatial transfer matrix in
$B_{\mathrm L}$. Its overlap with the raw cut tensors,
$\overline{\rho}_{\mathrm a}^{(k)}$ and
$\overline{\rho}_{\mathrm f}^{(k)}$, is therefore unchanged
by the contraction through this region, yielding
\begin{equation}
\tr\!\left[
\rho_{t+1,k}(g)^\dagger\overline{\rho}_{\mathrm a}^{(k)}
\right]
=
\tr\!\left[
\rho_{t+1,k}(g)^\dagger\overline{\rho}_{\mathrm f}^{(k)}
\right],
\qquad
g\in S_k.
\label{eq:original_eq_fixed_alpha}
\end{equation}

For a fixed positive integer $k$, define the channel kernel
\begin{equation}
K_k(g,g')=\frac{1}{d^k}\tr\!\left[
\rho_{1,k}(g)^\dagger
\Ncal^{\otimes k}\!\left(\rho_{1,k}(g')\right)
\right].
\label{eq:du-channel-kernel}
\end{equation}
Factorizing Eq.~\eqref{eq:original_eq_fixed_alpha}
across the $t$ unaffected legs and the channel leg gives
\begin{equation}
\begin{aligned}
&\sum_{g'\in S_k}G_t^{(k)}(g,g')K_k(g,g')\\
&\qquad=\sum_{g'\in S_k}
G_{t+1}^{(k)}(g,g')\alpha_{\mathrm f}(g'),
\qquad g\in S_k.
\end{aligned}
\label{eq:du-finite-linear-system}
\end{equation}
For $d^{t+1}\geq k$, the permutation operators are linearly
independent and $G_{t+1}^{(k)}$ is nonsingular, so
Eq.~\eqref{eq:du-finite-linear-system} uniquely determines all coefficients.

At fixed $k$, both $G_t^{(k)}$ and $G_{t+1}^{(k)}$ approach the identity
as $t\to\infty$. Equation~\eqref{eq:du-finite-linear-system} then gives
\begin{equation}\label{eq:alpha_solu_infinity_t}
\lim_{t\to\infty}\alpha_{\mathrm f}(g)=K_k(g,g).
\end{equation}

On the two-qudit Choi space $C=C^1\otimes C^2$, the replica permutation
is $\rho_{2,k}(g)$ in the notation of
Eq.~\eqref{eq:du-permutation-operator}. Hence, using the normalized Choi
state $M_{\mathcal{N}}$ defined in
Eq.~\eqref{eq:normalized-choi-state}, the diagonal channel contraction is
\begin{equation}
\begin{aligned}
K_k(g,g)
&=\tr\!\left[M_{\Ncal}^{\otimes k}\rho_{2,k}(g^{-1})\right]\\
&=\prod_{c\in\mathcal C(g)}\tr\!\left(M_{\Ncal}^{|c|}\right)
\equiv w_{M_{\Ncal}}(g),
\end{aligned}
\label{eq:du-choi-weight}
\end{equation}
where $|c|$ is the length of the cycle.
The limiting coefficients therefore depend on $M_{\Ncal}$ only through
its spectrum. The isometric map $W$ onto the retained region \(A\) then replaces each \(\rho_{t+1,k}(g)\) by \(\rho_{N_A,k}(g)\), leaving its coefficient unchanged.

The same coefficients arise in the CIE. As we explain below, for any
operator $O$ on the $k$ replicated two-qudit space $C^{\otimes k}$, including $M_{\Ncal}^{\otimes k}$, Wick's theorem and the Ginibre
normalization in Sec.~\ref{sec:setup} give
\begin{equation*}
\begin{aligned}
&\int d\mu_G(X)\,
X^{\otimes k}O(X^\dagger)^{\otimes k}\\
&\quad=D_A^{-k}\sum_{g\in S_k}
\tr\!\left[O\rho_{2,k}(g^{-1})\right]\rho_{N_A,k}(g).
\end{aligned}
\end{equation*}
A Wick pairing $g$ contracts the $m$th factor of $X$ with the $g(m)$th
factor of $X^*$, producing the output permutation $\rho_{N_A,k}(g)$,
the input contraction $\tr[O\rho_{2,k}(g^{-1})]$, and the factor
$D_A^{-k}$. Setting $O=M_{\Ncal}^{\otimes k}$, the left-hand side
becomes the Ginibre average of
$\widetilde\rho_{M_{\Ncal}}(X)^{\otimes k}$, while
Eq.~\eqref{eq:du-choi-weight} identifies each input contraction with
$w_{M_{\Ncal}}(g)$. Hence,
\begin{equation}
\int d\mu_G(X)\,
\widetilde\rho_{M_{\Ncal}}(X)^{\otimes k}
=D_A^{-k}\sum_{g\in S_k}
w_{M_{\Ncal}}(g)\rho_{N_A,k}(g).
\label{eq:du-ginibre-wick}
\end{equation}
Since the circuit coefficients $\alpha_{\mathrm f}(g)$ converge
to $w_{M_{\Ncal}}(g)$, the circuit and CIE raw moments
agree in the limit after dividing each by its trace.
This holds at every positive integer order, so
Lemma~\ref{lem:raw-to-physical-moments} gives
\begin{equation}
\lim_{t\to\infty}\rho_{N_A}^{(k)}
=\rho_{M_{\Ncal}}^{(k)},
\label{eq:du-normalized-moment-result}
\end{equation}
establishing Proposition~\ref{prop:central}
for the dual-unitary protocol.

\subsection{Finite-time corrections}
Finite-time deviations arise from the nonorthogonality of the
permutation sectors. We first solve the exact linear system
Eq.~\eqref{eq:du-finite-linear-system} for the raw coefficients
$\alpha_{\mathrm f}(g)$ and then use the replica trick to continue the result
to the physical Born-weighted moments. At fixed $k$ and $d\geq2$, let
$s_{ij}$ denote the transposition of replicas $i$ and $j$. To leading
nontrivial order at large $t$, the Gram matrix and its inverse have
matrix elements
\begin{equation}
\begin{aligned}
G_t^{(k)}(g,g')
&=\delta_{g,g'}
+d^{-t}\sum_{i<j}\delta_{g,g's_{ij}}
+\mathcal O_k(d^{-2t}),\\
\bigl[(G_{t+1}^{(k)})^{-1}\bigr]_{g,g'}
&=\delta_{g,g'}
-d^{-(t+1)}\sum_{i<j}\delta_{g,g's_{ij}}
+\mathcal O_k(d^{-2t}).\label{eq:SECIIIexpansion_of_G}
\end{aligned}
\end{equation}
Thus, at this order, only permutation sectors differing by one
transposition matter. Substituting these expansions into
Eq.~\eqref{eq:du-finite-linear-system} gives
\begin{equation}
\begin{aligned}
\alpha_{\mathrm f}(g)
={}&K_k(g,g)
+d^{-t}\sum_{1\leq i<j\leq k}K_k(g,gs_{ij})\\
&-d^{-(t+1)}\sum_{1\leq i<j\leq k}
K_k(gs_{ij},gs_{ij})+\mathcal O_k(d^{-2t})\\
={}&w_{M_{\Ncal}}(g)
+d^{-t}\sum_{1\leq i<j\leq k}K_k(g,gs_{ij})\\
&-d^{-(t+1)}\sum_{1\leq i<j\leq k}
w_{M_{\Ncal}}(gs_{ij})
+\mathcal O_k(d^{-2t}).
\end{aligned}
\label{eq:du-leading-finite-time-correction}
\end{equation}
The limiting weights $w_{M_{\Ncal}}(g)$ in
Eq.~\eqref{eq:du-choi-weight} depend only on the spectral moments
$\tr(M_{\Ncal}^s)$ and hence only on the eigenvalues of the Choi state.
By contrast, the off-diagonal contractions $K_k(g,gs_{ij})$ involve
partial traces of $M_{\Ncal}^s$ over its input or output factor and can
therefore also depend on its eigenvectors relative to this bipartition.
Consequently, isospectral Choi states, namely Choi states with the same
eigenvalues, select the same limiting ensemble but can approach it
differently at finite times.

Equation~\eqref{eq:du-leading-finite-time-correction} gives the
coefficients entering the trace-normalized raw moments at each positive
integer replica order.
To obtain the convergence rate of the physical Born-weighted moments, we use the $q$ additional replicas of Sec.~\ref{sec:du-replica-construction}.
At order $k+q$, we trace out these $q$ copies, expand the circuit and reference trace normalizations, and analytically continue the resulting polynomial coefficient to $q=1-k$.
We write $\mathcal D_k(g)$ for the coefficient of $d^{-t}\rho_{N_A,k}(g)/D_A^k$ in the expansion of $\rho_{N_A}^{(k)}-\rho_{M_{\Ncal}}^{(k)}$ below. Its explicit expression after the continuation is given in Eq.~\eqref{eq:appendix-D-continuation} of Appendix~\ref{sec:replica-identities}.
We assume that this continuation can be performed coefficient-wise at
fixed $k$ without enhancing the omitted terms. We then obtain
\begin{equation}
\begin{aligned}
\rho_{N_A}^{(k)}-\rho_{M_{\Ncal}}^{(k)}
&=\frac{1}{D_A^k}\sum_{g\in S_k}
\left[d^{-t}\mathcal D_k(g)+\mathcal R_k(g)\right]
\rho_{N_A,k}(g),\\
\max_{g\in S_k}|\mathcal R_k(g)|
&=\mathcal O_{d,k,\Ncal}\!\left(
d^{-2t}+d^{-t}D_A^{-1}\right).
\end{aligned}
\label{eq:du-physical-finite-time-correction}
\end{equation}
Since $\rho_{N_A,k}(g)$ is unitary on a
$D_A^k$-dimensional space,
$\|\rho_{N_A,k}(g)\|_\xi=D_A^{k/\xi}$ and, in particular,
$\|\rho_{N_A,k}(g)\|_1=D_A^k$.
This estimate holds for sufficiently large $t$ and $D_A$ at fixed $d$,
$k$, and $\Ncal$, in the regime $t+1\geq N_A$.
Taking the norms gives
\begin{equation}
\begin{aligned}
\Delta_1^{(k)}
&\leq d^{-t}\sum_{g\in S_k}|\mathcal D_k(g)|
+\mathcal O_{d,k,\Ncal}\!\left(
d^{-2t}+d^{-t}D_A^{-1}\right),\\
\Delta_2^{(k)}
&=d^{-t}
{\left[\displaystyle\sum_{g\in S_k}
|\mathcal D_k(g)|^2\right]^{1/2}}\big/
{\left[\displaystyle\sum_{g\in S_k}
|w_{M_{\Ncal}}(g)|^2\right]^{1/2}}\\
&\quad+\mathcal O_{d,k,\Ncal}\!\left(
d^{-2t}+d^{-t}D_A^{-1}\right).
\end{aligned}
\label{eq:du-finite-time-distances}
\end{equation}
The first line is an upper bound obtained from the triangle inequality,
whereas the second gives the leading Hilbert--Schmidt coefficient.

Equation~\eqref{eq:appendix-swap-coefficient} shows that every transposition $s_{ij}$ has the same contribution $\mathcal{D}_k(s_{ij})$. When transpositions dominate the correction sums in Eq.~\eqref{eq:du-finite-time-distances}, the trace-norm bound adds the absolute values of these $\binom{k}{2}$ coefficients together, while the Hilbert--Schmidt distance takes the square root of their squared sum. Hence
\begin{equation}
\begin{aligned}
\Delta_1^{(k)}
&\lesssim \binom{k}{2}|\mathcal D_k(s_{12})|d^{-t},\\
\Delta_2^{(k)}
&\simeq |\mathcal D_k(s_{12})|d^{-t}
\frac{\sqrt{\binom{k}{2}}}
{\left[\sum_{g\in S_k}|w_{M_{\Ncal}}(g)|^2\right]^{1/2}}.
\end{aligned}
\label{eq:du-moment-order-time-scales}
\end{equation}
The denominator in the second line comes from the Hilbert--Schmidt norm of the CIE moment $\rho_{M_{\Ncal}}^{(k)}$.

For complete depolarization, Eq.~\eqref{eq:complete-depolarization}, transpositions give the leading correction at fixed $k$ and large $d$. The identity permutation dominates the denominator sum, with
\[
\sum_{g\in S_k}|w_{M_{\Ncal}}(g)|^2
=1+\mathcal O_k(d^{-4}),
\]
as follows from Eq.~\eqref{eq:du-choi-weight}, since all nonidentity permutation weights are at most $d^{-2}$. Substituting $\mathcal D_k(s_{ij})=-(d^2-1)/d^5$ into Eq.~\eqref{eq:du-moment-order-time-scales} reproduces the moment-order dependence found in Ref.~\cite{YuHoKos2025Mixed}. In such cases, the deep thermalization time scales as $\log k$, as can be seen from Eq.~\eqref{eq:du-moment-order-time-scales}.

For partial reset, the transpositions contribute at the same order as the identity in the denominator of $\Delta_2^{(k)}$ in Eq.~\eqref{eq:du-moment-order-time-scales}. Consider
\[
\Ncal(\rho)=(1-\theta)\rho+\theta\,\tr(\rho)\ket0\!\bra0,
\]
with fixed reset probability $0<\theta<1$. Equation~\eqref{eq:appendix-partial-reset-purity} gives $w_{M_{\Ncal}}(s_{12})\to(1-\theta)^2$ as $d\to\infty$. The identity and transposition terms in the denominator sum give
\[
\sum_{g\in S_k}|w_{M_{\Ncal}}(g)|^2
\geq1+\binom{k}{2}|w_{M_{\Ncal}}(s_{12})|^2.
\]
Equation~\eqref{eq:appendix-partial-reset-corrections} also gives $\mathcal D_4(s_{12})\to\theta^2$ and $\mathcal D_4(s_{12}s_{34})\to2\theta^2(1-\theta)^2$. Both permutations contribute at the same order, and the dependence on $k$ is more complicated, as now it involves the full sums in Eq.~\eqref{eq:du-finite-time-distances}.

\begin{figure*}[t]
\centering
\begin{tikzpicture}[scale=0.50]
\foreach \x in {0,2,6,8,10,12,16,18,20}{
  \foreach \y in {0,2,4}{\Wgategreen{\x}{\y}}
}
\foreach \x in {-1,1,3,5,7,9,11,13,15,17,19,21}{
  \foreach \y in {1,3}{\Wgategreen{\x}{\y}}
}
\foreach \x in {1,3,5,7,9,11,13,15,17,19,21}{
  \Measurement{\x}{5}
}
\foreach \x in {-2,0,2,4,6,8,10,12,14,16,18,20}{
  \draw[very thick] (\x+0.5,-0.5) to[out=-45,in=225]
  (\x+1.5,-0.5);
}
\foreach \y in {-1,1,3}{
  \draw[very thick] (21.5,\y+0.5) to[out=45,in=315]
  (21.5,\y+1.5);
  \draw[very thick] (-1.5,\y+0.5) to[out=135,in=225]
  (-1.5,\y+1.5);
}
\foreach \x in {4,14}{
  \foreach \y in {0,2,4}{
    \node[fill=white,inner sep=1pt] at (\x,\y) {$\bm{\cdots}$};
  }
}
\draw[decorate,decoration={brace,amplitude=8pt},very thick]
  (-1.5,5.35)--(0.2,5.35);
\draw[decorate,decoration={brace,amplitude=8pt},very thick]
  (0.5,5.35)--(2.15,5.35);
\draw[decorate,decoration={brace,amplitude=8pt},very thick]
  (5.5,5.35)--(8.15,5.35);
\draw[decorate,decoration={brace,amplitude=8pt},very thick]
  (9.5,5.35)--(12.5,5.35);
\draw[decorate,decoration={brace,amplitude=8pt},very thick]
  (17.5,5.35)--(20.5,5.35);
\node at (-0.65,6.25) {$A$};
\node at (1.325,6.25) {$B_n$};
\node at (6.825,6.25) {$B_{i+1}$};
\node at (11.0,6.25) {$B_i$};
\node at (19.0,6.25) {$B_0$};
\foreach \x in {2,8,16}{
  \draw[red,line width=2.4pt] (\x+0.3,1.7)--(\x+0.8,1.2);
}
\foreach \x/\y in {2/2,3/1,8/2,9/1,16/2,17/1}{
  \draw[thick,fill=mygreen!70,rounded corners=2pt]
    (\x-0.35,\y+0.35) rectangle (\x+0.35,\y-0.35);
  \node at (\x,\y) {\scriptsize$k$};
}
\draw[very thick,dashed,blue] (0.4,-1.15)--(0.4,5.0);
\node at (0.4,-1.55) {$\overline{\rho}_n^{(k)}$};
\draw[very thick,dashed,blue] (5.4,-1.15)--(5.4,5.0);
\node at (5.4,-1.55) {$\overline{\rho}_{i+1}^{(k)}$};
\draw[very thick,dashed,blue] (9.4,-1.15)--(9.4,5.0);
\node at (9.4,-1.55) {$\overline{\rho}_i^{(k)}$};
\draw[very thick,dashed,blue] (17.4,-1.15)--(17.4,5.0);
\node at (17.4,-1.55) {$\overline{\rho}_0^{(k)}$};
\node[red,font=\fontsize{6}{7}\selectfont,scale=0.8] at (3.0,2.05) {$\Ncal_n$};
\node[red,font=\fontsize{6}{7}\selectfont,scale=0.8] at (8.9,2.05) {$\Ncal_{i+1}$};
\node[red,font=\fontsize{6}{7}\selectfont,scale=0.8] at (16.9,2.05) {$\Ncal_1$};
\draw[->,very thick] (22.25,-0.8)--(22.25,5.8);
\node[right] at (22.25,2.5) {time $t$};
\draw[->,very thick] (21.5,-2.35)--(-0.8,-2.35);
\node at (10.35,-2.85) {spatial contraction};
\end{tikzpicture}
\caption{Folded $k$-replica dual-unitary circuit with dilute local
channels. Green gates, orange measurement tensors, and blue dashed raw
cuts follow the conventions of Fig.~\ref{fig:du-spatial-contraction};
each red segment labeled $\Ncal_j$ denotes $\Ncal_j^{\otimes k}$.
Reading from right to left, the central unit maps
$\overline{\rho}_i^{(k)}$ to $\overline{\rho}_{i+1}^{(k)}$ through
$\Ncal_{i+1}^{\otimes k}$ and $B_{i+1}$; the ellipses denote omitted
channel--scrambling units. For $n=1$, only the endpoint unit remains,
reproducing Fig.~\ref{fig:du-spatial-contraction}.}
\label{fig:du-dilute-recursion}
\end{figure*}
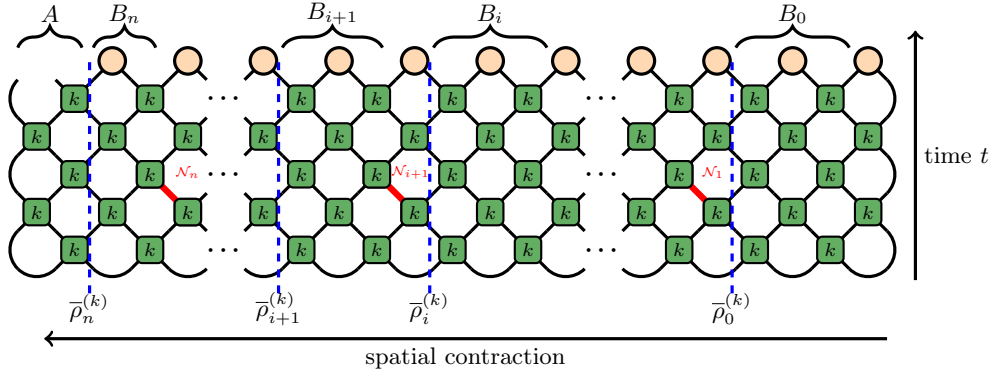

\subsection{Dilute well-separated errors}

We finally consider $n$ local channels $\{\Ncal_i\}_{i=1}^n$ separated
by $n+1$ measured scrambling regions. We label these regions
$B_0,\ldots,B_n$ in the order encountered by the right-to-left spatial
contraction (see
Fig.~\ref{fig:du-dilute-recursion}). The blue cut after $B_i$ carries the
raw projected time-like tensor $\overline{\rho}_i^{(k)}$, which lies
immediately before $\Ncal_{i+1}$ for $i<n$;
$\overline{\rho}_n^{(k)}$ is the final raw tensor. In the following, we consider the dilute limit, in which the channels are well separated: each internal scrambling region is taken to be much longer than the mixing length of the corresponding $k$-replica spatial transfer matrix, while the two boundary regions are likewise sufficiently long to implement the initial and final projections.

At the blue cuts in Fig.~\ref{fig:du-dilute-recursion}, we expand each
raw tensor in the permutation basis as
\begin{equation}
\overline{\rho}_i^{(k)}
=\sum_{g\in S_k}a_i(g)\rho_{t+1,k}(g).
\label{eq:du-dilute-permutation-expansion}
\end{equation}
Applying $\Ncal_{i+1}$ to the channel leg and repeating the overlap
calculation leading to Eq.~\eqref{eq:du-finite-linear-system} gives
\begin{equation}
\begin{aligned}
&\sum_{g'\in S_k}G_{t+1}^{(k)}(g,g')a_{i+1}(g')\\
&\qquad=\sum_{g'\in S_k}G_t^{(k)}(g,g')
K_{k,i+1}(g,g')a_i(g'),\\
&\hspace{4.0cm}g\in S_k,\quad i=0,\ldots,n-1,
\end{aligned}
\label{eq:du-dilute-recursion}
\end{equation}
where $K_{k,i}$ is the kernel in Eq.~\eqref{eq:du-channel-kernel} with
$\Ncal$ replaced by $\Ncal_i$. Thus each additional error inserts one
channel kernel into the same projection equation as in the single-error
calculation.

At fixed $k$, the large-$t$ expansion diagonalizes this recursion:
\begin{equation}
a_{i+1}(g)=w_{M_{\Ncal_{i+1}}}(g)a_i(g)
+\mathcal O_k\!\left(
d^{-t}\max_{g'\in S_k}|a_i(g')|
\right).
\label{eq:du-dilute-diagonal-recursion}
\end{equation}
The first long measured region produces equal permutation coefficients.
Because their common scalar cancels from the trace-normalized replicated
tensor, we choose $a_0(g)=1$ for every $g$. For a fixed number of errors,
iteration gives
\begin{equation}
a_n(g)=\prod_{i=1}^n w_{M_{\Ncal_i}}(g)
+\mathcal O_{k,n}(d^{-t}).
\label{eq:du-dilute-iterated-recursion}
\end{equation}
The product of the individual channel weights equals the cycle weight of their tensor-product Choi state:
\begin{equation}
\begin{aligned}
\prod_{i=1}^n w_{M_{\Ncal_i}}(g)
&=\prod_{i=1}^n
\prod_{c\in\mathcal C(g)}
\tr\!\left(M_{\Ncal_i}^{|c|}\right)\\
&=\prod_{c\in\mathcal C(g)}
\tr\!\left[
\left(\bigotimes_{i=1}^nM_{\Ncal_i}\right)^{|c|}
\right]\\
&=w_{\bigotimes_{i=1}^nM_{\Ncal_i}}(g).
\end{aligned}
\label{eq:du-dilute-factorization}
\end{equation}

This coincides with the single-channel result for the effective Choi state
\begin{equation}
M:=\bigotimes_{i=1}^nM_{\Ncal_i},
\label{eq:du-effective-choi-state}
\end{equation}
and thus proves Proposition~\ref{prop:central} in the dual-unitary circuit case.
Here, $M$ is an effective object that retains the Choi data of each
error separately. It should not be identified with the Choi state
$M_{\Ncal_n\circ\cdots\circ\Ncal_1}$ of the sequentially composed
channel. After each scrambling
region, the tensor is projected back onto a common permutation label
$g$. In the large-$t$ limit, the channel kernel is diagonal in this
label, so successive channel insertions multiply the same permutation
coefficient.

\section{Global Haar dynamics}

\label{sec:global-haar}

To isolate the role of scrambling, we replace the unitary dynamics
before and after a single-site error channel by two independent global
Haar-random unitaries $U_1$ and $U_2$ (each acting on the union of $A$ and $B$). Let $D_B=d^{N_B}$ and $D=D_AD_B$. Starting
from an arbitrary fixed pure state $\ket{\psi_0}$, we apply $U_1$, a
single-site channel $\Ncal$, and then $U_2$.
We then measure all of $B$
in the fixed basis of Sec.~\ref{sec:setup} and denote the realized
moment by $\rho_{N_A}^{(k)}(U_1,U_2)$. Throughout this section,
$d,D_A,k$, and $\Ncal$ are fixed while $D_B\to\infty$.

Our argument has two steps. We first average $\rho_{N_A}^{(k)}(U_1,U_2)$ over both independent Haar unitaries and show that this averaged moment converges to the CIE prediction. We then use concentration of measure to show that, for a typical pair $(U_1,U_2)$, the realized moment is close to this average. This establishes convergence in probability for the MSPE itself.

The first Haar unitary makes the input at the error site nearly maximally mixed.
We show below that the post-channel state is then close to the normalized Choi state $M_{\Ncal}$ embedded into the physical system.
The independent unitary $U_2$ makes this embedding Haar random, and a fixed measurement block approaches the Ginibre matrix $X$ of Sec.~\ref{sec:setup}.

\subsection{Haar-averaged projected moments}

We first control the change in the Born-weighted moment when the full-system state changes. For states $\rho$ and $\rho'$ measured in the same basis of $B$, Appendix~\ref{app:global-haar} proves
\begin{equation}
\left\|\rho_{N_A}^{(k)}-\rho_{N_A}^{\prime\,(k)}\right\|_1
\leq(2k-1)\|\rho-\rho'\|_1.
\label{eq:global-haar-recorded-outcome-contraction}
\end{equation}

Let $S$ denote the error site and $\overline S$ its complement. Define
the state of $S$ immediately before the channel and the full-system
state immediately after the channel by
\begin{equation}
\begin{aligned}
\rho_S
&:=
\tr_{\overline S}\!\left(
U_1\ket{\psi_0}\!\bra{\psi_0}U_1^\dagger
\right),\\
\sigma
&:=
(\Ncal\otimes\operatorname{id}_{\overline S})\!\left(
U_1\ket{\psi_0}\!\bra{\psi_0}U_1^\dagger
\right).\label{eq:SecIVdefinitionofsigma}
\end{aligned}
\end{equation}
To compare $\sigma$ with the Choi state, we use a
Stinespring dilation of $\Ncal$~\cite{Stinespring1955Positive}: the system interacts unitarily
with an initially pure environment, and tracing out the
environment gives the channel $\Ncal$. For the maximally mixed
input $I_d/d$, the resulting environment state has the same
nonzero eigenvalues as $M_{\Ncal}$. Taking the environment to
have dimension $d^2$, we choose its basis so that this state
equals $M_{\Ncal}$.

For the actual input \(\rho_S\), let \(\tau_E\) be the resulting environment state. Since the Stinespring map is an isometry followed by a partial trace,
\[
\|\tau_E-M_{\Ncal}\|_1
\leq
\left\|\rho_S-\frac{I_d}{d}\right\|_1.
\]
The states $\tau_E$ and $\sigma$ are reductions of the same pure state and hence have the same nonzero eigenvalues.
Hence, identifying the environment with $C$, there exists an isometry \(\mathsf V:C\to A\otimes B\) such that
\[
\sigma=\mathsf V\tau_E\mathsf V^\dagger.
\]
Therefore
\begin{equation}
\left\|
\sigma-\mathsf V M_{\Ncal}\mathsf V^\dagger
\right\|_1
\leq
\left\|\rho_S-\frac{I_d}{d}\right\|_1.
\label{eq:global-haar-choi-embedding}
\end{equation}

The Haar-averaged purity of $\rho_S$, evaluated in Appendix~\ref{app:global-haar}, then gives
\begin{equation}
\mathbb E_{U_1}
\left\|
\sigma-\mathsf V M_{\Ncal}\mathsf V^\dagger
\right\|_1
\leq
\sqrt{\frac{d^2-1}{D+1}}.
\label{eq:global-haar-choi-embedding-average}
\end{equation}

Equations~\eqref{eq:global-haar-recorded-outcome-contraction} and \eqref{eq:global-haar-choi-embedding-average} bound the difference between the Haar-averaged MSPE moment and the moment obtained from $U_2\mathsf V M_{\Ncal}\mathsf V^\dagger U_2^\dagger$.
Conditioned on $U_1$, the matrix $U_2\mathsf V$ is a Haar-distributed $D\times d^2$ isometry.
For a fixed outcome $\alpha_0$, we define its $D_A\times d^2$ scaled Haar block as 
\begin{equation}
X_{\mathrm H}
:=
\sqrt{D_B}\,
(I_A\otimes{}_B\!\bra{\alpha_0})U_2\mathsf V.
\end{equation}
For the state $U_2\mathsf V M_{\Ncal}\mathsf V^\dagger U_2^\dagger$,
the unnormalized conditional state at outcome $\alpha_0$ is
\begin{equation}
\widetilde\rho_{N_A}^{\mathrm{iso}}(\alpha_0)
=D_B^{-1}X_{\mathrm H} M_{\Ncal}X_{\mathrm H}^\dagger.
\end{equation}
All fixed measurement outcomes give the same probability distribution
for $X_{\mathrm H}$ under Haar-random $U_2$.
After summing the Born-weighted contributions and averaging over $U_2$, we denote the moment of the embedded Choi state by $\rho_{N_A,\mathrm{iso}}^{(k)}$: 
\[
\begin{aligned}
\rho_{N_A,\mathrm{iso}}^{(k)}
&:=\mathbb E_{U_2}\sum_{\alpha=1}^{D_B}
\frac{[\widetilde\rho_{N_A}^{\mathrm{iso}}(\alpha)]^{\otimes k}}
{[\tr\widetilde\rho_{N_A}^{\mathrm{iso}}(\alpha)]^{k-1}}\\
&=D_B\,\mathbb E_{U_2}
\frac{(D_B^{-1}X_{\mathrm H} M_{\Ncal}X_{\mathrm H}^\dagger)^{\otimes k}}
{[\tr(D_B^{-1}X_{\mathrm H} M_{\Ncal}X_{\mathrm H}^\dagger)]^{k-1}}\\
&=\mathbb E_{U_2}
\frac{(X_{\mathrm H} M_{\Ncal}X_{\mathrm H}^\dagger)^{\otimes k}}
{[\tr(X_{\mathrm H} M_{\Ncal}X_{\mathrm H}^\dagger)]^{k-1}}.
\end{aligned}
\]
The first equality uses the identical distribution of the outcome blocks.
This average is independent of $U_1$, since $U_2\mathsf V$ has the same
Haar distribution for every fixed $U_1$.

By Eq.~\eqref{eq:choi-ensemble-moment}, the CIE moment has the
corresponding form
\[
\rho_{M_{\Ncal}}^{(k)}
=\mathbb E_X
\frac{(XM_{\Ncal}X^\dagger)^{\otimes k}}
{[\tr(XM_{\Ncal}X^\dagger)]^{k-1}},
\]
where $X$ follows the Ginibre distribution defined in Sec.~\ref{sec:setup}.
We therefore need to compare the moments formed from $X_{\mathrm H}$ and $X$.
Appendix~\ref{app:global-haar-blocks} constructs these matrices from the
same Gaussian matrix and proves
\begin{equation}
\mathbb E\|X_{\mathrm H}-X\|_2^2
=\mathcal O_{d,D_A}(D^{-1}).
\label{eq:global-haar-coefficient-bound}
\end{equation}
Here $\|X_{\mathrm H}-X\|_2^2=\tr[(X_{\mathrm H}-X)^\dagger(X_{\mathrm H}-X)]$.
Using Eq.~\eqref{eq:global-haar-positive-lipschitz}, Appendix~\ref{app:global-haar-moment-comparison} converts the matrix estimate in Eq.~\eqref{eq:global-haar-coefficient-bound} into the following bound on the moment difference:
\begin{equation}
\|\rho_{N_A,\mathrm{iso}}^{(k)}-\rho_{M_{\Ncal}}^{(k)}\|_1
=\mathcal O_{d,D_A,k}(D^{-1/2}).\label{eq:main_text_GaussianvsUnitary}
\end{equation}
We can now add the error from replacing the post-channel state and the
error from replacing $X_{\mathrm H}$ by $X$:
\begin{equation}
\begin{aligned}
b_{D_B}^{(k)}
&:=\left\|\mathbb E_{U_1,U_2}\rho_{N_A}^{(k)}
-\rho_{M_{\Ncal}}^{(k)}\right\|_1\\
&\leq\left\|\mathbb E_{U_1,U_2}\rho_{N_A}^{(k)}
-\rho_{N_A,\mathrm{iso}}^{(k)}\right\|_1+\|\rho_{N_A,\mathrm{iso}}^{(k)}-\rho_{M_{\Ncal}}^{(k)}\|_1\\
&\leq(2k-1)\sqrt{\frac{d^2-1}{D+1}}
+\mathcal O_{d,D_A,k}(D^{-1/2})\\
&=\mathcal O_{d,D_A,k}(D_B^{-1/2}).
\end{aligned}
\label{eq:global-haar-average}
\end{equation}
The first inequality is the triangle inequality.
The next bounds the first difference using
Eqs.~\eqref{eq:global-haar-recorded-outcome-contraction}
and~\eqref{eq:global-haar-choi-embedding-average}, as shown in
Eq.~\eqref{eq:global-haar-recorded-measurement}, and the second using
Eq.~\eqref{eq:main_text_GaussianvsUnitary}, also derived in detail in Eq.~\eqref{eq:SECIVerrorpart2}.
The last equality uses $D=D_AD_B$ with $d,D_A,k$ fixed.
The bounds in Eqs.~\eqref{eq:global-haar-recorded-outcome-contraction} and~\eqref{eq:global-haar-positive-lipschitz}, the Haar-purity calculation used in
Eq.~\eqref{eq:global-haar-choi-embedding-average}, and the matrix estimate
in Eq.~\eqref{eq:global-haar-coefficient-bound} are proved in
Appendices~\ref{app:global-haar-stability},
\ref{app:global-haar-purity}, and~\ref{app:global-haar-blocks}, respectively.

Since $D_B=d^{N_B}$, the averaged moment approaches the CIE moment exponentially in $N_B$ at fixed $d,D_A,k$.

\subsection{Concentration and typicality}

The averaged estimate in Eq.~\eqref{eq:global-haar-average} controls
the finite-size bias, but does not by itself imply that a typical pair
$(U_1,U_2)$ realizes the reference moment. We now use concentration of
measure on $\mathsf U(D)\times\mathsf U(D)$ to control fluctuations
around the Haar mean. The required product-group version of L\'evy's
lemma is the following~\cite{MeckesHaarBookPDF}.
\begin{lemma}[L\'evy concentration for two Haar unitaries]
\label{lem:global-haar-levy}
Let $F:\mathsf U(D)\times\mathsf U(D)\to\mathbb R$ be $L$-Lipschitz
with respect to the product Hilbert--Schmidt distance, namely,
\[
\begin{aligned}
&\left|F(U_1,U_2)-F(U_1',U_2')\right|\\
&\qquad\leq
L\left(
\|U_1-U_1'\|_2^2+\|U_2-U_2'\|_2^2
\right)^{1/2}.
\end{aligned}
\]
If $U_1$ and $U_2$ are independently Haar distributed and $D\geq3$,
then, for every $x>0$,
\begin{equation}
\Pr\!\left[
\left|F-\mathbb EF\right|\geq x
\right]
\leq
2\exp\!\left[
-\frac{(D-2)x^2}{24L^2}
\right].
\label{eq:global-haar-levy}
\end{equation}
\end{lemma}
Here and below, probabilities and expectations are taken over the
independent Haar unitaries $U_1$ and $U_2$.

For fixed basis states $\ket{\bm i},\ket{\bm j}$ of $A^{\otimes k}$,
define the real-valued function
\begin{equation}
F(U_1,U_2)
:=\operatorname{Re}\!\left[
\bra{\bm i}\rho_{N_A}^{(k)}(U_1,U_2)\ket{\bm j}
\right]. \label{eq:maintext_function_used_in_Levy}
\end{equation}
The same argument applies when the real part is replaced by the imaginary
part. To determine the Lipschitz constant, let $\rho$ and $\rho'$ denote
the final full-system states generated by $(U_1,U_2)$ and
$(U_1',U_2')$, respectively. Namely, $\rho=U_2 \sigma U_2^\dagger$ with $\sigma$ defined in Eq.~\eqref{eq:SecIVdefinitionofsigma} and $\rho'$ is defined similarly by substituting $U_1\to U_1'$, $U_2\to U_2'$. Here $\|\cdot\|_\infty$ denotes the operator norm, the largest singular value.
 Changing one unitary at a time and applying Eq.~\eqref{eq:global-haar-recorded-outcome-contraction} gives
\begin{equation}
\begin{aligned}
& \left\lVert\rho^{(k)}_{N_A}(U_1,U_2)-\rho^{(k)}_{N_A}(U_1',U_2')\right\rVert_1 \\
 & \leq(2k-1)\lVert\rho-\rho'\rVert_{1}\\
 & \leq 2(2k-1)\left(\lVert U_{1}-U_{1}'\rVert_{\infty}+\lVert U_{2}-U_{2}'\rVert_{\infty}\right).
\end{aligned}
\label{eq:global-haar-coordinate-lipschitz}
\end{equation}
The first inequality follows from
Eq.~\eqref{eq:global-haar-recorded-outcome-contraction}.
The second changes one unitary at a time and uses trace-distance contractivity of $\mathcal N$. 
Finally, $\|U-V\|_\infty\leq\|U-V\|_2$ and Cauchy--Schwarz give the product Hilbert--Schmidt distance.
Each real or imaginary matrix element changes by at most the trace-norm difference, so both functions are $2\sqrt2(2k-1)$-Lipschitz. Therefore, by applying Lemma~\ref{lem:global-haar-levy} to Eq.~\eqref{eq:maintext_function_used_in_Levy} and its imaginary analogue, we obtain
\begin{equation}
\begin{aligned}
&\Pr\!\left[
\left|
\bra{\bm i}
\left(
\rho_{N_A}^{(k)}-\mathbb E\rho_{N_A}^{(k)}
\right)
\ket{\bm j}
\right|
\geq x
\right]\\
&\qquad\leq
4\exp\!\left[
-\frac{(D-2)x^2}{96(4k-2)^2}
\right].
\end{aligned}
\label{eq:global-haar-entry-concentration}
\end{equation}
The prefactor $4$ combines the real and imaginary estimates. 

To bound the full moment operator, we expand it in the $D_A^k$-dimensional basis and apply the triangle inequality:
\[
\begin{aligned}
&\left\|
\rho_{N_A}^{(k)}-\mathbb E\rho_{N_A}^{(k)}
\right\|_1%
\leq
\sum_{\bm i,\bm j}
\left|
\bra{\bm i}
\left(
\rho_{N_A}^{(k)}-\mathbb E\rho_{N_A}^{(k)}
\right)
\ket{\bm j}
\right|.
\end{aligned}
\]
Consequently, a trace-norm fluctuation of at least $\eta$ forces at
least one of the $D_A^{2k}$ matrix elements to fluctuate by at least
$\eta/D_A^{2k}$. Equation~\eqref{eq:global-haar-entry-concentration}
and a union bound give, for every $\eta>0$,
\begin{equation}
\begin{aligned}
&\Pr\!\left[
\left\|
\rho_{N_A}^{(k)}-\mathbb E\rho_{N_A}^{(k)}
\right\|_1\geq\eta
\right]\\&\qquad\leq
4D_A^{2k}\exp\!\left[
-\frac{(D-2)\eta^2}
{96(4k-2)^2D_A^{4k}}
\right].
\end{aligned}
\label{eq:global-haar-fluctuation-tail}
\end{equation}
Since the CIE moment has unit trace, the triangle inequality gives $\Delta_1^{(k)}\leq\|\rho_{N_A}^{(k)}-\mathbb E\rho_{N_A}^{(k)}\|_1+b_{D_B}^{(k)}$.
For $\varepsilon>b_{D_B}^{(k)}$, Eq.~\eqref{eq:global-haar-fluctuation-tail} therefore gives
\begin{equation}
\begin{aligned}
&\Pr\!\left[\Delta_1^{(k)}\geq\varepsilon\right]\\
&\quad\leq4D_A^{2k}
\exp\!\left[
-\frac{(D-2)(\varepsilon-b_{D_B}^{(k)})^2}
{96(4k-2)^2D_A^{4k}}
\right].
\end{aligned}
\label{eq:global-haar-bias-tail}
\end{equation}
At fixed $d,D_A,k$, the bias tends to zero and this probability vanishes for every $\varepsilon>0$ as $D_B\to\infty$.
Thus the realized MSPE moment converges in probability to $\rho_{M_{\Ncal}}^{(k)}$, establishing Proposition~\ref{prop:central} for global Haar dynamics.

As a corollary, if $0<\varepsilon\leq1$, 
$0<\delta<1$, and $N_B\geq 2\log_d\varepsilon^{-1}$ such that $b_{D_B}^{(k)}\leq\varepsilon/2$, Eq. (\ref{eq:global-haar-bias-tail}) gives the parametric sufficient condition
\[
N_B
\gtrsim
(4k-1)N_A
+2\log_d\varepsilon^{-1}
+\log_d\log\!\left(\frac{4D_A^{2k}}{\delta}\right),
\]
for failure probability at most $\delta$, which concludes the proof.

\subsection{Fixed-number errors with fresh Haar scramblers}

The same mechanism extends to a fixed number $n$ of single-site
channels if a fresh independent global Haar unitary acts before the
first channel, between adjacent channels, and after the last channel.
The complementary channel $\Ncal_i^c$ maps the channel input to the environment state.
Choose the padded Stinespring environment of $\Ncal_i$ to be the
previously defined space $C_i$, with its environment basis chosen so that
$\Ncal_i^c(I_d/d)=M_{\Ncal_i}$. Retaining the environments of the
earlier channels keeps the physical system and
$C_1\otimes\cdots\otimes C_{i-1}$ in a joint pure state. Conditional on
the preceding history, the fresh Haar unitary decouples the next error
site $S_i$ from this fixed-dimensional reference. Let
$\rho_{S_iC_1\cdots C_{i-1}}$ denote their joint state after this Haar
unitary and $\rho_{C_1\cdots C_{i-1}}$ its reduced reference marginal. Then
\begin{equation}
\mathbb E
\left\|
\rho_{S_iC_1\cdots C_{i-1}}
-\frac{I_d}{d}\otimes\rho_{C_1\cdots C_{i-1}}
\right\|_1
=
\mathcal O_{n,d}(D^{-1/2}).
\label{eq:global-haar-reference-decoupling}
\end{equation}
Applying
$\operatorname{id}_{C_1\cdots C_{i-1}}\otimes\Ncal_i^c$ and using
trace-distance contractivity results in
\[
\mathbb{E}\lVert\rho_{C_1\cdots C_{i}}-\rho_{C_1\cdots C_{i-1}}\otimes M_{\mathcal{N}_i}\rVert_1 = \mathcal{O}_{n,d}(D^{-1/2}).
\]
Iterating and telescoping the above procedure gives
\begin{equation}
\mathbb E
\left\|
\rho_{C_1\cdots C_n}
-\bigotimes_{i=1}^nM_{\Ncal_i}
\right\|_1
=
\mathcal O_{n,d}(D^{-1/2}).
\end{equation}
Thus the accumulated environment approaches the joint Choi state
$M=\bigotimes_{i=1}^nM_{\Ncal_i}$ defined in
Sec.~\ref{sec:setup}. The final Haar region and the single-channel
block comparison yield
\begin{equation}
\left\|
\mathbb{E}\rho_{N_A}^{(k)}
-\rho_M^{(k)}
\right\|_1
=
\mathcal O_{n,d,D_A,k}\!\left(D_B^{-1/2}\right).
\label{eq:global-haar-dilute-average}
\end{equation}

Appendix~\ref{app:global-haar} derives
Eq.~\eqref{eq:global-haar-reference-decoupling} from the second Haar
moment at each channel. The result holds for fixed $n$ and
sufficiently large $D$.
Changing one of the $n+1$ Haar unitaries at a time, as in Eq.~\eqref{eq:global-haar-coordinate-lipschitz}, bounds the trace-norm difference of the moment operators by $(4k-2)\sum_{j=1}^{n+1}\|U_j-U_j'\|_2$.
Cauchy--Schwarz gives $\sum_{j=1}^{n+1}\|U_j-U_j'\|_2\leq\sqrt{n+1}\bigl(\sum_{j=1}^{n+1}\|U_j-U_j'\|_2^2\bigr)^{1/2}$, so the Lipschitz constant is $\sqrt{n+1}(4k-2)$.
Applying concentration on $\mathsf U(D)^{n+1}$ then proves typicality for every fixed $n$. Here $M$ stores the individual errors
separated by fresh scrambling regions. Equation~\eqref{eq:global-haar-dilute-average}
reproduces the tensor-product rule derived independently for
dual-unitary circuits in Sec.~\ref{sec:dual-unitary}.

The global Haar calculation uses full unitary invariance and does not
address spatial locality. We next show that local Haar-random circuits recover the same CIE through a large-$d$ domain-wall mechanism.

\section{Local Haar-random circuits}
\label{sec:local-haar}

We consider an open one-dimensional chain, initialized in a product
state and evolved by independent Haar-random two-site gates on
alternating nearest-neighbor bonds. As in Sec.~\ref{sec:dual-unitary},
$t$ counts individual gate layers. The retained region $A$ consists of
the first $N_A$ sites at the left end, and all $N_B$ sites of its
complement are measured in a fixed product basis. At fixed moment
order $k$, we take $N_B>kN_A$ so that the number of measurement outcomes
is large enough to suppress finite-sample corrections. We denote the
average over the gates by $\mathbb E_U$.

We take the large-$d$ limit at fixed replica order and circuit
geometry. The channel is placed in the circuit bulk, subject to the
placement condition described below.

\subsection{Moment distance and replica construction}

To establish the CIE prediction for typical circuits, we calculate the mean squared Hilbert--Schmidt distance defined in Eq.~\eqref{eq:moment-distance}, with reference moment $\rho_{M_{\Ncal}}^{(k)}$. Its denominator is deterministic, so
\begin{equation}
\mathbb E_U[(\Delta_2^{(k)})^2]
=\frac{\mathbb E_U\|\rho_{N_A}^{(k)}-\rho_{M_{\Ncal}}^{(k)}\|_2^2}
{\tr[(\rho_{M_{\Ncal}}^{(k)})^2]}.
\label{eq:local-haar-distance-squared}
\end{equation}
The numerator contains three overlaps:
\begin{equation}
\begin{aligned}
&\mathbb E_U\|\rho_{N_A}^{(k)}-\rho_{M_{\Ncal}}^{(k)}\|_2^2%
=\mathbb E_U\tr[(\rho_{N_A}^{(k)})^2]
+\tr[(\rho_{M_{\Ncal}}^{(k)})^2]\\
&\qquad\quad-2\mathbb E_U\tr[
\rho_{N_A}^{(k)}\rho_{M_{\Ncal}}^{(k)}].
\end{aligned}
\label{eq:local-haar-distance-expansion}
\end{equation}
The first is the mean purity of the circuit moment, the second is the
reference purity, and the third compares the two ensembles.
We emphasize that the two circuit copies in the first term use the same realization of all random gates.

We will show that, in the saturated bulk regime specified below, all
three overlaps have the same leading value:
\begin{equation}
\begin{aligned}
\mathbb E_U\tr[(\rho_{N_A}^{(k)})^2]
&=D_A^{-k}\sum_{g\in S_k}w_{M_{\Ncal}}(g)^2[1+\mathcal O(d^{-1})],\\
\mathbb E_U\tr[\rho_{N_A}^{(k)}\rho_{M_{\Ncal}}^{(k)}]
&=D_A^{-k}\sum_{g\in S_k}w_{M_{\Ncal}}(g)^2[1+\mathcal O(d^{-1})],\\
\tr[(\rho_{M_{\Ncal}}^{(k)})^2]
&=D_A^{-k}\sum_{g\in S_k}w_{M_{\Ncal}}(g)^2[1+\mathcal O(d^{-1})].
\end{aligned}
\label{eq:local-haar-three-overlaps}
\end{equation}
Here $w_{M_{\Ncal}}(g)$ is the Choi cycle weight in
Eq.~\eqref{eq:du-choi-weight}, and
$\mathcal O(d^{-1})$ denotes a relative correction at fixed $k,t,N_A,N_B$.
Equality of these coefficients
cancels the leading contribution to the numerator of
Eq.~\eqref{eq:local-haar-distance-squared}. We derive this equality
by keeping the Born factors throughout the replica calculation.

We use the replica construction of Sec.~\ref{sec:du-replica-construction}. Each factor in a trace overlap has $k$ active copies and $q$ traced copies, with
$
r=k+q
$.
We first take integer $q\geq0$ and then continue to $q=1-k$ to obtain the Born factor $P^{1-k}$.
At integer $q$, the three scalar contractions are
\begin{equation}
\begin{aligned}
T_{cc}^{(k,q)}
&=\mathbb E_U\sum_{\alpha,\beta}P_\alpha^qP_\beta^q[\tr(\widetilde\rho_{N_A}(\alpha)
\widetilde\rho_{N_A}(\beta))]^k,\\
T_{cM}^{(k,q)}
&=\mathbb E_U\sum_\alpha\int d\mu_G(X)
P_\alpha^qP_{M_{\Ncal}}(X)^q\\
&\quad\times[\tr(\widetilde\rho_{N_A}(\alpha)
\widetilde\rho_{M_{\Ncal}}(X))]^k,\\
T_{MM}^{(k,q)}
&=\int d\mu_G(X)d\mu_G(Y)
P_{M_{\Ncal}}(X)^qP_{M_{\Ncal}}(Y)^q\\
&\quad\times[\tr(\widetilde\rho_{M_{\Ncal}}(X)
\widetilde\rho_{M_{\Ncal}}(Y))]^k.
\end{aligned}
\label{eq:local-haar-replicated-overlaps}
\end{equation}
The subscripts $c$ and $M$ denote a circuit factor and a reference
factor, respectively. Continuing $q$ to $1-k$ gives, in this order,
the circuit--circuit, circuit--reference, and reference--reference
physical overlaps in Eq.~\eqref{eq:local-haar-three-overlaps}.
The circuit--circuit term requires a $2r$-replica Haar average of each
local gate. The mixed term has $r$ circuit replicas and $r$ copies of the unnormalized conditional state $\widetilde\rho_{M_{\Ncal}}(X)=XM_{\Ncal}X^\dagger$, where $X$ is an independent Ginibre matrix. The reference term
has two independent Ginibre matrices and no circuit gates.

\subsection{Boundary permutations and Haar-gate contraction}

Each overlap in Eq.~\eqref{eq:local-haar-replicated-overlaps} contains two factors, each with $r=k+q$ copies. We label the copies of the first factor by $1,\ldots,r$ and those of the second by $r+1,\ldots,2r$. Within each factor, the first $k$ copies enter the trace overlap, while the other $q$ copies are traced separately. After averaging over the gates, every leg of this network carries a permutation of the copies, and each overlap becomes a sum over such labelings. Below we give the weight of a labeling: the boundaries fix or favor particular permutations, while each gate and the channel contribute a local factor.

On $A$, the trace $\tr(XY)=\sum_{a,b}X_{ab}Y_{ba}$ joins the $i$th active copy of the first factor to the $i$th active copy of the second, which is the same as swapping these two copies. Every retained site therefore carries the fixed permutation operator $\rho_{1,2r}(s_k)$ on its $2r$ copies, with
\begin{equation}
s_k=\prod_{i=1}^k(i,r+i).
\label{eq:local-haar-overlap-swap}
\end{equation}
Each of these swaps gives one trace-overlap factor in
Eq.~\eqref{eq:local-haar-replicated-overlaps}. The permutation $s_k$ acts as the identity on the traced copies, whose separate traces give the Born probabilities raised to the power $q$.

For $g\in S_m$ and $g'\in S_n$, $g\oplus g'$ acts as $g$ on the first $m$ copies and as $g'$ on the next $n$ copies:
\[
\begin{aligned}
(g\oplus g')(i)&=g(i),&&1\leq i\leq m,\\
(g\oplus g')(m+j)&=m+g'(j),&&1\leq j\leq n.
\end{aligned}
\]
We write $S_r\oplus S_r$ for all permutations $g_1\oplus g_2$ with $g_1,g_2\in S_r$, which permute the two sets of $r$ copies separately.

On a measured site of $T_{cc}^{(k,q)}$, the copies $1,\ldots,r$ share the outcome $a$, and the copies $r+1,\ldots,2r$ share the outcome $b$, with $a,b\in\{1,\ldots,d\}$ summed independently. Contracting with a permutation $h\in S_{2r}$ gives
\begin{equation}
\begin{aligned}
&\sum_{a,b=1}^d
\bra{a}^{\otimes r}\bra{b}^{\otimes r}
\rho_{1,2r}(h)
\ket{a}^{\otimes r}\ket{b}^{\otimes r}\\
&\qquad=\begin{cases}
d^2,&h\in S_r\oplus S_r,\\
d,&h\notin S_r\oplus S_r.
\end{cases}
\end{aligned}
\label{eq:local-haar-measured-boundary}
\end{equation}
For $h\in S_r\oplus S_r$, $a$ and $b$ are summed independently; if $h$ maps a label from one set into the other, the contraction forces $a=b$. The measured region therefore favors, with equal weight, every $h\in S_r\oplus S_r$. In $T_{cM}^{(k,q)}$, only the first $r$ copies come from the circuit and share a measurement outcome. At each measured site, summing this outcome gives
\[
\sum_{a=1}^{d}
\bra{a}^{\otimes r}\rho_{1,r}(h)\ket{a}^{\otimes r}
=d,\qquad h\in S_r.
\]
The other $r$ copies are averaged over the Ginibre matrix. In $T_{MM}^{(k,q)}$, both factors are Ginibre averages and there are no circuit measurement outcomes.  In all three overlaps the $A$ boundary is $s_k$.

At the beginning, a product state $\ket\psi$ on each site gives $\bra\psi^{\otimes2r}\rho_{1,2r}(g)\ket\psi^{\otimes2r}=1$ for every $g$, so the initial state weights all permutations equally.

Since regions $A$ and $B$ favor different permutations, the permutation label must change somewhere between the two boundaries; such an interface is a domain wall.
This permutation-spin description is part of the statistical
mechanics of random unitary circuits~\cite{ZhouNahum2019StatisticalMechanics}.
Haar averaging supplies the weights connecting these boundary
conditions through the circuit~\cite{ChanDeLuca2024Projected,YuHoKos2025Mixed}. A wall between labels $g_1$ and $g_2$ costs $d^{-|g_1^{-1}g_2|}$ for every gate it crosses, as shown below.
We use the permutation operators and Gram matrices of
Eqs.~\eqref{eq:du-permutation-operator} and
\eqref{eq:du-permutation-gram}. For a two-site gate $U$ and an operator
$O$ on its $r$ replicas, the normalized Haar integral is~\cite{CollinsSniady2006Haar}
\begin{equation}
\begin{aligned}
&\int dU\,U^{\otimes r}O(U^\dagger)^{\otimes r}\\
&\quad=d^{-2r}\sum_{g,g'\in S_r}
\rho_{2,r}(g)[(G_2^{(r)})^{-1}]_{gg'}\\
&\hspace{36mm}\times\tr[\rho_{2,r}(g')^\dagger O].
\end{aligned}
\label{eq:local-haar-gate-projector}
\end{equation}
This is the orthogonal projection onto the subspace spanned by the
replica permutations. The inverse exists for $d^2\geq r$, which holds at fixed $r$ and large $d$, and its elements satisfy $[(G_2^{(r)})^{-1}]_{gg'}=\mathcal O_r(d^{-2|g^{-1}g'|})$, i.e., $G_2^{(r)}$ is close to the identity.
The circuit--circuit calculation uses Eq.~\eqref{eq:local-haar-gate-projector} at order $2r$, and the mixed calculation at order $r$.

A leg joining two permutation labels gives
$\tr[\rho_{1,r}(g)^\dagger\rho_{1,r}(g')]
=d^rG_1^{(r)}(g,g')$. After removing the common factor $d^r$, Eq.~\eqref{eq:du-permutation-gram} gives the leg weight $G_1^{(r)}(g,g')=d^{-|g^{-1}g'|}$, so a mismatch $g\neq g'$ is suppressed by $|g^{-1}g'|$ powers of $d^{-1}$. Contracting the two incoming legs
of a Haar-averaged gate gives the complete local weight
\begin{equation}
\mathcal W_r(g;g_1,g_2)
=\sum_{g'\in S_r}[(G_2^{(r)})^{-1}]_{gg'}
G_1^{(r)}(g',g_1)G_1^{(r)}(g',g_2).
\label{eq:local-haar-triangle-weight}
\end{equation}
Here $g_1,g_2$ label the incoming legs and $g$ the outgoing pair. Both outgoing legs carry the same label $g$, because the output of Eq.~\eqref{eq:local-haar-gate-projector} is $\rho_{2,r}(g)=\rho_{1,r}(g)^{\otimes2}$. In
particular, we have
\begin{equation}
\mathcal W_r(g;h,h)=\delta_{g,h}.
\label{eq:local-haar-uniform-weight}
\end{equation}
It follows from the invariance of permutation operators, $U^{\otimes r}\rho_{2,r}(h)(U^\dagger)^{\otimes r}=\rho_{2,r}(h)$.

In the large-$d$ limit, neighboring sites favor the same permutation label. For incoming labels $g_1,g_2$, the leading term of Eq.~\eqref{eq:local-haar-triangle-weight} is $d^{-|g^{-1}g_1|-|g^{-1}g_2|}\leq d^{-|g_1^{-1}g_2|}$, with equality when $g$ lies on a shortest path between $g_1$ and $g_2$. A mismatch between neighboring labels therefore costs $d^{-|g_1^{-1}g_2|}$. We describe this interface as $|g_1^{-1}g_2|$ elementary domain walls, each contributing one power of
$d^{-1}$ per step. For a fixed circuit, the leading contributions
therefore come from the shortest allowed domain-wall configurations.

Equation~\eqref{eq:local-haar-uniform-weight} shows that a gate with
equal incoming labels keeps the same outgoing label, so it cannot
create a wall in a uniform region. When the incoming labels differ,
the gate contraction determines how the existing walls continue
through the circuit. These walls can move left or right, and their
interactions are included in the full weight of Eq.~\eqref{eq:local-haar-triangle-weight}. At the bottom product-state
boundary, all permutation labels have the same weight, so a wall
can end there without an additional boundary cost.

The channel replaces one ordinary leg contraction by the
kernel $K_r(g,g')$ of Eq.~\eqref{eq:du-channel-kernel}. Its upper (output) label $g$ and
lower (input) label $g'$ can be different. In terms of the input and output
Choi factors $C^1,C^2$, the local vertex is
\begin{equation}
\begin{aligned}
K_r(g,g')=\tr\bigl[M_{\Ncal}^{\otimes r}
\bigl(&\rho_{1,r}(g'^{-1})_{C^1}\otimes
\rho_{1,r}(g^{-1})_{C^2}\bigr)\bigr].
\end{aligned}
\label{eq:local-haar-choi-vertex}
\end{equation}
For the identity channel, $M_{\Ncal}=\ket{\phi_0}\!\bra{\phi_0}$ and $K_r(g,g')=G_1^{(r)}(g,g')$, so the ordinary leg is recovered. If $g=g'$, the two permutation operators act identically on $C^1$ and $C^2$ and therefore permute complete Choi copies. Each cycle
then closes into a trace of a power of $M_{\Ncal}$, giving
$K_r(g,g)=w_{M_{\Ncal}}(g)$. In Appendix~\ref{subsec:correction_regimes} we show that this diagonal weight dominates.

\subsection{Dominant domain walls}

First consider the circuit--circuit contraction $T_{cc}^{(k,q)}$, which has a physical
$2r$-replica permutation field. Its $A$ boundary is fixed to $s_k$, the permutation of Eq.~\eqref{eq:local-haar-overlap-swap},
whereas its leading measured boundary can choose any element of
$S_r\oplus S_r$.

\begin{lemma}
\label{lem:local-haar-minimum-distance}
Let $s_k=\prod_{i=1}^k(i,r+i)$ and $q=r-k$.
Then
\begin{equation}
\min_{g_1,g_2\in S_r}
\left|s_k^{-1}(g_1\oplus g_2)\right|
=k.
\label{eq:local-haar-minimum-distance}
\end{equation}
Let $e_m$ be the identity permutation on $m$ labels. The minimizers are exactly
\begin{equation}
g_1\oplus g_2=(g\oplus e_q)\oplus(g^{-1}\oplus e_q),
\qquad g\in S_k,
\label{eq:local-haar-optimal-endpoints}
\end{equation}
where $e_q$ is the identity permutation on the $q$ traced replicas.
For every such minimizer,
\begin{equation}
s_k^{-1}(g_1\oplus g_2)=
\prod_{i=1}^k(i,r+g(i)).
\label{eq:local-haar-endpoint-cycles}
\end{equation}
\end{lemma}
The proof is given in Appendix~\ref{app:local-haar-minimum-distance}.

Equation~\eqref{eq:local-haar-endpoint-cycles} contains $k$ swaps, so the mismatch between a minimizing bulk label and the $A$ boundary consists of $k$ elementary domain walls. There is one minimizer for each $g\in S_k$, i.e., $k!$ in total.

Given a fixed element $h=g_1\oplus g_2$ in such minimizing endpoints,
a shortest wall ending at the left spatial boundary crosses $N_A$ gates.
If the final layer has no gate
across the $A$--$B$ cut, the backward domain-wall path starts at layer
$t-1$. We therefore use
\begin{equation}
t\geq N_A+2
\label{eq:local-haar-saturation-time}
\end{equation}
as a sufficient condition that covers both choices of the final layer.
Under this condition, at large $d$ each wall costs $d^{-N_A}$, and the $k$ walls together give $d^{-kN_A}=D_A^{-k}$.

We assume the channel is in the bulk on the $B$ side of the
saturated walls in Fig.~\ref{fig:local-haar-walls}, well separated from
these walls and the circuit boundaries. In Appendix~\ref{subsec:correction_regimes}, we show that its incoming and outgoing labels coincide
with the endpoint $h$ in Eq.~\eqref{eq:local-haar-optimal-endpoints}. For a minimizing $h$, $g$ permutes the active labels $1,\ldots,k$, $g^{-1}$ permutes the active labels $r+1,\ldots,r+k$, and all $2q$ traced labels are fixed. The channel therefore gives
\begin{equation}
\begin{aligned}
K_{2r}(h,h)
&=w_{M_{\Ncal}}(g)w_{M_{\Ncal}}(g^{-1})
[\tr M_{\Ncal}]^{2q}\\
&=w_{M_{\Ncal}}(g)^2.
\end{aligned}
\label{eq:local-haar-saturated-channel-factor}
\end{equation}
The equality uses $\tr M_{\Ncal}=1$ and the identical cycle lengths
of $g$ and $g^{-1}$. Thus the spatial walls supply $D_A^{-k}$, while
the error supplies the squared Choi weight. Summing over the $k!$
allowed endpoints gives the coefficient announced in
Eq.~\eqref{eq:local-haar-three-overlaps}.

\begin{figure}[t]
\centering
\begin{tikzpicture}[scale=0.88, every node/.style={font=\small}]
\draw[thick] (0,0)--(4.4,0);
\draw[thick] (0,0)--(0,4.1);
\draw[thick] (0,4.1)--(4.4,4.1);
\draw[very thick,red] (1.55,4.1)--(1.2,3.75)--(0.8,3.35)--(0.4,2.95)--(0,2.55);
\draw[very thick,blue,dashed] (1.55,4.1)--(1.85,3.5)--(1.35,2.6)--(1.8,1.7)--(1.35,0.85)--(1.65,0);
\draw[decorate,decoration={brace,amplitude=4pt},thick] (0,4.3)--(1.55,4.3);
\draw[decorate,decoration={brace,amplitude=4pt},thick] (1.65,4.3)--(4.4,4.3);
\node at (0.78,4.73) {$A:\ s_k$};
\node at (3.05,4.73) {$B:\ S_r\oplus S_r$};
\node[red,anchor=east] at (0.63,3.6) {$N_A$};
\node[blue] at (2.05,2.5) {$t$};
\node at (3.25,3.4) {$h$};
\fill[red] (3.3,1.4) circle (2.1pt);
\node[right] at (3.4,1.4) {$\Ncal$};
\node at (2.3,-0.35) {initial product state};
\draw[->,thick] (4.75,0.1)--(4.75,3.9);
\node[rotate=90] at (5.03,2) {time};
\end{tikzpicture}
\caption{Domain-wall configurations for the circuit--circuit overlap.
The retained boundary carries $s_k$. The measured boundary favors
$S_r\oplus S_r$, while permutations that map a replica of one circuit factor into the other are suppressed at large $d$. The bulk label $h$ is one of the minimizers in Eq.~\eqref{eq:local-haar-optimal-endpoints}.
One representative of the $k$ elementary walls is shown:
the solid red path crosses $N_A$ gates before exiting through
the spatial boundary. The dashed blue alternative crosses $t$ gates
before reaching the initial boundary if the final layer crosses the
$A$--$B$ cut, and $t-1$ gates otherwise. The initial product boundary leaves the
permutation unpinned. The red dot marks the channel in the uniform
domain on the $B$ side of the saturated wall.}
\label{fig:local-haar-walls}
\end{figure}
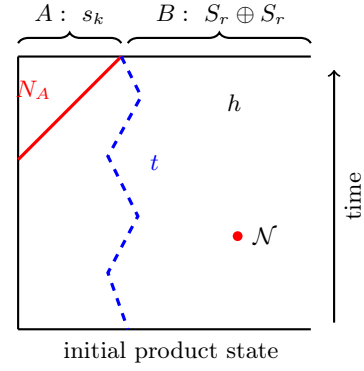

In the mixed overlap $T_{cM}^{(k,q)}$, the first $r$ copies come from the circuit conditional state and the other $r$ copies come from the Ginibre conditional state. Only the circuit factor has a measurement outcome in $B$; the Ginibre factor has no circuit gates or measured sites. The saturated uniform sectors of the single-circuit $r$-replica tensor give
\begin{equation}
\begin{aligned}
\!\!\!\mathbb E_U \!\!\sum_\alpha
\widetilde\rho_{N_A}(\alpha)^{\otimes r}
\!\simeq{}\!D_B^{1-r}D_A^{-r}%
\!\!\sum_{g\in S_r}w_{M_{\Ncal}}(g)\rho_{N_A,r}(g).
\end{aligned}
\label{eq:local-haar-open-circuit-tensor}
\end{equation}
Here, the uncontracted $A$ legs retain $\rho_{N_A,r}(g)$, and the channel supplies $w_{M_{\Ncal}}(g)$.
Each circuit copy contributes the state normalization $(D_AD_B)^{-1}$, as can be seen from the Weingarten coefficient in Eq.~\eqref{eq:local-haar-gate-projector}, while summing the common measurement outcome gives $D_B$, giving the prefactor above.
Averaging the $r$ copies of the Ginibre conditional state supplies another $D_A^{-r}$ by Eq.~\eqref{eq:du-ginibre-wick}.
On $A$, the trace pairs each of the $k$ active circuit copies with the corresponding Ginibre copy, while the other $q$ copies in each factor are traced separately. These contractions are represented by $\rho_{N_A,2r}(s_k)$, with $s_k$ defined in Eq.~\eqref{eq:local-haar-overlap-swap}, and give
\begin{equation}
\begin{aligned}
T_{cM}^{(k,q)}
=&D_A^{-2r}D_B^{1-r}\sum_{g,g'\in S_r}
w_{M_{\Ncal}}(g)w_{M_{\Ncal}}(g')\\
&\times\tr\!\Bigl[
\bigl(\rho_{N_A,r}(g)\otimes\rho_{N_A,r}(g')\bigr)
\rho_{N_A,2r}(s_k)\Bigr]\\
=&D_B^{1-r}\sum_{g,g'\in S_r}
w_{M_{\Ncal}}(g)w_{M_{\Ncal}}(g') D_A^{-|s_k^{-1}(g\oplus g')|}.
\end{aligned}
\label{eq:local-haar-reference-wick-overlap}
\end{equation}
Here we used
\begin{equation}
\begin{aligned}
&\tr\!\left[(\rho_{N_A,r}(g)\otimes\rho_{N_A,r}(g'))
\rho_{N_A,2r}(s_k)\right]\\
&\qquad=D_A^{2r-|s_k^{-1}(g\oplus g')|}.
\end{aligned}
\end{equation}

For the reference overlap $T_{MM}^{(k,q)}$, both factors are Ginibre conditional states. Their averages have the form in Eq.~\eqref{eq:local-haar-open-circuit-tensor}, with the same factor $D_A^{-r}$ and permutation sum, but without the circuit measurement factor $D_B^{1-r}$. Hence, $T_{MM}^{(k,q)}\simeq T_{cM}^{(k,q)}/D_B^{1-r}$.

The leading contribution of $D_A^{-|s_k^{-1}(g\oplus g')|}$ is given by Lemma~\ref{lem:local-haar-minimum-distance}, which selects the endpoints in Eq.~\eqref{eq:local-haar-optimal-endpoints}, with exponent $k$ and coefficient $w_{M_{\Ncal}}(g)w_{M_{\Ncal}}(g^{-1})=w_{M_{\Ncal}}(g)^2$. The remaining terms in the sum carry at least one extra power of $D_A^{-1}$ and can be neglected in the large-$d$ limit.

Combining this with the circuit--circuit result gives
\begin{equation}
\begin{aligned}
T_{cc}^{(k,q)}
&\simeq D_B^{2(1-r)}D_A^{-k}
\sum_{g\in S_k}w_{M_{\Ncal}}(g)^2,\\
T_{cM}^{(k,q)}
&\simeq D_B^{1-r}D_A^{-k}
\sum_{g\in S_k}w_{M_{\Ncal}}(g)^2,\\
T_{MM}^{(k,q)}
&\simeq D_A^{-k}\sum_{g\in S_k}w_{M_{\Ncal}}(g)^2.
\end{aligned}
\label{eq:local-haar-overlap-leading-factors}
\end{equation}
There are two circuit outcome sums in $T_{cc}^{(k,q)}$, one in $T_{cM}^{(k,q)}$, and none in $T_{MM}^{(k,q)}$, giving two, one, and zero factors of $D_B^{1-r}$, respectively.
The continuation $q\to1-k$ sets $r=1$ in the scalar normalization factors, making all three factors equal to each other. Following the standard replica argument, and assuming, as in Sec.~\ref{sec:dual-unitary}, that the continuation can be taken term by term in the large-$d$ expansion, this gives the three physical overlap formulas announced in Eq.~\eqref{eq:local-haar-three-overlaps}.

The identity permutation has $w_{M_{\Ncal}}(e_k)=1$, so the common
coefficient is at least one.
The cancellation of the three leading terms and Cauchy--Schwarz therefore give
\begin{equation}
\lim_{d\to\infty}\mathbb E_U\Delta_2^{(k)}=0.
\label{eq:local-haar-moment-convergence}
\end{equation}
Markov's inequality gives convergence in probability at every fixed
$k$ in this saturated bulk regime. For a unitary channel,
$w_{M_{\Ncal}}(g)=1$ and the common coefficient becomes $k!$, recovering
the pure-state Haar result. A generic channel changes this coefficient
through the spectral moments of its Choi state.

The leading-order result relies on four approximations, discussed
in detail in Appendix~\ref{subsec:correction_regimes}. First, the channel
is assumed to lie in the bulk, far from domain walls and circuit boundaries,
with both labels equal to the surrounding label $h$.
Second, we assume that the leading domain walls end at the left edge rather than on the initial state, and use the
depth condition in Eq.~\eqref{eq:local-haar-saturation-time}.
Third, the measured region is assumed to select only labels in
$S_r\oplus S_r$, requiring $N_B>kN_A$.
Finally, the Haar weights are approximated by their leading terms in $1/d$.
At fixed $k,t,N_A,N_B$, these approximations give relative corrections of order $d^{-1}$ to Eq.~\eqref{eq:local-haar-three-overlaps}.
More specifically, taking these corrections into Eq.~\eqref{eq:local-haar-distance-squared} gives
\begin{equation}
\begin{aligned}
\mathbb E_U\Delta_2^{(k)}
&\leq\sqrt{\mathbb E_U[(\Delta_2^{(k)})^2]}\\
&=\mathcal O\!\Bigl(\bigl[D_A^{-1}\!+\!d^{-2}\\
&\qquad\!+2^{kt}d^{-(t-1-N_A)}
\!+\!d^{-(N_B-kN_A)}\bigr]^{\frac{1}{2}}\Bigr)\\
&=\mathcal O(d^{-\frac{1}{2}}).
\end{aligned}
\label{eq:local-haar-error-scale}
\end{equation}
The first inequality is Cauchy--Schwarz, and the last equality uses fixed $k,t,N_A,N_B$ with $t\geq N_A+2$ and $N_B>kN_A$.
Although the above analytical bound decreases with $t$ at fixed $d$ only when $d>2^k$, numerical results show good convergence even for small $d$.
The behavior at $d=2$ is examined numerically in Sec.~\ref{sec:numerics}.

\subsection{Dilute well-separated errors}

For a fixed number of errors, the leading bulk region contains every channel, so all channel vertices carry the same label. When the errors are separated from
one another, the boundaries, and the saturated walls by long
scrambling regions, a correction connecting different channel
vertices has an extra wall cost. Using the channel kernels $K_{r,i}$
defined in Eq.~\eqref{eq:du-dilute-recursion}, the circuit--circuit channel factor at the minimizing label $h$ of Eq.~\eqref{eq:local-haar-optimal-endpoints} is
\begin{equation}
\prod_i K_{2r,i}(h,h)
=\prod_i w_{M_{\Ncal_i}}(g)^2
=w_{\bigotimes_iM_{\Ncal_i}}(g)^2.
\label{eq:local-haar-dilute-factorization}
\end{equation}
The second equality follows from Eq.~\eqref{eq:du-dilute-factorization}. Thus the three-overlap calculation repeats with
$M=\bigotimes_iM_{\Ncal_i}$. This is the same tensor product of the
individual Choi states as in
Secs.~\ref{sec:dual-unitary} and \ref{sec:global-haar}.

\section{Numerical verification in qubit systems}
\label{sec:numerics}
We test the CIE prediction at $d=2$ in three local dynamical models, varying the measured size $N_B$.
The evolution time is fixed and lasts 80 even--odd layer pairs or Floquet periods, which lies in the regime $t\gg N_A+N_B$. We consider an open chain initialized in
$\ket{+}^{\otimes(N_A+N_B)}$, where
$\ket{+}=(\ket0+\ket1)/\sqrt2$. After the evolution, we retain the first $N_A=2$ qubits,
and measure the remaining $N_B=2,4,\ldots,22$ qubits in the
computational basis. A single-qubit channel acts halfway through
the evolution on site $(N_A+N_B)/2+1$, with sites numbered from one.
The same channel is used for every system size and dynamical model.

The first two models are random dual-unitary and local Haar circuits,
as in Ref.~\cite{YuHoKos2025Mixed}.
Both have alternating even and odd nearest-neighbor layers, and their
gates are sampled independently in space and time. In the random dual-unitary case, we write
$\sigma^{x,y,z}$ for the Pauli operators, and fix 
the interacting part of the gate to $\exp[i(\frac{\pi}{4}\sigma^x\otimes\sigma^x+\frac{\pi}{4}\sigma^y\otimes\sigma^y+J\sigma^z\otimes\sigma^z)]$, while sampling $J$ uniformly from $[0,2\pi)$, and applying
independent single-qubit Haar rotations on all four legs.
In the local Haar case, we use Haar-random two-qubit gates in the circuit.
The third model is kicked Ising dynamics. Writing $\sigma_j^{y,z}$
for the Pauli operators on site $j$, its Floquet operator is
\begin{equation}
\begin{aligned}
U_F={}&
\exp\left[-ih \sum_j\sigma_j^y\right]
\exp\left[-iJ \sum_j\sigma_j^z\right]\\
&\times\exp\left[-i\gamma \sum_j\sigma_j^z\sigma_{j+1}^z\right].
\end{aligned}
\label{eq:numerics-kicked-ising}
\end{equation}
 The choice of parameters $(h,J,\gamma)=(0.9, 0.6, 0.7)$ lies away from the dual-unitary point. 

We compare three channels: the partial reset
$\Ncal(\rho)=0.8\rho+0.2\,\tr(\rho)\ket0\!\bra0$;
a random-unitary channel formed from four independent Haar
single-qubit unitaries, with weights drawn uniformly from the
probability simplex; and a generic channel with four Kraus operators.
For the latter, we draw four independent $2\times2$ complex Gaussian
matrices and multiply them on the right by a common positive matrix
that enforces trace preservation.
Each random channel is drawn once and then held fixed throughout
the comparison.

\begin{figure*}[t]
\centering
\includegraphics[width=\textwidth]{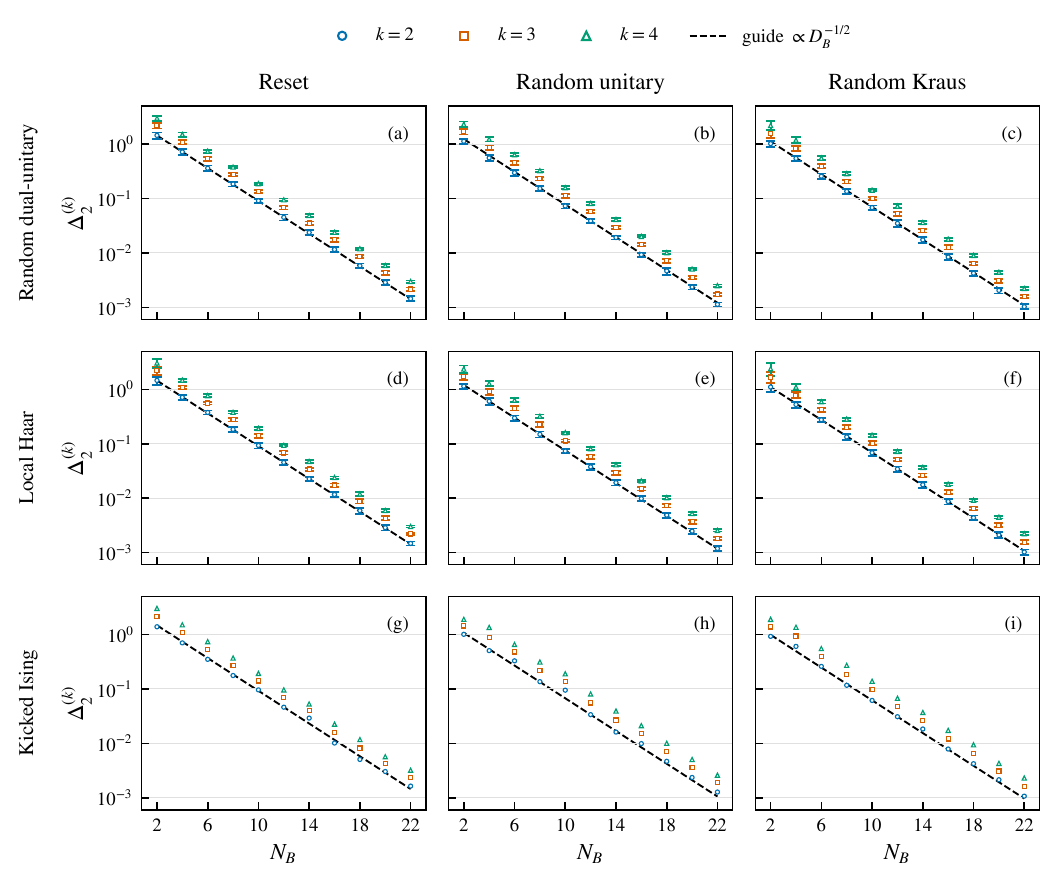}
\caption{Numerical moment distances from the CIE at $d=2$ and $N_A=2$. Rows show random dual-unitary circuits, local
Haar circuits, and kicked Ising dynamics; columns show the partial
reset, random-unitary, and generic random Kraus channels.
Circles, squares, and triangles denote $k=2,3,4$, respectively.
The evolution consists of 80 even--odd layer pairs for the random
circuits and 80 periods for kicked Ising, with the channel applied halfway.
In both random circuit cases, we show means of the distance $\Delta_2^{(k)}$ over 25 realizations, with error
bars indicating the standard deviation. The kicked Ising dynamics use
the evolution in Eq.~\eqref{eq:numerics-kicked-ising}.
Black dashed lines show the scaling $D_B^{-1/2}\propto2^{-N_B/2}$. In each panel, the line passes through the $k=2$ value at $N_B=12$.}
\label{fig:numerics-size-scaling}
\end{figure*}

For each evolution, we construct the Born-weighted moments in
Eq.~\eqref{eq:mspe-moment} for $k=2,3,4$ and evaluate
$\Delta_2^{(k)}$ in Eq.~\eqref{eq:moment-distance}, using
$\rho_{M_{\Ncal}}^{(k)}$ as the reference.
In the upper two rows of Fig.~\ref{fig:numerics-size-scaling}, markers show the mean over 25 circuit realizations, and vertical error bars show one standard deviation.
The kicked Ising evolution is fixed.
We enumerate the computational-basis outcomes of $B$ and sum their contributions. The CIE moments are evaluated numerically using Eq.~\eqref{eq:choi-ensemble-moment}.

Figure~\ref{fig:numerics-size-scaling} shows an exponential decay to the CIE of the form $D_B^{-1/2}\propto2^{-N_B/2}$ over the studied sizes of $B$. This behavior is consistent with the global-Haar bound of Sec.~\ref{sec:global-haar}. 
The dual-unitary simulations use a product initial state and single-site measurements, beyond the exactly solvable boundary conditions of Sec.~\ref{sec:dual-unitary}.
The similar size dependence in all three models supports the CIE prediction in generic chaotic systems, including a fixed kicked Ising evolution away from the dual-unitary point.

\section{Discussion and outlook}

\label{sec:discussion}
We have shown that dilute local errors select the CIE as the limiting mixed-state projected ensemble. This ensemble is fixed entirely by the spectrum of the joint normalized Choi state [Eq.~\eqref{eq:du-effective-choi-state}] and is insensitive to the microscopic unitary dynamics. This answers the question posed in the Introduction: errors that differ only by unitary rotations select the same ensemble, whereas errors with equal Choi purity but different spectra, such as suitably chosen partial depolarization and partial reset, select different ones. Assuming the errors are well separated, we establish this result analytically for dual-unitary circuits, global Haar-random dynamics, and local Haar-random circuits at large local dimension. Numerical simulations indicate that it also holds for other generic chaotic dynamics. In dual-unitary circuits, we show that the approach is exponential in time, and its dependence on the moment order varies with the channel.

A finite density of errors poses a further challenge. Error channels that act close in space or time need not be separated by the long scrambling regions assumed in the dilute limit, so our derivation does not apply directly. An open question is whether their combined effect can still be captured by a correlated effective Choi state. Answering it will likely require more powerful numerical and analytical methods.

Errors are unavoidable in noisy intermediate-scale quantum (NISQ) devices, so the pure-state Haar ensemble that emerges from deep thermalization in closed systems describes their projected ensembles only approximately. In the dilute-error regime studied here, the CIE gives a more realistic description of the random mixed states that scrambling and measurement generate on such hardware, and is hence the appropriate reference for benchmarking based on emergent randomness~\cite{ChoiEtAl2023PreparingRandom,MarkEtAl2023BenchmarkingErgodic}. Beyond describing these states, the CIE suggests a way to characterize the noise itself from the measured distribution of conditional states. Because this distribution depends only on the spectrum of the normalized Choi state after sufficient scrambling, it could be used to infer that spectrum even when the times and locations of the errors are unknown, complementing noise learning from bit-string distributions~\cite{ShawEtAl2025NoiseLearning,SauliereLamiBoyerDeNardisDeLuca2026NoisyAnticoncentration}. For several well-separated errors, the same approach would yield the spectrum of their joint Choi state without requiring the individual channels to be identified. Engineered channels, such as mid-circuit partial resets, would allow a direct test, and the finite-time corrections would further probe the input--output structure of the Choi state.

On noisy hardware, applications that assume emergent Haar randomness, such as shadow tomography~\cite{McGinleyFava2023ShadowTomography} and randomness conversion~\cite{MokEtAl2025ClassicalRandomness}, should use the CIE moments instead. The CIE may also be useful for quantum information tasks such as data hiding~\cite{AubrunLancien2015} and resource hiding~\cite{Bansal2025}, which use random mixed states. With the number of errors controlling the rank of the conditional states, scrambling dynamics with local errors could provide a tunable physical route to preparing states for these tasks. It would be interesting to identify the spectra of the normalized Choi state for which the CIE has these hiding properties.

Symmetries and conservation laws offer another natural direction. In closed systems with conservation laws, the limiting projected ensemble is the generalized Scrooge ensemble~\cite{ChangShrotriyaHoIppoliti2025Charge,MarkEtAl2024MaximumEntropy} rather than the Haar ensemble. Such Scrooge ensembles, which minimize the accessible information for a given average state~\cite{JozsaRobbWootters1994Accessible}, are emerging as a universal description of many-body randomness under physical constraints~\cite{MokHaugHoPreskill2026NatureStingy,LiuZhang2026SYKScrooge,RussottoKlobasCalabreseBertini2026GaussianScrooge,SarmaHaugPreskillMok2026EquilibriumMagic,PawlikSierantZakrzewski2026RestrictedTypicality,MarkEndresIppolitiHoChoi2026Review}.

Our approach could be extended to such constrained settings, where local errors would replace the Scrooge ensemble by a mixed-state counterpart, as they replace the Haar ensemble by the CIE. This raises the question of how conserved charges and local noise together determine the limiting ensemble once errors are present. Denoting $\rho_A:=\rho_{N_A}^{(1)}$ as the average reduced state on $A$, a natural candidate replaces $XMX^\dagger$ in Eq.~\eqref{eq:correlated-wishart-state} by $D_A\,\rho_A^{1/2}XMX^\dagger\rho_A^{1/2}$, with an effective
positive-semidefinite matrix $M$ of unit trace. This combines the Gaussian form of the Scrooge ensemble of $\rho_A$~\cite{GoldsteinLebowitzTumulkaZanghi2006GAP} with the CIE. It reduces to the former for a unitary error and to the latter for $\rho_A=I_A/D_A$. The channel can affect both $\rho_A$ and $M$, and $M$ may not
coincide with the channel's normalized Choi state.
Determining this effective matrix, and understanding how the limiting
ensemble depends on whether the channel respects the symmetry, and
on the measurement basis, are interesting questions.

More fundamentally, it would be interesting to derive the CIE from a general principle such as maximum entropy. Such a principle has already been formulated for pure-state deep thermalization~\cite{MarkEtAl2024MaximumEntropy} and for thermal wave-function ensembles~\cite{WillsonHeelwegWillard2026MaxEnt}. Extending it to the mixed-state setting raises the question of which information about the noise must enter. The candidate above suggests that such a principle should account
for both the average reduced state and an effective matrix $M$. Such a principle would also connect the CIE with other universal mixed-state structures in chaotic dynamics~\cite{SherryRoy2025Mixed,MandalClaeysRoy2025PartialProjected,HuangMokHaugHo2026ProjectedKraus}.

\begin{acknowledgments}

We thank Wen Wei Ho for collaboration on related work~\cite{YuHoKos2025Mixed} and for insightful discussions.
X.Y. acknowledges the Peterson postdoctoral fellowship from the Department of Computing and Mathematical Sciences at Caltech. X.Y. acknowledges the Simons Targeted Grant in Mathematics and Physical Sciences on Moir\'e Materials Magic, Award No. 896630.
P.K. acknowledges funding from the European Union’s Horizon Europe research and innovation programme under the Marie Skłodowska-Curie grant agreement No. 101200860 (FAQ-QuantuM2D) and the Slovenian Research and Innovation Agency (ARIS) through grant J1-70061.

\emph{AI disclosure}---We acknowledge the use of GPT (OpenAI) and Claude Opus (Anthropic) for assistance with technical derivations, literature search, and proofreading of the manuscript. The authors take full responsibility for the accuracy and integrity of the manuscript and its conclusions.
\end{acknowledgments} 
\clearpage
\appendix
\onecolumngrid

\section{Dual-unitary circuit calculations}
\label{sec:replica-identities}

We give the replica identities and finite-time coefficients used in
Sec.~\ref{sec:dual-unitary}.

\subsection{Channel contractions for the $d^{-t}$ correction \label{Appendix:Sec:A:first:order}}

Let $r\geq1$ be the integer replica order and let $g,g'\in S_r$,
where $S_r$ is the group of permutations of the $r$ replicas.
In Eq.~\eqref{eq:du-leading-finite-time-correction}, the permutation coefficient $\alpha_{\mathrm f}(g)$ differs from its infinite-time value $w_{M_{\Ncal}}(g)$ by terms of order $d^{-t}$. To evaluate this correction, we calculate $K_r(g,gs_{ij})$, where $s_{ij}$ exchanges replicas $i$ and $j$.
We calculate this kernel for the channel $\Ncal$ on a
$d$-dimensional site, with normalized Choi state $M_{\Ncal}$.
All component indices below run from $1$ to $d$.
For input indices $\mu,\nu$ and output indices $\alpha,\gamma$,
the normalized Choi state satisfies
\begin{equation}
\bra\alpha\Ncal(\ket\mu\!\bra\nu)\ket\gamma
=d\,\bra{\mu,\alpha}M_{\Ncal}\ket{\nu,\gamma}.
\label{eq:appendix-choi-components}
\end{equation}
The factor $d$ comes from using the normalized Choi state, $\tr M_{\Ncal}=1$.
Substituting into Eq.~\eqref{eq:du-channel-kernel}, the factor
$d^r$ cancels the kernel normalization. The permutation matrix
elements then give
\begin{equation}
\begin{aligned}
K_r(g,g')={}&\sum_{\mu_1,\ldots,\mu_r}
\sum_{\alpha_1,\ldots,\alpha_r}\prod_{m=1}^r
(M_{\Ncal})_{(\mu_{g'(m)},\alpha_{g(m)}),
(\mu_m,\alpha_m)}.
\end{aligned}
\label{eq:appendix-kernel-components}
\end{equation}
The indices $\mu_m$ belong to $C^1$, the reference copy of the channel input, while $\alpha_m$ belong to $C^2$, the channel output. In Eq.~\eqref{eq:appendix-kernel-components}, $g'$ specifies the pairing of the $\mu$ indices, and $g$ specifies the pairing of the $\alpha$ indices.
For $g'=g$, the input
and output indices are summed in the same order along each cycle.
A cycle $c$ of length $|c|$ therefore gives $\tr M_{\Ncal}^{|c|}$.
For integer $s\geq1$, we denote these traces and their partial
traces by
\begin{equation}
p_s=\tr M_{\Ncal}^s,\qquad
Q_s=\tr_2M_{\Ncal}^s,\qquad
H_s=\tr_1M_{\Ncal}^s.
\label{eq:appendix-choi-invariants}
\end{equation}
The traces $\tr_1$ and $\tr_2$ are over the reference copy of the channel input and the channel output of the Choi
factors, respectively. Trace preservation gives $Q_1=I_d/d$,
where $I_d$ is the identity on one site, and normalization gives
$p_1=1$. The diagonal kernel is therefore
$K_r(g,g)=\prod_{c\in\mathcal C(g)}p_{|c|}
=w_{M_{\Ncal}}(g)$, where $\mathcal C(g)$ is the set of cycles of
$g$.

We use the composition $(gs_{ij})(m)=g(s_{ij}(m))$. Setting
$g'=gs_{ij}$ changes only the input indices in factors $i$ and
$j$ of the product:
\[
\begin{aligned}
m=i:&\quad
(M_{\Ncal})_{(\mu_{g(j)},\alpha_{g(i)}),(\mu_i,\alpha_i)},\\
m=j:&\quad
(M_{\Ncal})_{(\mu_{g(i)},\alpha_{g(j)}),(\mu_j,\alpha_j)}.
\end{aligned}
\]
All other factors are unchanged, so every cycle not containing
$i$ or $j$ still contributes $p_{|c|}$.

If $i$ and $j$ belong to distinct cycles $c_i,c_j$ of lengths
$a,b$, we have
\begin{equation}
\begin{aligned}
K_r(g,gs_{ij})
&=\sum_{\mu_1,\ldots,\mu_r}\sum_{\alpha_1,\ldots,\alpha_r}
\prod_{m=1}^r
(M_{\Ncal})_{(\mu_{g(s_{ij}(m))},\alpha_{g(m)}),(\mu_m,\alpha_m)}\\
&=\sum_{\mu_{g(i)},\mu_{g(j)}}
\sum_{\alpha_{g(i)},\alpha_{g(j)}}
(M_{\Ncal}^{a})_{(\mu_{g(j)},\alpha_{g(i)}),(\mu_{g(i)},\alpha_{g(i)})}
(M_{\Ncal}^{b})_{(\mu_{g(i)},\alpha_{g(j)}),(\mu_{g(j)},\alpha_{g(j)})}
\prod_{c\notin\{c_i,c_j\}}p_{|c|}\\
&=\sum_{\mu_{g(i)},\mu_{g(j)}}
(Q_a)_{\mu_{g(j)},\mu_{g(i)}}
(Q_b)_{\mu_{g(i)},\mu_{g(j)}}
\prod_{c\notin\{c_i,c_j\}}p_{|c|}\\
&=\tr(Q_aQ_b)\prod_{c\notin\{c_i,c_j\}}p_{|c|}.
\end{aligned}
\label{eq:appendix-different-cycle-contraction}
\end{equation}
The first equality uses Eq.~\eqref{eq:appendix-kernel-components} with $g'=gs_{ij}$. The third equality uses the definition $Q_s=\tr_2M_{\Ncal}^s$. The second equality is more involved.
To obtain the second equality, consider the factors belonging to the original cycle $c_i$ of $g$. Summing their intermediate indices gives
\[
\begin{aligned}
&\sum_{\substack{\mu_m,\alpha_m\\m\in c_i\setminus\{g(i)\}}}
\prod_{m\in c_i}
(M_{\Ncal})_{(\mu_{g'(m)},\alpha_{g(m)}),(\mu_m,\alpha_m)}\\
&\qquad=
(M_{\Ncal}^{a})_{(\mu_{g(j)},\alpha_{g(i)}),(\mu_{g(i)},\alpha_{g(i)})}.
\end{aligned}
\]
Summing the intermediate indices of the \(a\) factors on \(c_i\) gives the displayed matrix element of \(M_{\mathcal N}^{a}\). Their remaining input row and column indices are $\mu_{g(j)}$ and $\mu_{g(i)}$, respectively, while both output indices are $\alpha_{g(i)}$. The cycle $c_j$ gives the displayed factor $M_{\Ncal}^{b}$ in the same way. Multiplying these two expressions and summing the remaining indices gives the second equality in Eq.~\eqref{eq:appendix-different-cycle-contraction}.

Since $s_{ij}$ joins the two cycles, we have
\[
w_{M_{\Ncal}}(gs_{ij})=
p_{a+b}\prod_{c\in\mathcal C(g)\setminus\{c_i,c_j\}}p_{|c|}.
\]

If $i$ and $j$ belong to the same cycle $c$ of length $\ell$, define
$1\leq a_{ij}\leq\ell-1$ by $g^{a_{ij}}(i)=j$. Then
\begin{equation}
\begin{aligned}
K_r(g,gs_{ij})
&=\sum_{\mu_1,\ldots,\mu_r}\sum_{\alpha_1,\ldots,\alpha_r}
\prod_{m=1}^r
(M_{\Ncal})_{(\mu_{g(s_{ij}(m))},\alpha_{g(m)}),(\mu_m,\alpha_m)}\\
&=\sum_{\alpha_{g(i)},\alpha_{g(j)}}
\sum_{\mu_{g(i)},\mu_{g(j)}}
(M_{\Ncal}^{a_{ij}})_{(\mu_{g(i)},\alpha_{g(j)}),(\mu_{g(i)},\alpha_{g(i)})}
(M_{\Ncal}^{\ell-a_{ij}})_{(\mu_{g(j)},\alpha_{g(i)}),(\mu_{g(j)},\alpha_{g(j)})}
\prod_{c'\ne c}p_{|c'|}\\
&=\sum_{\alpha_{g(i)},\alpha_{g(j)}}
(H_{a_{ij}})_{\alpha_{g(j)},\alpha_{g(i)}}
(H_{\ell-a_{ij}})_{\alpha_{g(i)},\alpha_{g(j)}}
\prod_{c'\ne c}p_{|c'|}\\
&=\tr(H_{a_{ij}}H_{\ell-a_{ij}})\prod_{c'\ne c}p_{|c'|}.
\end{aligned}
\label{eq:appendix-same-cycle-contraction}
\end{equation}
The first equality again uses Eq.~\eqref{eq:appendix-kernel-components}.
For the second equality, consider the factors with replica labels $g(i),g^2(i),\ldots,g^{a_{ij}}(i)=j$. These $a_{ij}$ factors give the first matrix power in the second line: the remaining row indices are $(\mu_{g(i)},\alpha_{g(j)})$, and the column indices are $(\mu_{g(i)},\alpha_{g(i)})$. The remaining $\ell-a_{ij}$ factors have replica labels $g(j),g^2(j),\ldots,i$ and give the second matrix power.
The third equality uses $H_s=\tr_1M_{\Ncal}^s$ to sum
these input indices. 

Since $s_{ij}$ splits the cycle,
\[
w_{M_{\Ncal}}(gs_{ij})=
p_{a_{ij}}p_{\ell-a_{ij}}
\prod_{c'\in\mathcal C(g)\setminus\{c\}}p_{|c'|}.
\]
Thus the first-order correction depends on both partial traces
of the Choi powers, in addition to their full traces.

\subsection{Finite-time correction}
We calculate the finite-time difference between the physical moment
$\rho_{N_A}^{(k)}$ and the reference moment
$\rho_{M_{\Ncal}}^{(k)}$ for fixed integer $k\geq1$ at the order of $\mathcal{O}(d^{-t})$. The quantities
of interest are
\[
\Delta_\xi^{(k)}=
\frac{\|\rho_{N_A}^{(k)}-\rho_{M_{\Ncal}}^{(k)}\|_\xi}
{\|\rho_{M_{\Ncal}}^{(k)}\|_\xi},\qquad \xi=1,2.
\]
Here $D_A=d^{N_A}$ is the dimension of the retained region, and
$\|O\|_\xi=(\tr[(O^\dagger O)^{\xi/2}])^{1/\xi}$ denotes the
trace norm for $\xi=1$ and the Hilbert--Schmidt norm for $\xi=2$.
We first obtain the coefficients of this difference from the raw
replica moments, and then evaluate the two norms.

For integer replica order $r$, the raw coefficient of a permutation
$g\in S_r$ obeys Eq.~\eqref{eq:du-leading-finite-time-correction}:
\begin{equation}
\alpha_{\mathrm f}(g)
=w_{M_{\Ncal}}(g)+d^{-t}\sum_{1\leq i<j\leq r}
\left[K_r(g,gs_{ij})-\frac1d w_{M_{\Ncal}}(gs_{ij})\right]
+\mathcal O_r(d^{-2t}).
\label{eq:appendix-raw-coefficient-expansion}
\end{equation}
We denote
the sum in the square brackets by
\begin{equation}
F_r(g)=\sum_{1\leq i<j\leq r}
\left[K_r(g,gs_{ij})-\frac1d w_{M_{\Ncal}}(gs_{ij})\right].
\label{eq:appendix-F-definition}
\end{equation}
Thus $F_r(g)$ is the first-order correction to the raw coefficient.

To calculate the difference $\rho_{N_A}^{(k)}-\rho_{M_{\Ncal}}^{(k)}$, we take an integer $q\geq0$ and set $r=k+q$, as in Sec.~\ref{sec:du-replica-construction}.
The first $k$ copies remain open and are called active replicas; the other $q$ copies are traced out. For a permutation $\pi\in S_r$, let $\bar\pi\in S_k$ be obtained by deleting the traced labels $k+1,\ldots,k+q$ from its cycles. Let $c_{\mathrm{tr}}(\pi)$ count the cycles containing only the traced labels $k+1,\ldots,k+q$.  Write
$\rho_{N_A,r}(\pi)$ for the permutation operator on $r$ copies of
the retained region. Then
\begin{equation}
\tr_{k+1,\ldots,k+q}\rho_{N_A,r}(\pi)
=D_A^{c_{\mathrm{tr}}(\pi)}\rho_{N_A,k}(\bar\pi).
\label{eq:appendix-spectator-trace}
\end{equation}
Each cycle containing only traced labels gives a summed basis index and hence a factor $D_A$. A cycle containing active labels becomes
the corresponding cycle of $\bar\pi$.

For $t+1\geq N_A$, the retained-region map carries the raw
permutation expansion to $A$ with unchanged coefficients. We
normalize these expansions, with weights $\alpha_{\mathrm f}$ for
the circuit and $w_{M_{\Ncal}}$ for the reference, by their traces.
Tracing the additional $q$ copies gives
\begin{equation}
\begin{aligned}
\tr_{k+1,\ldots,k+q}\overline\rho_{N_A}^{(r)}
&=\frac{\displaystyle\sum_{\pi\in S_r}\alpha_{\mathrm f}(\pi)
\tr_{k+1,\ldots,k+q}\rho_{N_A,r}(\pi)}
{\displaystyle\sum_{\pi\in S_r}\alpha_{\mathrm f}(\pi)
D_A^{|\mathcal C(\pi)|}}=\frac{\displaystyle\sum_{\pi\in S_r}\alpha_{\mathrm f}(\pi)
D_A^{c_{\mathrm{tr}}(\pi)}\rho_{N_A,k}(\bar\pi)}
{\displaystyle\sum_{\pi\in S_r}\alpha_{\mathrm f}(\pi)
D_A^{|\mathcal C(\pi)|}},\\
\tr_{k+1,\ldots,k+q}\overline\rho_{M_{\Ncal}}^{(r)}
&=\frac{\displaystyle\sum_{\pi\in S_r}w_{M_{\Ncal}}(\pi)
D_A^{c_{\mathrm{tr}}(\pi)}\rho_{N_A,k}(\bar\pi)}
{\displaystyle\sum_{\pi\in S_r}w_{M_{\Ncal}}(\pi)
D_A^{|\mathcal C(\pi)|}}.
\end{aligned}
\label{eq:appendix-traced-raw-moments}
\end{equation}
The first equality uses $\tr\rho_{N_A,r}(\pi)=D_A^{|\mathcal C(\pi)|}$, and the same for the
reference expression.

We now subtract the circuit and reference expressions. Let $g\in S_k$.
The permutation $g\oplus e_q\in S_{k+q}$ acts as $g$ on labels $1,\ldots,k$ and leaves labels $k+1,\ldots,k+q$ unchanged.
The subtraction gives 
\begin{equation}\label{Eq:appendix:A:different_row}
\begin{aligned}
&\tr_{k+1,\ldots,k+q}
\left(\overline\rho_{N_A}^{(r)}-\overline\rho_{M_{\Ncal}}^{(r)}\right)\\
&\quad=d^{-t}\sum_{\pi\in S_r}
\left[
\frac{F_r(\pi)}
{\displaystyle\sum_{\nu\in S_r}w_{M_{\Ncal}}(\nu)D_A^{|\mathcal C(\nu)|}}
-\frac{\displaystyle w_{M_{\Ncal}}(\pi)
\sum_{\nu\in S_r}F_r(\nu)D_A^{|\mathcal C(\nu)|}}
{\displaystyle\left[\sum_{\nu\in S_r}w_{M_{\Ncal}}(\nu)
D_A^{|\mathcal C(\nu)|}\right]^2}
\right]D_A^{c_{\mathrm{tr}}(\pi)}\rho_{N_A,k}(\bar\pi)
+\frac1{D_A^k}\sum_{g\in S_k}\mathcal O_r(d^{-2t})\rho_{N_A,k}(g)\\
&\quad=\frac1{D_A^k}\sum_{g\in S_k}
\left\{
d^{-t}\left[F_r(g\oplus e_q)-F_r(e_r)w_{M_{\Ncal}}(g)\right]
+\mathcal O_r(d^{-2t}+d^{-t}D_A^{-1})
\right\}\rho_{N_A,k}(g).
\end{aligned}
\end{equation}
The first equality inserts
$\alpha_{\mathrm f}(\pi)=w_{M_{\Ncal}}(\pi)+d^{-t}F_r(\pi)
+\mathcal O_r(d^{-2t})$ and expands the circuit denominator.
The zeroth-order terms cancel exactly. The first term in brackets
comes from the numerator correction; the second comes from the
trace normalization.
The second equality keeps the leading powers of $D_A$ in this
first-order difference. In the numerator,
$\pi=g\oplus e_q$ gives $c_{\mathrm{tr}}(\pi)=q$ and
$w_{M_{\Ncal}}(g\oplus e_q)=w_{M_{\Ncal}}(g)$ because $p_1=1$.
Every permutation that moves a traced label has $c_{\mathrm{tr}}(\pi)\leq q-1$. In the denominator, $e_r$ gives
$D_A^r$ and $w_{M_{\Ncal}}(e_r)=1$, while every other permutation
has $|\mathcal C(\nu)|\leq r-1$.
These discarded contributions to the first-order bracket are
therefore suppressed by at least one factor of $D_A^{-1}$, giving the error
$d^{-t}D_A^{-1}$ in the last line.

We now evaluate the two $F$ terms in this difference. The label
pairs in $F_{k+q}(g\oplus e_q)$ fall into three sets:
\begin{equation}
\begin{aligned}
F_{k+q}(g\oplus e_q)
={}&\sum_{1\leq i<j\leq k}
\left[K_{k+q}(g\oplus e_q,(g\oplus e_q)s_{ij})
-\frac{w_{M_{\Ncal}}((g\oplus e_q)s_{ij})}{d}\right]\\
&+\sum_{i=1}^k\sum_{j=k+1}^{k+q}
\left[K_{k+q}(g\oplus e_q,(g\oplus e_q)s_{ij})
-\frac{w_{M_{\Ncal}}((g\oplus e_q)s_{ij})}{d}\right]\\
&+\sum_{k+1\leq i<j\leq k+q}
\left[K_{k+q}(g\oplus e_q,(g\oplus e_q)s_{ij})
-\frac{w_{M_{\Ncal}}((g\oplus e_q)s_{ij})}{d}\right].
\end{aligned}
\label{eq:appendix-first-order-pair-sums}
\end{equation}
The three sums pair two active labels, one active and one traced label, and two traced labels, respectively. In the first sum, all cycles containing only traced labels are unchanged and contribute $p_1=1$.
For two active labels $i,j$, the summand therefore reduces to the
one at replica order $k$. If their cycles $c_i,c_j$ are distinct,
write $a=|c_i|$ and $b=|c_j|$. If they lie in the same cycle $c$,
write $\ell=|c|$ and let $a_{ij}$ satisfy $g^{a_{ij}}(i)=j$,
with $1\leq a_{ij}\leq\ell-1$. The two contractions in
Appendix~\ref{Appendix:Sec:A:first:order} give
\begin{equation}
\begin{aligned}
&K_{k+q}(g\oplus e_q,(g\oplus e_q)s_{ij})
-\frac{w_{M_{\Ncal}}((g\oplus e_q)s_{ij})}{d}=K_k(g,gs_{ij})-\frac{w_{M_{\Ncal}}(gs_{ij})}{d}\\
&\qquad=\begin{cases}
\displaystyle\left[\tr(Q_aQ_b)-\frac{p_{a+b}}d\right]
\prod_{c\in\mathcal C(g)\setminus\{c_i,c_j\}}p_{|c|},
&c_i\ne c_j,\\[2mm]
\displaystyle\left[\tr(H_{a_{ij}}H_{\ell-a_{ij}})
-\frac{p_{a_{ij}}p_{\ell-a_{ij}}}d\right]
\prod_{c'\in\mathcal C(g)\setminus\{c\}}p_{|c'|},
&c_i=c_j=c.
\end{cases}
\end{aligned}
\label{eq:appendix-active-pair-contraction}
\end{equation}

For an active label $i$ and a traced label $j$, let $c_i\in\mathcal C(g)$ be the active cycle containing $i$. The different-cycle contraction
in Appendix~\ref{Appendix:Sec:A:first:order} gives
\begin{equation}
\begin{aligned}
&K_{k+q}(g\oplus e_q,(g\oplus e_q)s_{ij})
-\frac{w_{M_{\Ncal}}((g\oplus e_q)s_{ij})}{d}=\left[\tr(Q_{|c_i|}Q_1)-\frac{p_{|c_i|+1}}{d}\right]
\prod_{c\in\mathcal C(g)\setminus\{c_i\}}p_{|c|}\\
&\qquad=\frac{p_{|c_i|}-p_{|c_i|+1}}{d}
\prod_{c\in\mathcal C(g)\setminus\{c_i\}}p_{|c|}.
\end{aligned}
\label{eq:appendix-active-traced-contraction}
\end{equation}
The first equality uses Eq.~\eqref{eq:appendix-different-cycle-contraction}. The second
uses $Q_1=I_d/d$ and $\tr Q_s=p_s$. The result is independent of which traced copy is chosen. This is the contribution of one active and one traced copy to $F_{k+q}(g\oplus e_q)$, which we denote by $\eta_i(g)$.
\begin{equation}
\eta_i(g)=\frac{p_{|c_i|}-p_{|c_i|+1}}{d}
\prod_{c\in\mathcal C(g)\setminus\{c_i\}}p_{|c|}.
\label{eq:appendix-chi-eta-definition}
\end{equation}
It is a contribution per label pair, so a cycle of length $a$
contributes $a$ terms to $\sum_i\eta_i(g)$.

For two traced labels $i,j$, the same contraction gives
\begin{equation}
\begin{aligned}
K_{k+q}(g\oplus e_q,(g\oplus e_q)s_{ij})
-\frac{w_{M_{\Ncal}}((g\oplus e_q)s_{ij})}{d}
&=\left[\tr(Q_1^2)-\frac{p_2}{d}\right]w_{M_{\Ncal}}(g)=\frac{1-p_2}{d}w_{M_{\Ncal}}(g).
\end{aligned}
\label{eq:appendix-traced-pair-contraction}
\end{equation}
In the first equality, all
active cycles remain unchanged and give $w_{M_{\Ncal}}(g)$.
The second equality uses $Q_1=I_d/d$. We denote the common scalar by $\chi=(1-p_2)/d$, so each pair of traced copies contributes $\chi w_{M_{\Ncal}}(g)$.

There are $q$ choices of traced copy for each active label and $\binom q2$ pairs of traced copies. Thus
\begin{equation}
\begin{aligned}
F_{k+q}(g\oplus e_q)
&=F_k(g)+q\sum_{i=1}^k\eta_i(g)
+\binom q2\chi w_{M_{\Ncal}}(g),\\
F_{k+q}(e_{k+q})&=\binom{k+q}{2}\chi.
\end{aligned}
\label{eq:appendix-F-spectator}
\end{equation}
The first equality sums the three pair classes above. The second
uses the two-fixed-point contribution $\chi$ for every pair of
labels in the identity permutation.

We denote the terms in square brackets in Eq.~\eqref{Eq:appendix:A:different_row} by
\begin{equation}
\begin{aligned}
\mathcal D_k^{(q)}(g)
&=F_{k+q}(g\oplus e_q)
-F_{k+q}(e_{k+q})w_{M_{\Ncal}}(g)\\
&=F_k(g)+q\sum_{i=1}^k\eta_i(g)
-\left[kq+\binom k2\right]\chi w_{M_{\Ncal}}(g).
\end{aligned}
\label{eq:appendix-D-q}
\end{equation}
The second equality
uses Eq.~\eqref{eq:appendix-F-spectator} and
$\binom q2-\binom{k+q}2=-kq-\binom k2$.

The physical Born-weighted moment is obtained by continuing to
$q=1-k$, as in Lemma~\ref{lem:raw-to-physical-moments}. Define
$\mathcal D_k(g)=\mathcal D_k^{(1-k)}(g)$. Then
\begin{equation}
\begin{aligned}
\mathcal D_k(g)
&=F_k(g)+(1-k)\sum_{i=1}^k\eta_i(g)
-\left[k(1-k)+\binom k2\right]\chi w_{M_{\Ncal}}(g)\\
&=F_k(g)+(1-k)\sum_{i=1}^k\eta_i(g)
+\binom k2\chi w_{M_{\Ncal}}(g).
\end{aligned}
\label{eq:appendix-D-continuation}
\end{equation}
Here, we continue each permutation coefficient
separately and assume that the omitted terms retain their stated
bounds. For fixed $d,k,\Ncal$, large $t,D_A$, and $t+1\geq N_A$,
continuing the raw-moment difference above gives
\begin{equation}
\begin{aligned}
\rho_{N_A}^{(k)}-\rho_{M_{\Ncal}}^{(k)}
&=\frac1{D_A^k}\sum_{g\in S_k}
\left[d^{-t}\mathcal D_k(g)+\mathcal R_k(g)\right]\rho_{N_A,k}(g),\\
\max_{g\in S_k}|\mathcal R_k(g)|
&=\mathcal O_{d,k,\Ncal}(d^{-2t}+d^{-t}D_A^{-1}).
\end{aligned}
\end{equation}
This is Eq.~\eqref{eq:du-physical-finite-time-correction}.

We now return to the two distances. The reference moment is a
positive operator of trace one, so
$\|\rho_{M_{\Ncal}}^{(k)}\|_1=1$. Since each permutation is
unitary, $\|\rho_{N_A,k}(g)\|_1=D_A^k$. Hence
\begin{equation}
\begin{aligned}
\Delta_1^{(k)}
&\leq\frac1{D_A^k}\sum_{g\in S_k}
\left[d^{-t}|\mathcal D_k(g)|+|\mathcal R_k(g)|\right]
\|\rho_{N_A,k}(g)\|_1\\
&=d^{-t}\sum_{g\in S_k}|\mathcal D_k(g)|
+\sum_{g\in S_k}|\mathcal R_k(g)|\\
&=d^{-t}\sum_{g\in S_k}|\mathcal D_k(g)|
+\mathcal O_{d,k,\Ncal}(d^{-2t}+d^{-t}D_A^{-1}).
\end{aligned}
\label{eq:appendix-trace-norm-bound}
\end{equation}

For the Hilbert--Schmidt distance, the permutation overlaps are
\begin{equation}
\tr[\rho_{N_A,k}(g)^\dagger\rho_{N_A,k}(h)]
=D_A^{k-|g^{-1}h|},\qquad g,h\in S_k.
\label{eq:appendix-permutation-HS-gram}
\end{equation}
Here $|g^{-1}h|$ is the minimum number of transpositions whose
product is $g^{-1}h$. Each cycle gives a factor $D_A$, and the
number of cycles is $k-|g^{-1}h|$. These overlaps give
\begin{equation}
\begin{aligned}
&\left\|\frac{d^{-t}}{D_A^k}\sum_{g\in S_k}
\mathcal D_k(g)\rho_{N_A,k}(g)\right\|_2^2
=\frac{d^{-2t}}{D_A^{2k}}\sum_{g,h\in S_k}
\mathcal D_k(g)^*\mathcal D_k(h)D_A^{k-|g^{-1}h|}=\frac{d^{-2t}}{D_A^k}
\left[\sum_{g\in S_k}|\mathcal D_k(g)|^2
+\mathcal O_k(D_A^{-1})\right],\\
&\|\rho_{M_{\Ncal}}^{(k)}\|_2^2
=\frac1{D_A^{2k}}\sum_{g,h\in S_k}
w_{M_{\Ncal}}(g)^*w_{M_{\Ncal}}(h)D_A^{k-|g^{-1}h|}
+\mathcal O_k(D_A^{-k-1})=\frac1{D_A^k}\left[\sum_{g\in S_k}|w_{M_{\Ncal}}(g)|^2
+\mathcal O_k(D_A^{-1})\right].
\end{aligned}
\label{eq:appendix-HS-norms}
\end{equation}
The
second equality retains $g=h$, since every off-diagonal overlap has at
least one fewer power of $D_A$. The first equality for the reference
uses its leading coefficients $w_{M_{\Ncal}}(g)$, with its
finite-$D_A$ correction included in the remainder. 

The remainder  satisfies
\begin{equation}
\begin{aligned}
\frac{\left\|D_A^{-k}\sum_{g\in S_k}\mathcal R_k(g)
\rho_{N_A,k}(g)\right\|_2}
{\|\rho_{M_{\Ncal}}^{(k)}\|_2}
&\leq\frac{D_A^{-k/2}\sum_{g\in S_k}|\mathcal R_k(g)|}
{\|\rho_{M_{\Ncal}}^{(k)}\|_2}\\
&\leq\sum_{g\in S_k}|\mathcal R_k(g)|
=\mathcal O_{d,k,\Ncal}(d^{-2t}+d^{-t}D_A^{-1}).
\end{aligned}
\label{eq:appendix-HS-remainder}
\end{equation}
The first inequality uses the triangle inequality and
$\|\rho_{N_A,k}(g)\|_2=D_A^{k/2}$. The second uses
$\|\rho_{M_{\Ncal}}^{(k)}\|_2\geq D_A^{-k/2}$, which holds
for any density operator on a space of dimension $D_A^k$.
The equality uses the bound on $\mathcal R_k$. Taking the square root of
the ratio in Eq.~\eqref{eq:appendix-HS-norms} therefore yields
\begin{equation}
\Delta_2^{(k)}=d^{-t}
\frac{\left(\sum_{g\in S_k}|\mathcal D_k(g)|^2\right)^{1/2}}
{\left(\sum_{g\in S_k}|w_{M_{\Ncal}}(g)|^2\right)^{1/2}}
+\mathcal O_{d,k,\Ncal}(d^{-2t}+d^{-t}D_A^{-1}).
\label{eq:2resultapp}
\end{equation}
The reference coefficient sum is at least one because
$w_{M_{\Ncal}}(e_k)=1$. This gives both distances in
Eq.~\eqref{eq:du-finite-time-distances}. The leading power of
$D_A$ cancels in each normalized distance.

As discussed after Eq.~\eqref{eq:du-moment-order-time-scales}, when (i) transpositions dominate the correction sums in Eq.~\eqref{eq:du-finite-time-distances} and (ii) the identity dominates the denominator, the deep thermalization time will scale as $t\sim\log(k)$, illustrating a separation among different moments. We first evaluate the identity and transposition coefficients.
For the identity permutation, Eqs.~\eqref{eq:appendix-F-spectator} and
\eqref{eq:appendix-chi-eta-definition} give
$F_k(e_k)=\binom k2\chi$ and $\sum_{i=1}^k\eta_i(e_k)=k\chi$.
Substituting these values and $w_{M_{\Ncal}}(e_k)=1$ into
Eq.~\eqref{eq:appendix-D-continuation} gives
\[
\mathcal D_k(e_k)=0.
\]

For a transposition $g=s_{ij}$ with $k\geq2$, there is one cycle
of length two and $k-2$ fixed points. Substituting this cycle structure
into Eqs.~\eqref{eq:appendix-F-definition} and
\eqref{eq:appendix-chi-eta-definition}, with the contractions of
Appendix~\ref{Appendix:Sec:A:first:order}, gives
\begin{equation}
\begin{aligned}
F_k(s_{ij})
&=\tr(H_1^2)-\frac1d
+\frac{2(k-2)(p_2-p_3)}d+\binom{k-2}{2}\chi p_2,\\
\sum_{m=1}^k\eta_m(s_{ij})
&=\frac{2(p_2-p_3)}d+(k-2)\chi p_2.
\end{aligned}
\label{eq:appendix-transposition-inputs}
\end{equation}
With $w_{M_{\Ncal}}(s_{ij})=p_2$ and $\chi=(1-p_2)/d$,
Eq.~\eqref{eq:appendix-D-continuation} then gives
\begin{equation}
\mathcal D_k(s_{ij})
=\tr(H_1^2)-\frac1d-\frac{2(p_2-p_3)}d
+\frac{p_2(1-p_2)}d.
\label{eq:appendix-swap-coefficient}
\end{equation}
All $\binom{k}{2}$ transpositions have the same coefficient, independent of $k$. For complete depolarization, substituting
$H_1=I_d/d$ and $p_s=d^{2(1-s)}$ into
Eq.~\eqref{eq:appendix-swap-coefficient} gives
\[
\mathcal D_k(s_{ij})=-\frac{d^2-1}{d^5}.
\]
At fixed $k$ and large $d$, complete depolarization satisfies both assumptions (i), (ii), giving the moment-order dependence discussed after Eq.~\eqref{eq:du-moment-order-time-scales}.

However, these assumptions need not hold for a general channel. We consider the partial reset channel discussed after Eq.~\eqref{eq:du-moment-order-time-scales}, with fixed reset probability $0<\theta<1$. Its normalized Choi state is
\[
M_{\Ncal}=(1-\theta)\ket{\phi_0}\!\bra{\phi_0}
+\theta\frac{I_d}{d}\otimes\ket0\!\bra0,
\]
where $\ket{\phi_0}$ is defined in Eq.~\eqref{eq:maximally-entangled-state}. To evaluate the denominator in Eq.~\eqref{eq:du-finite-time-distances}, we first calculate the transposition weight $w_{M_{\Ncal}}(s_{12})=p_2$:
\begin{equation}
p_2=\tr M_{\Ncal}^2
=(1-\theta)^2+\frac{\theta^2}{d}+\frac{2\theta(1-\theta)}{d^2}.
\label{eq:appendix-partial-reset-purity}
\end{equation}
The cross term uses $\bra{\phi_0}(I_d/d\otimes\ket0\!\bra0)\ket{\phi_0}=d^{-2}$. Thus $w_{M_{\Ncal}}(s_{12})\to(1-\theta)^2$ as $d\to\infty$, and transpositions contribute at the same order in $d$ as the identity to the denominator.

We next compare the correction coefficients of $s_{12}$ and $s_{12}s_{34}$ at $k=4$. The required partial traces are
\begin{equation}
\begin{aligned}
H_1&=(1-\theta)\frac{I_d}{d}+\theta\ket0\!\bra0,
&\tr H_1^2&=\theta^2+\frac{1-\theta^2}{d},\\
Q_2&=\left(\frac{(1-\theta)^2}{d}+\frac{\theta^2}{d^2}\right)I_d
+\frac{2\theta(1-\theta)}{d^2}\ket0\!\bra0.
\end{aligned}
\label{eq:appendix-partial-reset-matrices}
\end{equation}
For $g=s_{12}s_{34}$ containing two cycles, the sum defining $F_4(g)$ in
Eq.~\eqref{eq:appendix-F-definition} includes six choices of $s_{ij}$.
The choices $s_{12}$ and $s_{34}$ exchange labels within the two cycles. The other four choices,
$s_{13},s_{14},s_{23},s_{24}$, each exchange one label from each cycle.
Thus the first two terms use Eq.~\eqref{eq:appendix-same-cycle-contraction},
while the remaining four use
Eq.~\eqref{eq:appendix-different-cycle-contraction}. Together with Eqs.~\eqref{eq:appendix-chi-eta-definition} and~\eqref{eq:appendix-D-continuation}, this gives, as $d\to\infty$,
\begin{equation}
\begin{aligned}
\mathcal D_4(s_{12})
&=\tr H_1^2-\frac1d
+\frac{2p_3-p_2-p_2^2}{d}
=\theta^2+\mathcal O(d^{-1}),\\
\mathcal D_4(s_{12}s_{34})
&=2p_2\left(\tr H_1^2-\frac1d\right)
+4\left(\tr Q_2^2-\frac{p_4}{d}\right)
-\frac{12p_2(p_2-p_3)}d+6\chi p_2^2\\
&=2\theta^2(1-\theta)^2+\mathcal O(d^{-1}).
\end{aligned}
\label{eq:appendix-partial-reset-corrections}
\end{equation}
The last equalities use Eqs.~\eqref{eq:appendix-partial-reset-purity} and~\eqref{eq:appendix-partial-reset-matrices}, together with $p_s\leq1$ and $\chi=(1-p_2)/d$.
Both permutations therefore contribute at order $d^{-t}$, and the moment-order dependence involves the full coefficient sums in Eq.~\eqref{eq:du-finite-time-distances}.

\section{Global Haar calculations}
\label{app:global-haar}

We derive the finite-size estimates for the global Haar model of
Sec.~\ref{sec:global-haar}. Let $D_B$ be the Hilbert-space
dimension of the measured region $B$, and $D=D_AD_B$ the total
dimension. We keep $d,D_A$, the moment order $k$, and the number
of channels fixed as $D_B\to\infty$.

This appendix is organized as follows.
Appendix~\ref{app:global-haar-stability} proves the stability bounds for the Born-weighted moment, including Eq.~\eqref{eq:global-haar-recorded-outcome-contraction}. Appendix~\ref{app:global-haar-purity} uses the Haar-averaged purity to bound the error from replacing the post-channel state by an embedded Choi state [Eq.~\eqref{eq:global-haar-choi-embedding-average}]. Appendix~\ref{app:global-haar-blocks} bounds the error from replacing the resulting scaled Haar block by a Ginibre matrix [Eq.~\eqref{eq:global-haar-coefficient-bound}]. Appendix~\ref{app:global-haar-moment-comparison} uses the stability bounds of Appendix~\ref{app:global-haar-stability} to convert these two estimates into moment errors and adds them to obtain Eq.~\eqref{eq:global-haar-average}. Finally, Appendix~\ref{app:global-haar-decoupling} derives Eqs.~\eqref{eq:global-haar-reference-decoupling} and~\eqref{eq:global-haar-dilute-average} for several channels.

\subsection{Stability of the Born-weighted moment}
\label{app:global-haar-stability}

For a positive conditional operator $Z$ on $A$, its contribution to the Born-weighted moment is
\begin{equation}
h_k(Z)=
Z^{\otimes k}/(\tr Z)^{k-1}.
\end{equation}
Then $h_k(cZ)=ch_k(Z)$ for $c\geq0$,
$\|h_k(Z)\|_1=\tr Z$, and
\begin{equation}
\rho_{N_A}^{(k)}
=
\sum_\alpha
h_k\!\left(\widetilde\rho_{N_A}(\alpha)\right).\label{eq:SecIVexpressingwithh}
\end{equation}
For another positive conditional operator $Z'$, we first prove
\begin{equation}
\left\|h_k(Z)-h_k(Z')\right\|_1
\leq
(2k-1)\|Z-Z'\|_1.
\label{eq:global-haar-positive-lipschitz}
\end{equation}
Interpolating between positive operators
$Z,Z'$ as $Z_s=(1-s)Z+sZ'$, $0\leq s\leq1$, gives, whenever
$\tr Z_s>0$,
\begin{equation}
\begin{aligned}
\frac{d}{ds}h_k(Z_s)
={}&\frac{1}{(\tr Z_s)^{k-1}}
\sum_{\ell=1}^k Z_s^{\otimes(\ell-1)}
\otimes(Z'-Z)%
\otimes Z_s^{\otimes(k-\ell)}
-(k-1)\frac{\tr(Z'-Z)}{(\tr Z_s)^k}Z_s^{\otimes k}.
\end{aligned}
\end{equation}
The trace norm of the derivative obeys
\[
\begin{aligned}
\left\|\frac{d}{ds}h_k(Z_s)\right\|_1
&\leq\frac{k\|Z_s\|_1^{k-1}\|Z'-Z\|_1}{(\tr Z_s)^{k-1}}
+\frac{(k-1)|\tr(Z'-Z)|\|Z_s\|_1^k}{(\tr Z_s)^k}\\
&=k\|Z'-Z\|_1+(k-1)|\tr(Z'-Z)|\leq(2k-1)\|Z'-Z\|_1.
\end{aligned}
\]
The equality uses
$\|Z_s\|_1=\tr Z_s$ for positive $Z_s$. 
Integration gives
\[
\begin{aligned}
\|h_k(Z')-h_k(Z)\|_1
&=\left\|\int_0^1ds\,\frac{d}{ds}h_k(Z_s)\right\|_1\leq\int_0^1ds\,\left\|\frac{d}{ds}h_k(Z_s)\right\|_1\leq(2k-1)\|Z'-Z\|_1.
\end{aligned}
\]
This proves
Eq.~\eqref{eq:global-haar-positive-lipschitz}.

To sum this bound over measurement outcomes, we retain the outcome
of the measurement on $B$ in an orthogonal classical register.
For density matrices $\rho,\rho'$ on $A\otimes B$, the recorded
measurement is the trace-preserving map
$\rho\mapsto\bigoplus_\alpha{}_B\!\braket{\alpha|\rho|\alpha}_B$.
Its output differences satisfy
\begin{equation}
\begin{aligned}
\sum_\alpha
\left\|{}_B\!\braket{\alpha|(\rho-\rho')|\alpha}_B\right\|_1
&=\left\|\bigoplus_\alpha
{}_B\!\braket{\alpha|(\rho-\rho')|\alpha}_B\right\|_1\leq\|\rho-\rho'\|_1.
\end{aligned}
\end{equation}
The inequality uses trace-distance contractivity of the recorded measurement.
Applying Eq.~\eqref{eq:global-haar-positive-lipschitz} to each conditional operator gives
\begin{equation}
\begin{aligned}
\left\|\rho_{N_A}^{(k)}-\rho_{N_A}^{\prime\,(k)}\right\|_1
&=
\left\|
\sum_\alpha\left[
h_k\!\left({}_B\!\braket{\alpha|\rho|\alpha}_B\right)
-h_k\!\left({}_B\!\braket{\alpha|\rho'|\alpha}_B\right)
\right]
\right\|_1\\
&\leq
(2k-1)\sum_\alpha
\left\|
{}_B\!\braket{\alpha|(\rho-\rho')|\alpha}_B
\right\|_1\leq
(2k-1)\|\rho-\rho'\|_1.
\end{aligned}
\label{eq:appendix-recorded-outcome-stability}
\end{equation}
The first inequality uses the triangle inequality and Eq.~\eqref{eq:global-haar-positive-lipschitz}; the second uses the recorded-measurement bound above.
The final bound is Eq.~\eqref{eq:global-haar-recorded-outcome-contraction} in the main text.

\subsection{Haar-purity calculation}
\label{app:global-haar-purity}

The estimate in Eq.~\eqref{eq:global-haar-choi-embedding-average}
uses the Haar-averaged purity of $\rho_S$ in
Eq.~\eqref{eq:SecIVdefinitionofsigma}. Write
$\ket\psi=U_1\ket{\psi_0}$ for the state before the channel.
The site $S$ has dimension $d$, and the remaining system
$\overline S$ has dimension $D/d$. Let $\operatorname{SWAP}_R$
exchange two copies of a space $R$ and act as the identity on
the other factors. The Haar second moment is
\begin{equation}
\mathbb E_{U_1}\bigl(\ket\psi\!\bra\psi\bigr)^{\otimes2}
=\frac{I+\operatorname{SWAP}_{S\overline S}}{D(D+1)},
\end{equation}
where $I$ is the identity on the two full-system copies. Thus
\begin{equation}
\begin{aligned}
\mathbb E_{U_1}\tr\rho_S^2
&=\tr\left[\operatorname{SWAP}_S
\frac{I+\operatorname{SWAP}_{S\overline S}}{D(D+1)}\right]=\frac{d(D/d)^2+d^2(D/d)}{D(D+1)}=\frac{d+D/d}{D+1}.
\end{aligned}
\end{equation}
The first equality uses the swap identity for purity and the
Haar second moment. The second evaluates the two traces:
$\tr\operatorname{SWAP}_S=d(D/d)^2$ and
$\tr(\operatorname{SWAP}_S\operatorname{SWAP}_{S\overline S})
=d^2(D/d)$.
We now use this purity to bound the error from replacing the post-channel state $\sigma$ by $\mathsf V M_{\Ncal}\mathsf V^\dagger$. Here $\sigma$ is defined in Eq.~\eqref{eq:SecIVdefinitionofsigma}, and $\mathsf V:C\to A\otimes B$ is the isometry in Eq.~\eqref{eq:global-haar-choi-embedding}, with $\dim C=d^2$. Thus
\begin{equation}
\begin{aligned}
\mathbb E_{U_1}\|\sigma-\mathsf V M_{\Ncal}\mathsf V^\dagger\|_1
&\leq\mathbb E_{U_1}\left\|\rho_S-\frac{I_d}{d}\right\|_1\\
&\leq\left[d\,\mathbb E_{U_1}\tr\rho_S^2-1\right]^{1/2}
=\sqrt{\frac{d^2-1}{D+1}}.
\end{aligned}
\label{eq:appendix-global-haar-choi-embedding-average}
\end{equation}
The first inequality is Eq.~\eqref{eq:global-haar-choi-embedding}. The second uses $\|O\|_1\leq\sqrt d\|O\|_2$ on $S$ and Jensen's inequality. The equality substitutes the purity above. This proves Eq.~\eqref{eq:global-haar-choi-embedding-average}.

\subsection{Finite Haar blocks and Ginibre matrices}
\label{app:global-haar-blocks}

We prove Eq.~\eqref{eq:global-haar-coefficient-bound}, which bounds the difference between the scaled Haar block $X_{\mathrm H}$ and the Ginibre matrix $X$. Both matrices have size $D_A\times d^2$. For a fixed measurement outcome $\alpha_0$ in $B$,
\[
X_{\mathrm H}:=\sqrt{D_B}\,(I_A\otimes{}_B\!\bra{\alpha_0})U_2\mathsf V,
\]
where $U_2\mathsf V$ is a Haar-distributed $D\times d^2$ isometry. The entries of $X$ are independent complex Gaussian variables with mean zero and $\mathbb E|X_{a\mu}|^2=1/D_A$. As $D_B\to\infty$ at fixed $d,D_A$, the scaled block $X_{\mathrm H}$ obtained from the Haar-random unitary $U_2$ approaches the Gaussian matrix $X$ in distribution. To control this approximation at finite $D_B$, we construct the two matrices from the same Gaussian matrix and bound $\mathbb E\|X_{\mathrm H}-X\|_2^2$. 

To construct $X_{\mathrm H}$ and $X$ from the same Gaussian
variables, let $Y$ be a $D\times d^2$ matrix.
The real and imaginary parts of all entries are independent Gaussian
variables with mean zero and variance $1/2$, so $\mathbb E|Y_{j\mu}|^2=1$.
Expectations and probabilities in this subsection are over $Y$.
For $D\geq d^2$, a Haar-random isometry can be constructed from $Y$ as~\cite{MeckesHaarBookPDF}
\[
U_2\mathsf V\overset{\mathrm d}{=}Y(Y^\dagger Y)^{-1/2},
\]
where $\overset{\mathrm d}{=}$ denotes equality in distribution.
Define
\begin{equation}
X:=\frac{(I_A\otimes{}_B\!\bra{\alpha_0})Y}{\sqrt{D_A}},\qquad
S_D:=\frac{Y^\dagger Y}{D}.
\end{equation}
The projection selects the $D_A$ rows of $Y$ corresponding to
outcome $\alpha_0$. Their entries remain independent Gaussian
variables. Dividing by $\sqrt{D_A}$ gives
$\mathbb E[X_{a\mu}]=0$ and $\mathbb E|X_{a\mu}|^2=1/D_A$,
so $X$ has exactly the distribution specified in
Eq.~\eqref{eq:ginibre-covariance}.

Using $D=D_AD_B$, we can therefore construct the scaled Haar
block as
\begin{equation}
X_{\mathrm H}=XS_D^{-1/2}.
\label{eq:global-haar-block-coupling}
\end{equation}
The matrices $X$ and $S_D$ are dependent because they are constructed
from the same $Y$.

Since $X_{\mathrm H}-X=X(S_D^{-1/2}-I_{d^2})$, we estimate
$S_D-I_{d^2}$. Here $\|O\|_\infty$ is the largest singular value of $O$.
We split the mean squared difference according to whether
$\|S_D-I_{d^2}\|_\infty$ is at most $1/2$.
The indicator $\mathbf1_{\{\cdots\}}$ is one when the condition
in braces holds and zero otherwise. Thus
\[
\begin{aligned}
\mathbb E\|X_{\mathrm H}-X\|_2^2
&=\mathbb E\left[\|X_{\mathrm H}-X\|_2^2
\mathbf1_{\{\|S_D-I_{d^2}\|_\infty\leq1/2\}}\right]
+\mathbb E\left[\|X_{\mathrm H}-X\|_2^2
\mathbf1_{\{\|S_D-I_{d^2}\|_\infty>1/2\}}\right].
\end{aligned}
\]
This split is needed because $S_D^{-1/2}$ can be large when an eigenvalue of $S_D$ is close to zero. In the first term, the condition on $S_D$ keeps its eigenvalues away from zero, so we can bound $S_D^{-1/2}-I_{d^2}$ by $S_D-I_{d^2}$. For the second term, we bound the probability that the condition fails.
Write each entry of
$S_D-I_{d^2}$, for $1\leq\mu,\nu\leq d^2$, as
\[
(S_D-I_{d^2})_{\mu\nu}
=\frac1D\sum_{j=1}^D
\left(Y_{j\mu}^*Y_{j\nu}-\delta_{\mu\nu}\right).
\]
The fourth moment will control the mean squared matrix difference
through Cauchy--Schwarz. The eighth moment will bound the probability
of large fluctuations through Markov's inequality. They obey
\begin{equation}
\begin{aligned}
\mathbb E\|S_D-I_{d^2}\|_\infty^4&=\mathcal O_d(D^{-2}),\\
\mathbb E\|S_D-I_{d^2}\|_\infty^8&=\mathcal O_d(D^{-4}).
\end{aligned}
\label{eq:global-haar-wishart-moments}
\end{equation}
Each row contribution has mean zero, and different rows are
independent. In the fourth-moment expansion, a row appearing once
therefore gives zero, leaving at most two distinct row indices and
$\mathcal O(D^2)$ terms with prefactor $D^{-4}$. The eighth moment
has at most four distinct rows and $\mathcal O(D^4)$ terms with
prefactor $D^{-8}$. The number of columns is fixed, so these
entrywise estimates also bound the operator norm.
If $\|S_D-I_{d^2}\|_\infty\leq1/2$, every eigenvalue $\lambda$ of $S_D$ satisfies
\[
\begin{aligned}
|\lambda-1|
&\leq\|S_D-I_{d^2}\|_\infty\leq\tfrac12,\qquad
\tfrac12&\leq\lambda\leq\tfrac32.
\end{aligned}
\]
On this interval, $|\lambda^{-1/2}-1|\leq\sqrt2|\lambda-1|$. Applying this scalar bound to each eigenvalue gives
\[
\|S_D^{-1/2}-I_{d^2}\|_\infty
\leq\sqrt2\,\|S_D-I_{d^2}\|_\infty.
\]
The Gaussian entries give
$\mathbb E\|X\|_2^4=d^4+d^2/D_A$. Thus
\begin{equation}
\begin{aligned}
&\mathbb E\left[\|X_{\mathrm H}-X\|_2^2
\mathbf1_{\{\|S_D-I_{d^2}\|_\infty\leq1/2\}}\right]\\
&\quad=\mathbb E\left[\|X(S_D^{-1/2}-I_{d^2})\|_2^2
\mathbf1_{\{\|S_D-I_{d^2}\|_\infty\leq1/2\}}\right]\\
&\quad\leq2\mathbb E\left[\|X\|_2^2\|S_D-I_{d^2}\|_\infty^2
\mathbf1_{\{\|S_D-I_{d^2}\|_\infty\leq1/2\}}\right]\\
&\quad\leq
2\left(\mathbb E\|X\|_2^4\right)^{1/2}
\left(\mathbb E\|S_D-I_{d^2}\|_\infty^4\right)^{1/2}\\
&\quad=\mathcal O_{d,D_A}(D^{-1}).
\end{aligned}
\end{equation}
The first equality uses $X_{\mathrm H}=XS_D^{-1/2}$. The first inequality
uses $\|X(S_D^{-1/2}-I_{d^2})\|_2
\leq\|X\|_2\|S_D^{-1/2}-I_{d^2}\|_\infty$ and the
inverse-square-root bound. The second inequality uses Cauchy--Schwarz and bounds the
indicator by one. The last equality uses the fourth moments of
$X$ and $S_D-I_{d^2}$.
The probability that $\|S_D-I_{d^2}\|_\infty>1/2$ is
\[
\begin{aligned}
\Pr[\|S_D-I_{d^2}\|_\infty>1/2]
&\leq2^8\mathbb E\|S_D-I_{d^2}\|_\infty^8=\mathcal O_d(D^{-4}).
\end{aligned}
\]
The inequality is Markov's inequality. For this part of the average,
\begin{equation}
\begin{aligned}
&\mathbb E\left[\|X_{\mathrm H}-X\|_2^2
\mathbf1_{\{\|S_D-I_{d^2}\|_\infty>1/2\}}\right]\\
&\quad\leq2\mathbb E\left[(\|X_{\mathrm H}\|_2^2+\|X\|_2^2)
\mathbf1_{\{\|S_D-I_{d^2}\|_\infty>1/2\}}\right]\\
&\quad\leq2D_Bd^2\Pr\left[\|S_D-I_{d^2}\|_\infty>\frac12\right]
+2\left(\mathbb E\|X\|_2^4\right)^{1/2}
\Pr\left[\|S_D-I_{d^2}\|_\infty>\frac12\right]^{1/2}\\
&\quad=\mathcal O_{d,D_A}(D^{-2}).
\end{aligned}
\end{equation}
The first inequality uses
$\|X_{\mathrm H}-X\|_2^2\leq2\|X_{\mathrm H}\|_2^2+2\|X\|_2^2$.
The second uses $\|X_{\mathrm H}\|_2^2\leq D_Bd^2$ and Cauchy--Schwarz.
The equality uses the probability bound and $D_B=D/D_A$ at fixed
$d,D_A$.
Adding the two contributions gives
\begin{equation}
\mathbb E\|{X_{\mathrm H}}-X\|_2^2=\mathcal O_{d,D_A}(D^{-1}).
\label{eq:appendix-global-haar-coefficient-bound}
\end{equation}
This proves Eq.~\eqref{eq:global-haar-coefficient-bound}.

\subsection{From Haar blocks to the CIE moment}
\label{app:global-haar-moment-comparison}

We now prove Eq.~\eqref{eq:global-haar-average}, which bounds
\[
b_{D_B}^{(k)}:=\left\|
\mathbb E_{U_1,U_2}\rho_{N_A}^{(k)}-\rho_{M_{\Ncal}}^{(k)}
\right\|_1.
\]
We first replace the post-channel state $\sigma$ by $\mathsf V M_{\Ncal}\mathsf V^\dagger$, and then replace $X_{\mathrm H}$ by $X$. The two errors will each be of order $D^{-1/2}$.
For measurement outcome $\alpha$, the state $U_2\mathsf V M_{\Ncal}\mathsf V^\dagger U_2^\dagger$ gives the unnormalized conditional state
\begin{equation}
\widetilde\rho_{N_A}^{\mathrm{iso}}(\alpha)
:=
{}_B\!\bra{\alpha}
U_2\mathsf V M_{\Ncal}\mathsf V^\dagger U_2^\dagger
\ket{\alpha}_B.
\end{equation}

Conditioned on $U_1$, $U_2\mathsf V$ is a Haar-distributed
$D\times d^2$ isometry.
Its blocks have the same distribution for every outcome, so the block $X_{\mathrm H}$ defined in Appendix~\ref{app:global-haar-blocks} obeys
\begin{equation}
\widetilde\rho_{N_A}^{\mathrm{iso}}(\alpha)
\overset{\mathrm d}{=}
D_B^{-1}X_{\mathrm H} M_{\Ncal}X_{\mathrm H}^\dagger
.
\end{equation}
Together with $h_k(cZ)=ch_k(Z)$, this gives the averaged moment
\begin{equation}
\begin{aligned}
\rho_{N_A,\mathrm{iso}}^{(k)}:=&\mathbb E_{U_2}
\sum_{\alpha=1}^{D_B}
h_k\!\left(
\widetilde\rho_{N_A}^{\mathrm{iso}}(\alpha)
\right)=
D_B\,\mathbb E_{U_2}
h_k\!\left(
D_B^{-1}X_{\mathrm H} M_{\Ncal}X_{\mathrm H}^\dagger
\right)=
\mathbb E_{U_2}
h_k\!\left(
X_{\mathrm H} M_{\Ncal}X_{\mathrm H}^\dagger
\right).
\end{aligned}
\label{eq:global-haar-exact-block-reduction}
\end{equation}
The sum over outcomes gives the factor $D_B$, because all outcome blocks have the same distribution. The last equality uses $h_k(cZ)=ch_k(Z)$.

This average is independent of $U_1$, because the Haar distribution of $U_2\mathsf V$ is the same for every fixed $U_1$. It is the intermediate moment $\rho_{N_A,\mathrm{iso}}^{(k)}$ used in Eq.~\eqref{eq:global-haar-average}. The triangle inequality separates the two errors:
\[
\begin{aligned}
b_{D_B}^{(k)}
&\leq\left\|\mathbb E_{U_1,U_2}\rho_{N_A}^{(k)}-\rho_{N_A,\mathrm{iso}}^{(k)}\right\|_1
+\left\|\rho_{N_A,\mathrm{iso}}^{(k)}-\rho_{M_{\Ncal}}^{(k)}\right\|_1.
\end{aligned}
\]

For the first error, convexity, the stability estimate of Appendix~\ref{app:global-haar-stability}, and the purity calculation of Appendix~\ref{app:global-haar-purity} give
\begin{equation}
\begin{aligned}
\left\|\mathbb E_{U_1,U_2}\rho_{N_A}^{(k)}-\rho_{N_A,\mathrm{iso}}^{(k)}\right\|_1
&\leq\mathbb E_{U_1,U_2}\left\|
\rho_{N_A}^{(k)}-\sum_\alpha h_k\!\left(\widetilde\rho_{N_A}^{\mathrm{iso}}(\alpha)\right)\right\|_1\\
&\leq(2k-1)\mathbb E_{U_1}\left\|\sigma-\mathsf V M_{\Ncal}\mathsf V^\dagger\right\|_1\\
&\leq(2k-1)\sqrt{\frac{d^2-1}{D+1}}.
\end{aligned}
\label{eq:global-haar-recorded-measurement}
\end{equation}
The first inequality uses convexity of the trace norm and the expression for $\rho_{N_A,\mathrm{iso}}^{(k)}$ above. The second applies Eq.~\eqref{eq:appendix-recorded-outcome-stability} to $U_2\sigma U_2^\dagger$ and $U_2\mathsf V M_{\Ncal}\mathsf V^\dagger U_2^\dagger$, and uses unitary invariance of the trace norm. The third uses Eq.~\eqref{eq:appendix-global-haar-choi-embedding-average}.

We next bound the second error, using $X_{\mathrm H}$ and $X$ constructed from the same $Y$ in Appendix~\ref{app:global-haar-blocks}. Since $M_{\Ncal}\geq0$ and $\tr M_{\Ncal}=1$,
\begin{equation}
\begin{aligned}
&\left\|
X_{\mathrm H} M_{\Ncal}X_{\mathrm H}^\dagger
-X M_{\Ncal}X^\dagger
\right\|_1\leq
(\|X_{\mathrm H}\|_2+\|X\|_2)\|X_{\mathrm H}-X\|_2.
\end{aligned}
\end{equation}
Moreover,
$\mathbb E\|X_{\mathrm H}\|_2^2=\mathbb E\|X\|_2^2=d^2$.
The matrix estimate proved in Appendix~\ref{app:global-haar-blocks} is
\[
\mathbb E\|X_{\mathrm H}-X\|_2^2=\mathcal O_{d,D_A}(D^{-1}).
\]
The Ginibre expectation $\mathbb E h_k(XM_{\Ncal}X^\dagger)$ equals the CIE moment $\rho_{M_{\Ncal}}^{(k)}$ by Eq.~\eqref{eq:choi-ensemble-moment}. Hence
\begin{equation}
\begin{aligned}
\|\rho_{N_A,\mathrm{iso}}^{(k)}-\rho_{M_{\Ncal}}^{(k)}\|_1
&\leq(2k-1)\mathbb E\!\left[(\|X_{\mathrm H}\|_2+\|X\|_2)\|X_{\mathrm H}-X\|_2\right]\\
&\leq2d(2k-1)\left(\mathbb E\|X_{\mathrm H}-X\|_2^2\right)^{1/2}\\
&=\mathcal O_{d,D_A,k}(D^{-1/2}).
\end{aligned}
\label{eq:SECIVerrorpart2}
\end{equation}
The first inequality uses Eq.~\eqref{eq:global-haar-positive-lipschitz} and the conditional-state bound above. The second uses Cauchy--Schwarz and $\mathbb E(\|X_{\mathrm H}\|_2+\|X\|_2)^2\leq4d^2$. The equality uses the matrix estimate.
Combining the two comparisons yields the detailed bound
\begin{equation}
\begin{aligned}
b_{D_B}^{(k)}
&\leq(2k-1)\sqrt{\frac{d^2-1}{D+1}}
+\frac{C_{d,D_A,k}}{\sqrt D}=\mathcal O_{d,D_A,k}(D_B^{-1/2}).
\end{aligned}
\label{eq:appendix-global-haar-average-bound}
\end{equation}
Here $C_{d,D_A,k}$ is finite and independent of $D_B$ and the fixed channel.
The first term is the error from replacing $\sigma$ by the embedded Choi state. The second is the error from replacing $X_{\mathrm H}$ by $X$. Their sum proves Eq.~\eqref{eq:global-haar-average}.

\subsection{Decoupling under independent Haar unitaries}
\label{app:global-haar-decoupling}

We derive Eq.~\eqref{eq:global-haar-reference-decoupling} from the
second Haar moment. For the earlier environments of
Sec.~\ref{sec:global-haar}, write
$C_{<i}:=\bigotimes_{j=1}^{i-1}C_j$, where each channel environment $C_j$ has dimension $d^2$. Conditional on the preceding
history, the physical system and $C_{<i}$ are in a joint pure state.
Let $U_i$ be the fresh Haar unitary before channel $i$.
After $U_i$, let $\rho_{S_iC_{<i}}$ be the reduced state on the
error site and earlier environments. Its reference marginal
$\rho_{C_{<i}}$ is unchanged by $U_i$.
All Haar averages below are conditional on this history,
and $\overline S_i$ denotes the physical system outside the error
site $S_i$.

The Haar average of the conjugated $\operatorname{SWAP}_{S_i}$
is a linear combination $aI+b\operatorname{SWAP}_{S_i\overline S_i}$,
where the coefficients satisfy
\[
\begin{aligned}
aD^2+bD&=d(D/d)^2,\qquad
aD+bD^2&=d^2(D/d).
\end{aligned}
\]
The first equation takes the ordinary trace; the second takes the
trace with the full-system swap. Hence
\begin{equation}
\begin{aligned}
&\mathbb E_{U_i}\left[
(U_i^\dagger)^{\otimes2}\operatorname{SWAP}_{S_i}U_i^{\otimes2}
\right]=aI+b\operatorname{SWAP}_{S_i\overline S_i}=\frac{d[(D/d)^2-1]}{D^2-1}I+\frac{(D/d)(d^2-1)}{D^2-1}
\operatorname{SWAP}_{S_i\overline S_i}.
\end{aligned}
\label{eq:SM_AppendixBhaarcalcu}
\end{equation}
The first equality uses unitary invariance of the Haar average;
the second solves the two trace equations for $a,b$.
The joint purity is
\begin{equation}
\begin{aligned}
\mathbb E_{U_i}\tr\rho_{S_iC_{<i}}^2
&=a\tr\rho_{C_{<i}}^2+b=\frac{d[(D/d)^2-1]}{D^2-1}\tr\rho_{C_{<i}}^2+\frac{(D/d)(d^2-1)}{D^2-1}.
\end{aligned}
\end{equation}
The first equality applies the swap identity for purity. The
identity term gives $\tr\rho_{C_{<i}}^2$, and the full-system
swap gives one because the system and its environments are in a
pure state. The second equality substitutes $a,b$.
The squared Hilbert--Schmidt error is
\begin{equation}
\begin{aligned}
&\mathbb E_{U_i}\left\|
\rho_{S_iC_{<i}}-\frac{I_d}{d}\otimes\rho_{C_{<i}}
\right\|_2^2
=\mathbb E_{U_i}\tr\rho_{S_iC_{<i}}^2
-\frac1d\tr\rho_{C_{<i}}^2
=\frac{(d^2-1)[D-\tr\rho_{C_{<i}}^2]}{d(D^2-1)}.
\end{aligned}
\label{eq:global-haar-reference-decoupling-two-norm}
\end{equation}
The first equality expands the squared norm and uses
$\tr[\rho_{S_iC_{<i}}(I_d/d\otimes\rho_{C_{<i}})]
=d^{-1}\tr\rho_{C_{<i}}^2$.
The second equality substitutes the purity above.
The space $S_i\otimes C_{<i}$ has dimension $d^{2i-1}$.
Thus
\begin{equation}
\begin{aligned}
&\mathbb E_{U_i}\left\|
\rho_{S_iC_{<i}}-\frac{I_d}{d}\otimes\rho_{C_{<i}}
\right\|_1\leq d^{i-1/2}\left(\mathbb E_{U_i}\left\|
\rho_{S_iC_{<i}}-\frac{I_d}{d}\otimes\rho_{C_{<i}}
\right\|_2^2\right)^{1/2}\\
&\quad=\left[
\frac{d^{2(i-1)}(d^2-1)[D-\tr\rho_{C_{<i}}^2]}{D^2-1}
\right]^{1/2}=\mathcal O_{d,i}(D^{-1/2}).
\end{aligned}
\label{eq:global-haar-reference-decoupling-bound}
\end{equation}
The inequality uses $\|O\|_1\leq d^{i-1/2}\|O\|_2$ and Jensen's
inequality. The first equality uses the squared error above, and
the second holds at fixed $d,i$ as $D\to\infty$.
For $i=1$, this reduces to $\sqrt{(d^2-1)/(D+1)}$.

Averaging over the earlier history proves
Eq.~\eqref{eq:global-haar-reference-decoupling}. The complementary-channel
contraction and iteration are given in Sec.~\ref{sec:global-haar}.
For $D\geq d^{2n}$, the final Haar block has $d^{2n}$ columns.
The same Gaussian calculation at fixed $n$ then gives
Eq.~\eqref{eq:global-haar-dilute-average}.

\section{Local Haar circuit contractions}
\label{app:local-haar}

Appendix~\ref{app:local-haar-minimum-distance} proves Lemma~\ref{lem:local-haar-minimum-distance}, which identifies the permutations that give the leading terms in Eq.~\eqref{eq:local-haar-three-overlaps}. Appendix~\ref{subsec:correction_regimes} lists the corrections to Eq.~\eqref{eq:local-haar-three-overlaps} and the conditions under which they are small. Appendix~\ref{app:local-haar-channel-bound} proves the bound on $K_r(g,g')$ used in Appendix~\ref{subsec:correction_regimes} to estimate the effect of changing the permutation labels at the channel.
\subsection{Minimum boundary distance}
\label{app:local-haar-minimum-distance}

\noindent\textbf{Lemma~\ref{lem:local-haar-minimum-distance} (restated).}
Let $k\geq1$ and $q\geq0$ be integers, set $r=k+q$, and define
$s_k=\prod_{i=1}^k(i,r+i)$.
For a permutation $\pi$, let $|\pi|$ be the minimum number of
transpositions whose product is $\pi$. Then
\begin{equation}
\min_{g_1,g_2\in S_r}
\left|s_k^{-1}(g_1\oplus g_2)\right|=k.
\label{eq:appendix-local-haar-minimum-distance}
\end{equation}
Moreover, the minimizers are exactly
\begin{equation}
g_1\oplus g_2=(g\oplus e_q)\oplus(g^{-1}\oplus e_q),
\qquad g\in S_k,
\label{eq:appendix-local-haar-optimal-endpoints}
\end{equation}
where $e_q$ is the identity permutation on the $q$ traced
replicas. For every such minimizer,
\begin{equation}
s_k^{-1}(g_1\oplus g_2)=\prod_{i=1}^k(i,r+g(i)).
\label{eq:appendix-local-haar-endpoint-cycles}
\end{equation}
\begin{proof}[Proof of Lemma~\ref{lem:local-haar-minimum-distance}]
The first trace factor has replica labels $1,\ldots,r$, and the second has labels $r+1,\ldots,2r$. Each factor has $k$ active copies followed by $q$ traced copies.
Set $\pi=s_k^{-1}(g_1\oplus g_2)$.
The permutation $g_1\oplus g_2$ maps $i$ to $g_1(i)$ and
$r+i$ to $r+g_2(i)$ for $1\leq i\leq r$, so it maps each set of labels
into itself. A label $i\leq r$ is mapped by $\pi$ to a label greater than
$r$ precisely when $g_1(i)\in\{1,\ldots,k\}$. Hence exactly $k$ labels
are mapped from the first set into the second, i.e.,
$|\{i\le r:\pi(i)>r\}|=k$.

Consider a cycle of $\pi$ containing $a$ labels from
$\{1,\ldots,r\}$ and $b$ labels from $\{r+1,\ldots,2r\}$.
The number of transitions from the first set to the second in this cycle
is at most $\min(a,b)$, while its contribution to the transposition
length is $a+b-1$. Therefore each cycle contributes at least as much to
$|\pi|$ as to the number of these transitions. Summing over all cycles gives $|\pi|\ge k$, proving
Eq.~\eqref{eq:appendix-local-haar-minimum-distance}.

Equality requires every nontrivial cycle containing labels
from both sets to have $a=b=1$, and forbids any nontrivial cycle lying
entirely within either set. Thus every nontrivial cycle exchanges one
label in $\{1,\ldots,r\}$ with one in $\{r+1,\ldots,2r\}$. Since $s_k$
moves only active labels, a transposition $(i,r+j)$ of $\pi$ requires
$g_1(i)=j\leq k$ and $i\leq k$. The $k$ transpositions therefore pair the
active labels $1,\ldots,k$ with $r+1,\ldots,r+k$,
so for some $g\in S_k$,
$$
\pi=\prod_{i=1}^k(i,r+g(i)),
$$
Here all traced labels are fixed.

Since $g_1\oplus g_2=s_k\pi$, we have
$i\xrightarrow{\pi}r+g(i)\xrightarrow{s_k}g(i)$, so $g_1$ acts as $g$ on
the active labels $1,\ldots,k$ of the first factor. Likewise,
$r+g(i)\xrightarrow{\pi}i\xrightarrow{s_k}r+i$, so $g_2=g^{-1}\oplus e_q$. Hence
$g_1\oplus g_2=(g\oplus e_q)\oplus(g^{-1}\oplus e_q)$, which proves
Eq.~\eqref{eq:appendix-local-haar-optimal-endpoints}. Equation
\eqref{eq:appendix-local-haar-endpoint-cycles} then follows directly.
\end{proof}

\subsection{Corrections and regime of validity\label{subsec:correction_regimes}}

We discuss the corrections corresponding to the four approximations in Sec.~\ref{sec:local-haar}. All orders below are relative to the leading overlaps in Eq.~\eqref{eq:local-haar-three-overlaps}.

\begin{enumerate}
\renewcommand{\labelenumi}{(\roman{enumi})}
\item Channel labels different from the surrounding label $h$.
The bound below holds at any integer replica order and is applied at order $2r=2(k+q)$ for the circuit--circuit overlap.
The channel can favor labels $(g,g')\neq(h,h)$, but only by a bounded amount: for any $g,g',h\in S_r$,
\begin{equation}
|K_r(g,g')|
\leq d^{|h^{-1}g|+|h^{-1}g'|}w_{M_{\Ncal}}(h),
\label{eq:local-haar-channel-bound}
\end{equation}
where $s=|h^{-1}g|+|h^{-1}g'|$ is the number of swaps that turn the labels $(h,h)$ into $(g,g')$. Appendix~\ref{app:local-haar-channel-bound} proves this bound by expanding each swap in a unitary operator basis and applying H\"older's inequality. The bound is tight: for complete depolarization, $K_r(e_r,e_r)/w_{M_{\Ncal}}(h)=d^{2|h|}$, with $s=2|h|$.

The ordinary gates surrounding the channel compensate for this enhancement. Consider first a single gate with label $g$, while its neighbors retain $h$. Then $g'=h$ and $s=|h^{-1}g|$. The three ordinary legs of this gate, shown by the empty box below, each contribute $G_1^{(r)}(g,h)$:

\[
\begin{tikzpicture}[baseline=(gate.center),font=\scriptsize,
  line width=0.4pt,
  gatebox/.style={draw,minimum width=3.8mm,minimum height=3.2mm,
    inner sep=0pt}]
\node[gatebox] (gate) at (0,0) {};
\node[gatebox] (nw) at (-0.8,0.6) {$h$};
\node[gatebox] (ne) at (0.8,0.6) {$h$};
\node[gatebox] (se) at (0.8,-0.6) {$h$};
\node[inner sep=0pt] (channel) at (-0.8,-0.6) {$\Ncal$};
\draw (gate.north west) -- node[midway,above left,inner sep=1pt]{}(nw.south east);
\draw (gate.north east) -- node[midway,above right,inner sep=1pt] {}
  (ne.south west);
\draw (gate.south east) -- node[midway,below right,inner sep=1pt] {}
  (se.north west);
\draw (gate.south west) -- node[midway,below left,inner sep=1pt] {}
  (channel.north east);
\node at (0.25,0.3) {$g$};
\node at (-0.25,0.3) {$g$};
\node at (-0.25,-0.3) {$g$};
\node at (0.25,-0.3) {$g$};
\end{tikzpicture}
\qquad
\bigl[G_1^{(r)}(g,h)\bigr]^3=d^{-3s}.
\]
The channel contribution can enhance the weight by at most $d^s$,
so the net relative factor is at most $d^{-2s}$.
The same estimate applies when the two permutations in a Haar pairing differ, since the inverse Gram matrix contributes two powers of $d^{-1}$ per transposition. Thus a finite group of changed labels around the channel gives a relative correction of $\mathcal O(d^{-2})$.

\item Domain walls ending on the initial state.
A wall reaching the initial state crosses $t$ gates if the
final layer crosses the cut, and $t-1$ otherwise. The shortest path to
the left edge crosses $N_A$ gates. We therefore use the bound~\cite{ChanDeLuca2024Projected}
\begin{equation}
C_k\,2^{kt}d^{-(t-1-N_A)},\qquad t\geq N_A+2,
\label{eq:local-haar-initial-correction}
\end{equation}
where $C_k$ is independent of $d$. The factor $2^{kt}$ accounts for the possible directions of the $k$ walls over $t$ layers. At fixed $k$ and $t$, this factor does not grow with $d$, so the extra wall length suppresses the correction as $d\to\infty$.
At fixed $d$, the bound
$C_kd^{N_A+1}(2^k/d)^t$ decreases with $t$ only when $d>2^k$. Nonetheless, this is only an upper bound. We still expect convergence to the CIE at smaller $d$, as suggested by
the numerical results at $d=2$ in Sec.~\ref{sec:numerics}, although the mechanism may differ from that described by the large-$d$ analysis.

\item Permutations that send a label in $\{1,\ldots,r\}$ to one in $\{r+1,\ldots,2r\}$ in the measured region $B$.
Consider the configuration with permutation $h=s_k$
throughout the circuit. It matches the boundary condition on $A$,
so the $k$ walls of length $N_A$ are no longer needed.
At each measured site, this label gives $d$ rather than $d^2$, as shown in Eq.~\eqref{eq:local-haar-measured-boundary}, so it contributes an extra factor $d^{-1}$. Its weight
is therefore $d^{-N_B}$, compared with $d^{-kN_A}$ for the
saturated configuration. The relative contribution is thus
\begin{equation}
\mathcal O\!\left(\frac{d^{-N_B}}{d^{-kN_A}}\right)
=\mathcal O\!\left(d^{-(N_B-kN_A)}\right).
\label{eq:local-haar-measured-correction}
\end{equation}
The condition $N_B>kN_A$ ensures that this competing configuration
is suppressed in the large-$d$ limit.

\item Terms omitted from the Haar weights. In Eq.~\eqref{eq:local-haar-triangle-weight}, we keep only the leading terms of the gate weight in $1/d$. At fixed replica order and circuit geometry, the omitted terms give a relative correction of $\mathcal O(d^{-2})$. Together with (i), these terms give the $d^{-2}$ term in Eq.~\eqref{eq:local-haar-error-scale}.
\end{enumerate}

There is also a correction from endpoint pairs $(g_1,g_2)$ whose distance $|s_k^{-1}(g_1\oplus g_2)|$ is larger than the minimum $k$ in Lemma~\ref{lem:local-haar-minimum-distance}. Their distance is at least $k+1$, so each contributes at least one extra factor of $D_A^{-1}$ compared with the leading terms in Eq.~\eqref{eq:local-haar-reference-wick-overlap}. This gives a relative correction of $\mathcal O(D_A^{-1})$, where $D_A=d^{N_A}$.

At fixed $k,t,N_A,N_B$ with $t\geq N_A+2$ and $N_B>kN_A$, all these relative corrections are of order $d^{-1}$ or smaller, as in Eq.~\eqref{eq:local-haar-three-overlaps}. After the term-by-term continuation $q\to1-k$, the common leading terms of the three overlaps cancel in Eq.~\eqref{eq:local-haar-distance-expansion}. Dividing by the reference purity gives the mean squared distance in Eq.~\eqref{eq:local-haar-distance-squared}, which is bounded by the sum of these relative corrections. Cauchy--Schwarz then gives the bound on the mean distance in Eq.~\eqref{eq:local-haar-error-scale}.

\subsection{Channel-kernel bound}
\label{app:local-haar-channel-bound}

To prove Eq.~\eqref{eq:local-haar-channel-bound}, we separate the
common permutation $h\in S_r$ in the Choi vertex of
Eq.~\eqref{eq:local-haar-choi-vertex}:
\begin{equation}
\begin{aligned}
&\rho_{1,r}(g'^{-1})_{C^1}\otimes\rho_{1,r}(g^{-1})_{C^2}=\rho_{2,r}(h^{-1})\bigl[
\rho_{1,r}(g'^{-1}h)_{C^1}\otimes
\rho_{1,r}(g^{-1}h)_{C^2}\bigr].
\end{aligned}
\label{eq:local-haar-common-wiring}
\end{equation}
The equality uses the composition rule
$\rho_{1,r}(g)\rho_{1,r}(g')=\rho_{1,r}(g'g)$.
Here $C^1$ and $C^2$ are the reference copy of the channel input and the channel output space, respectively,
and $\rho_{2,r}(h^{-1})$ permutes the complete Choi copies.
Substitution into the kernel gives
\[
\begin{aligned}
K_r(g,g')
&=\tr\Bigl[M_{\Ncal}^{\otimes r}\rho_{2,r}(h^{-1})
\bigl(\rho_{1,r}(g'^{-1}h)_{C^1}\otimes
\rho_{1,r}(g^{-1}h)_{C^2}\bigr)\Bigr]\\
&=\tr\Bigl[M_{\Ncal}^{\otimes r}
\bigl(\rho_{1,r}(g'^{-1}h)_{C^1}\otimes
\rho_{1,r}(g^{-1}h)_{C^2}\bigr)
\rho_{2,r}(h^{-1})\Bigr].
\end{aligned}
\]
The first equality uses Eq.~\eqref{eq:local-haar-common-wiring} and Eq.~\eqref{eq:local-haar-choi-vertex}.
The second uses cyclicity of the trace and
$[M_{\Ncal}^{\otimes r},\rho_{2,r}(h^{-1})]=0$.

The two remaining permutations require
$s=|h^{-1}g'|+|h^{-1}g|$ transpositions in total.
Choose shortest products of swap operators with endpoints
$1\leq u_m,v_m\leq r$, $u_m\ne v_m$, such that
\[
\begin{aligned}
\rho_{1,r}(g'^{-1}h)_{C^1}
&=\prod_{m=1}^{|h^{-1}g'|}\rho_{1,r}(s_{u_mv_m})_{C^1}, \qquad
\rho_{1,r}(g^{-1}h)_{C^2}
&=\prod_{m=|h^{-1}g'|+1}^{s}\rho_{1,r}(s_{u_mv_m})_{C^2}.
\end{aligned}
\]
All products below retain this left-to-right order of increasing
$m$. To expand the swaps, choose a unitary Weyl basis
$\{T_a\}_{a=1}^{d^2}$ on a $d$-dimensional Choi factor, normalized
by $\tr(T_a^\dagger T_b)=d\delta_{ab}$. The superscript in
$T_a^{(i)}$ means that $T_a$ acts on the selected Choi factor of
replica $i$ and as the identity elsewhere. Then using the completeness of the Weyl basis
\begin{equation}
\rho_{1,r}(s_{ij})
=\frac1d\sum_{a=1}^{d^2}T_a^{(i)}(T_a^\dagger)^{(j)}.
\label{eq:local-haar-swap-expansion}
\end{equation}
For $\bm a=(a_1,\ldots,a_s)$ with $1\leq a_m\leq d^2$, the
operator on Choi copy $j$ is explicitly
\[
\begin{aligned}
R_j(\bm a)
={}&\left[\prod_{m=1}^{|h^{-1}g'|}
T_{a_m}^{\delta_{j,u_m}}(T_{a_m}^\dagger)^{\delta_{j,v_m}}
\right]_{C^1}\otimes\left[\prod_{m=|h^{-1}g'|+1}^{s}
T_{a_m}^{\delta_{j,u_m}}(T_{a_m}^\dagger)^{\delta_{j,v_m}}
\right]_{C^2}.
\end{aligned}
\]
Here $\delta$ is the Kronecker delta,
$T_a^0=(T_a^\dagger)^0=I_d$, and an empty product is the identity.
Thus $R_j(\bm a)$ is unitary. Expanding the ordered swap products
and collecting the operators on each copy gives
\[
\rho_{1,r}(g'^{-1}h)_{C^1}\otimes\rho_{1,r}(g^{-1}h)_{C^2}
=d^{-s}\sum_{a_1,\ldots,a_s=1}^{d^2}
\bigotimes_{j=1}^rR_j(\bm a).
\]
Each swap supplies one Weyl index $a_m$ and a factor $d^{-1}$.
For each cycle $c=(j_1\,\ldots\,j_p)\in\mathcal C(h)$, choose the
index order so that $h(j_m)=j_{m+1}$ for $1\leq m<p$ and
$h(j_p)=j_1$. Then
\begin{equation}
\begin{aligned}
K_r(g,g')
&=d^{-s}\!\!\!\!\sum_{a_1,\ldots,a_s=1}^{d^2}
\!\!\!\tr\left[\left(\bigotimes_{j=1}^rM_{\Ncal}R_j(\bm a)\right)
\rho_{2,r}(h^{-1})\right]=d^{-s}\!\!\!\sum_{a_1,\ldots,a_s=1}^{d^2}
\prod_{c=(j_1\,\ldots\,j_p)\in\mathcal C(h)}
\!\!\!\!\tr\left[M_{\Ncal}R_{j_p}(\bm a)\cdots
M_{\Ncal}R_{j_1}(\bm a)\right].
\end{aligned}
\label{eq:local-haar-weyl-terms}
\end{equation}
The first equality substitutes the expanded permutations into the
kernel. The second contracts the replica indices around each cycle.
For the Schatten norm
$\|O\|_p=(\tr[(O^\dagger O)^{p/2}])^{1/p}$, each cycle obeys
\begin{equation}
\begin{aligned}
&\left|\tr\left[M_{\Ncal}R_{j_p}(\bm a)\cdots
M_{\Ncal}R_{j_1}(\bm a)\right]\right|\leq\prod_{m=1}^p\|M_{\Ncal}R_{j_m}(\bm a)\|_p=\prod_{m=1}^p[\tr(M_{\Ncal}^p)]^{1/p}=\tr(M_{\Ncal}^p).
\end{aligned}
\label{eq:local-haar-cycle-holder}
\end{equation}
The inequality is the H\"older inequality for Schatten norms.
The first equality uses $M_{\Ncal}\geq0$ and the unitarity of
$R_j$, which preserves the singular values of $M_{\Ncal}$. The bound includes fixed points ($p=1$).
Multiplying over the cycles and summing the Weyl terms gives
\begin{equation}
\begin{aligned}
|K_r(g,g')|
&\leq d^{-s}\sum_{a_1,\ldots,a_s=1}^{d^2}
\prod_{c\in\mathcal C(h)}\tr(M_{\Ncal}^{|c|})=d^{-s}d^{2s}\prod_{c\in\mathcal C(h)}\tr(M_{\Ncal}^{|c|})=d^s w_{M_{\Ncal}}(h).
\end{aligned}
\end{equation}
The inequality uses the triangle inequality and the bound for
each cycle. The first equality counts the $d^{2s}$ Weyl terms. The second uses
$w_{M_{\Ncal}}(h)=\prod_{c\in\mathcal C(h)}\tr(M_{\Ncal}^{|c|})$. This proves
Eq.~\eqref{eq:local-haar-channel-bound}.

\bigskip
\twocolumngrid
\bibliographystyle{apsrev4-2}
\bibliography{references}

@article{DAlessioEtAl2016ETH,
  author = {D'Alessio, Luca and Kafri, Yariv and Polkovnikov, Anatoli and Rigol, Marcos},
  title = {{From Quantum Chaos and Eigenstate Thermalization to Statistical Mechanics and Thermodynamics}},
  year = {2016},
  journal = {Advances in Physics},
  volume = {65},
  pages = {239--362},
  doi = {10.1080/00018732.2016.1198134},
  url = {https://arxiv.org/abs/1509.06411}
}

@article{VidmarRigol2016GGE,
  author = {Vidmar, Lev and Rigol, Marcos},
  title = {{Generalized Gibbs ensemble in integrable lattice models}},
  year = {2016},
  journal = {Journal of Statistical Mechanics: Theory and Experiment},
  volume = {2016},
  pages = {064007},
  doi = {10.1088/1742-5468/2016/06/064007},
  url = {https://arxiv.org/abs/1604.03990}
}

@article{HoChoi2022Exact,
  author = {Ho, Wen Wei and Choi, Soonwon},
  title = {Exact Emergent Quantum State Designs from Quantum Chaotic Dynamics},
  journal = {Physical Review Letters},
  volume = {128},
  pages = {060601},
  year = {2022},
  doi = {10.1103/PhysRevLett.128.060601}
}

@article{CotlerEtAl2023Designs,
  author = {Cotler, Jordan S. and Mark, Daniel K. and Huang, Hsin-Yuan and Hern\'{a}ndez, Felipe and Choi, Joonhee and Shaw, Adam L. and Endres, Manuel and Choi, Soonwon},
  title = {Emergent Quantum State Designs from Individual Many-Body Wave Functions},
  journal = {PRX Quantum},
  volume = {4},
  pages = {010311},
  year = {2023},
  doi = {10.1103/PRXQuantum.4.010311}
}

@article{IppolitiHo2023Purification,
  author = {Ippoliti, Matteo and Ho, Wen Wei},
  title = {Dynamical Purification and the Emergence of Quantum State Designs from the Projected Ensemble},
  journal = {PRX Quantum},
  volume = {4},
  pages = {030322},
  year = {2023},
  doi = {10.1103/PRXQuantum.4.030322}
}

@article{MarkEtAl2024MaximumEntropy,
  author = {Mark, Daniel K. and Surace, Federica and Elben, Andreas and Shaw, Adam L. and Choi, Joonhee and Refael, Gil and Endres, Manuel and Choi, Soonwon},
  title = {Maximum Entropy Principle in Deep Thermalization and in {Hilbert}-Space Ergodicity},
  journal = {Physical Review X},
  volume = {14},
  pages = {041051},
  year = {2024},
  doi = {10.1103/PhysRevX.14.041051}
}

@article{ChoiEtAl2023PreparingRandom,
  author = {Choi, Joonhee and Shaw, Adam L. and Madjarov, Ivaylo S. and Xie, Xin and Finkelstein, Ran and Covey, Jacob P. and Cotler, Jordan S. and Mark, Daniel K. and Huang, Hsin-Yuan and Kale, Anant and Pichler, Hannes and Brand{\~a}o, Fernando G. S. L. and Choi, Soonwon and Endres, Manuel},
  title = {{Preparing random states and benchmarking with many-body quantum chaos}},
  year = {2023},
  journal = {Nature},
  volume = {613},
  pages = {468--473},
  doi = {10.1038/s41586-022-05442-1},
  url = {https://arxiv.org/abs/2103.03535}
}

@article{YanEtAl2026Superconducting,
  author = {Yan, Zhiguang and Ge, Zi-Yong and Li, Rui and Zhang, Yu-Ran and Nori, Franco and Nakamura, Yasunobu},
  title = {{Characterizing many-body dynamics with projected ensembles on a superconducting quantum processor}},
  year = {2026},
  journal = {Science Advances},
  volume = {12},
  pages = {eaeb8213},
  doi = {10.1126/sciadv.aeb8213},
  url = {https://arxiv.org/abs/2506.21061}
}

@article{MarkEtAl2023BenchmarkingErgodic,
  author = {Mark, Daniel K. and Choi, Joonhee and Shaw, Adam L. and Endres, Manuel and Choi, Soonwon},
  title = {{Benchmarking Quantum Simulators Using Ergodic Quantum Dynamics}},
  year = {2023},
  journal = {Physical Review Letters},
  volume = {131},
  pages = {110601},
  doi = {10.1103/PhysRevLett.131.110601},
  url = {https://arxiv.org/abs/2205.12211}
}

@article{ZhangEtAl2025Holographic,
  author = {Zhang, Bingzhi and Xu, Peng and Chen, Xiaohui and Zhuang, Quntao},
  title = {{Holographic deep thermalization for secure and efficient quantum random state generation}},
  year = {2025},
  journal = {Nature Communications},
  volume = {16},
  pages = {6341},
  doi = {10.1038/s41467-025-61546-y},
  url = {https://arxiv.org/abs/2411.03587}
}

@article{ChangShrotriyaHoIppoliti2025Charge,
  author = {Chang, Rui-An and Shrotriya, Harshank and Ho, Wen Wei and Ippoliti, Matteo},
  title = {{Deep Thermalization under Charge-Conserving Quantum Dynamics}},
  year = {2025},
  journal = {PRX Quantum},
  volume = {6},
  pages = {020343},
  doi = {10.1103/PRXQuantum.6.020343},
  url = {https://journals.aps.org/prxquantum/abstract/10.1103/PRXQuantum.6.020343}
}

@article{MokHaugHoPreskill2026NatureStingy,
  author = {Mok, Wai-Keong and Haug, Tobias and Ho, Wen Wei and Preskill, John},
  title = {{Nature Is Stingy: Universality of Scrooge Ensembles in Quantum Many-Body Systems}},
  year = {2026},
  journal = {Physical Review X},
  volume = {16},
  pages = {041003},
  doi = {10.1103/tb52-jxmx},
  url = {https://journals.aps.org/prx/abstract/10.1103/tb52-jxmx}
}

@article{VarikutiBandyopadhyay2024Symmetry,
  author = {Varikuti, Naga Dileep and Bandyopadhyay, Soumik},
  title = {{Unraveling the emergence of quantum state designs in systems with symmetry}},
  year = {2024},
  journal = {Quantum},
  volume = {8},
  pages = {1456},
  doi = {10.22331/q-2024-08-29-1456},
  url = {https://arxiv.org/abs/2402.08949}
}

@article{MannaRoySreejith2025ConservedCharges,
  author = {Manna, Sandipan and Roy, Sthitadhi and Sreejith, G. J.},
  title = {{Projected ensemble in a system with locally supported conserved charges}},
  year = {2025},
  journal = {Physical Review B},
  volume = {111},
  pages = {144302},
  doi = {10.1103/PhysRevB.111.144302},
  url = {https://arxiv.org/abs/2501.01823}
}

@article{BhoreDesaulesPapic2023Constrained,
  author = {Bhore, Tanmay and Desaules, Jean-Yves and Papi{\'c}, Zlatko},
  title = {{Deep thermalization in constrained quantum systems}},
  year = {2023},
  journal = {Physical Review B},
  volume = {108},
  pages = {104317},
  doi = {10.1103/PhysRevB.108.104317},
  url = {https://arxiv.org/abs/2307.03769}
}

@misc{WilmingRoth2022HighTemperature,
  author = {Wilming, Henrik and Roth, Ingo},
  year = {2022},
  eprint = {2202.01669},
  archivePrefix = {arXiv},
  url = {https://arxiv.org/abs/2202.01669}
}

@article{LucasEtAl2023FreeFermions,
  author = {Lucas, Maxime and Piroli, Lorenzo and De Nardis, Jacopo and De Luca, Andrea},
  title = {{Generalized Deep Thermalization for Free Fermions}},
  year = {2023},
  journal = {Physical Review A},
  volume = {107},
  pages = {032215},
  doi = {10.1103/PhysRevA.107.032215},
  url = {https://arxiv.org/abs/2207.13628}
}

@article{LiuHuangHo2024Gaussian,
  author = {Liu, Chang and Huang, Qi Camm and Ho, Wen Wei},
  title = {{Deep Thermalization in Gaussian Continuous-Variable Quantum Systems}},
  year = {2024},
  journal = {Physical Review Letters},
  volume = {133},
  pages = {260401},
  doi = {10.1103/PhysRevLett.133.260401},
  url = {https://journals.aps.org/prl/abstract/10.1103/PhysRevLett.133.260401}
}

@article{LiuIppolitiHo2026CoherenceTransition,
  author = {Liu, Chang and Ippoliti, Matteo and Ho, Wen Wei},
  title = {{Coherence-Induced Deep Thermalization Transition in Random Permutation Quantum Dynamics}},
  year = {2026},
  journal = {Physical Review Letters},
  volume = {136},
  pages = {100404},
  doi = {10.1103/mjgs-y2z9},
  url = {https://journals.aps.org/prl/abstract/10.1103/mjgs-y2z9}
}

@misc{FengLiuChengHoIppoliti2026ResourceLocalizability,
  author = {Feng, Xiaozhou and Liu, Chang and Cheng, Zihan and Ho, Wen Wei and Ippoliti, Matteo},
  year = {2026},
  eprint = {2606.08756},
  archivePrefix = {arXiv},
  url = {https://arxiv.org/abs/2606.08756}
}

@article{SherryRoy2024MeasurementInvisible,
  author = {Sherry, Alan and Roy, Sthitadhi},
  title = {{Measurement-invisible quantum correlations in scrambling dynamics}},
  year = {2025},
  journal = {Physical Review B},
  volume = {111},
  pages = {L180301},
  doi = {10.1103/PhysRevB.111.L180301},
  url = {https://arxiv.org/abs/2410.24212}
}

@misc{MandalSherryRoy2026Locality,
  author = {Mandal, Saptarshi and Sherry, Alan and Roy, Sthitadhi},
  year = {2026},
  eprint = {2607.15276},
  archivePrefix = {arXiv},
  url = {https://arxiv.org/abs/2607.15276}
}

@article{SauliereLamiBoyerDeNardisDeLuca2026NoisyAnticoncentration,
  author = {Sauliere, Arman and Lami, Guglielmo and Boyer, Corentin and De Nardis, Jacopo and De Luca, Andrea},
  title = {{Universality in the Anticoncentration of Noisy Quantum Circuits at Finite Depths}},
  year = {2026},
  journal = {PRX Quantum},
  volume = {7},
  pages = {020365},
  doi = {10.1103/xl16-cdy9},
  url = {https://journals.aps.org/prxquantum/abstract/10.1103/xl16-cdy9}
}

@article{MokEtAl2025ClassicalRandomness,
  author = {Mok, Wai-Keong and Haug, Tobias and Shaw, Adam L. and Endres, Manuel and Preskill, John},
  title = {{Optimal Conversion from Classical to Quantum Randomness via Quantum Chaos}},
  year = {2025},
  journal = {Physical Review Letters},
  volume = {134},
  pages = {180403},
  doi = {10.1103/PhysRevLett.134.180403},
  url = {https://arxiv.org/abs/2410.05181}
}

@article{McGinleyFava2023ShadowTomography,
  author = {McGinley, Max and Fava, Michele},
  title = {{Shadow tomography from emergent state designs in analog quantum simulators}},
  year = {2023},
  journal = {Physical Review Letters},
  volume = {131},
  pages = {160601},
  doi = {10.1103/PhysRevLett.131.160601},
  url = {https://doi.org/10.1103/PhysRevLett.131.160601}
}

@misc{MarkEndresIppolitiHoChoi2026Review,
  author = {Mark, Daniel K. and Endres, Manuel and Ippoliti, Matteo and Ho, Wen Wei and Choi, Soonwon},
  year = {2026},
  eprint = {2609.30248},
  archivePrefix = {arXiv},
  url = {https://arxiv.org/abs/2609.30248}
}

@article{IppolitiHo2022DistinctDesignTimes,
  author = {Ippoliti, Matteo and Ho, Wen Wei},
  title = {{Solvable model of deep thermalization with distinct design times}},
  year = {2022},
  journal = {Quantum},
  volume = {6},
  pages = {886},
  doi = {10.22331/q-2022-12-29-886},
  url = {https://quantum-journal.org/papers/q-2022-12-29-886/}
}

@article{ShrotriyaHo2025Nonlocality,
  author = {Shrotriya, Harshank and Ho, Wen Wei},
  title = {{Nonlocality of deep thermalization}},
  year = {2025},
  journal = {SciPost Physics},
  volume = {18},
  pages = {107},
  doi = {10.21468/SciPostPhys.18.3.107},
  url = {https://scipost.org/SciPostPhys.18.3.107/}
}

@misc{HerrmannFritzschBacker2026Timescales,
  author = {Herrmann, Tabea and Fritzsch, Felix and B{\"a}cker, Arnd},
  year = {2026},
  eprint = {2604.27749},
  archivePrefix = {arXiv},
  url = {https://arxiv.org/abs/2604.27749}
}

@misc{ArvindHunterJones2026FiniteBath,
  author = {Arvind, Rahul and Hunter-Jones, Nicholas},
  year = {2026},
  eprint = {2610.06754},
  archivePrefix = {arXiv}
}

@article{ChristopoulosChanDeLuca2025UniversalOverlaps,
  author = {Christopoulos, Alexios and Chan, Amos and De Luca, Andrea},
  title = {{Universal distributions of overlaps from generic dynamics in quantum many-body systems}},
  year = {2025},
  journal = {Physical Review Research},
  volume = {7},
  pages = {043035},
  doi = {10.1103/kl64-xnsz},
  url = {https://arxiv.org/abs/2404.10057}
}

@article{YuHoKos2025Mixed,
  author = {Yu, Xie-Hang and Ho, Wen Wei and Kos, Pavel},
  title = {Mixed State Deep Thermalization},
  journal = {Physical Review Letters},
  volume = {135},
  pages = {260402},
  year = {2025},
  doi = {10.1103/t6zs-3f8k},
}

@article{MandalClaeysRoy2025PartialProjected,
  author = {Mandal, Saptarshi and Claeys, Pieter W. and Roy, Sthitadhi},
  title = {{Partial projected ensembles and spatiotemporal structure of information scrambling}},
  year = {2026},
  journal = {Physical Review B},
  volume = {113},
  pages = {024303},
  doi = {10.1103/h2q2-yfqs},
  url = {https://arxiv.org/abs/2508.05632}
}

@article{SherryMandalRoy2025InformationPhases,
  author = {Sherry, Alan and Mandal, Saptarshi and Roy, Sthitadhi},
  title = {{Information phases of partial projected ensembles generated from random quantum states and scrambling dynamics}},
  year = {2026},
  journal = {Physical Review B},
  volume = {113},
  pages = {104301},
  doi = {10.1103/dcdc-pxn3},
  url = {https://arxiv.org/abs/2511.10595}
}

@article{MilekhinMurciano2025ObservableProjected,
  author = {Milekhin, Alexey and Murciano, Sara},
  title = {{Observable-projected ensembles}},
  year = {2025},
  journal = {Quantum},
  volume = {9},
  pages = {1888},
  doi = {10.22331/q-2025-10-20-1888},
  url = {https://arxiv.org/abs/2410.21397}
}

@article{SherryRoy2025Mixed,
  author = {Sherry, Alan and Roy, Sthitadhi},
  title = {{Deep thermalization for mixed states}},
  year = {2026},
  journal = {Physical Review B},
  volume = {113},
  pages = {174309},
  doi = {10.1103/x397-3nk5},
  url = {https://arxiv.org/abs/2507.14135}
}

@misc{HuangMokHaugHo2026ProjectedKraus,
  author = {Huang, Qi Camm and Mok, Wai-Keong and Haug, Tobias and Ho, Wen Wei},
  year = {2026},
  eprint = {2609.12689},
  archivePrefix = {arXiv},
  url = {https://arxiv.org/abs/2609.12689}
}

@article{KosStyliaris2023SpaceTimeChannels,
  author = {Kos, Pavel and Styliaris, Georgios},
  title = {{Circuits of space and time quantum channels}},
  year = {2023},
  journal = {Quantum},
  volume = {7},
  pages = {1020},
  doi = {10.22331/q-2023-05-24-1020},
  url = {https://arxiv.org/abs/2206.12155}
}

@article{ShawEtAl2025NoiseLearning,
  author = {Shaw, Adam L. and Mark, Daniel K. and Choi, Joonhee and Finkelstein, Ran and Scholl, Pascal and Choi, Soonwon and Endres, Manuel},
  title = {{Experimental Signatures of Hilbert-Space Ergodicity: Universal Bitstring Distributions and Applications in Noise Learning}},
  year = {2025},
  journal = {Physical Review X},
  volume = {15},
  pages = {031001},
  doi = {10.1103/h6xy-zpx4},
  url = {https://arxiv.org/abs/2403.11971}
}

@article{ZyczkowskiSommers2001Induced,
  author = {\.{Z}yczkowski, Karol and Sommers, Hans-J\"urgen},
  title = {Induced Measures in the Space of Mixed Quantum States},
  journal = {Journal of Physics A: Mathematical and General},
  volume = {34},
  pages = {7111--7125},
  year = {2001},
  doi = {10.1088/0305-4470/34/35/335}
}

@article{SommersZyczkowski2004Statistical,
  author = {Sommers, Hans-J\"urgen and \.{Z}yczkowski, Karol},
  title = {Statistical Properties of Random Density Matrices},
  journal = {Journal of Physics A: Mathematical and General},
  volume = {37},
  pages = {8457--8466},
  year = {2004},
  doi = {10.1088/0305-4470/37/35/004}
}

@article{ZyczkowskiEtAl2011Generating,
  author = {\.{Z}yczkowski, Karol and Penson, Karol A. and Nechita, Ion and Collins, Beno{\^i}t},
  title = {{Generating random density matrices}},
  year = {2011},
  journal = {Journal of Mathematical Physics},
  volume = {52},
  pages = {062201},
  doi = {10.1063/1.3595693},
  url = {https://arxiv.org/abs/1010.3570}
}

@article{BertiniKosProsen2019Correlations,
  author = {Bertini, Bruno and Kos, Pavel and Prosen, Toma{\v{z}}},
  title = {{Exact Correlation Functions for Dual-Unitary Lattice Models in $1+1$ Dimensions}},
  year = {2019},
  journal = {Physical Review Letters},
  volume = {123},
  pages = {210601},
  doi = {10.1103/PhysRevLett.123.210601},
  url = {https://doi.org/10.1103/PhysRevLett.123.210601}
}

@article{SuzukiMitaraiFujii2022Computational,
  author = {Suzuki, Ryotaro and Mitarai, Kosuke and Fujii, Keisuke},
  title = {{Computational power of one- and two-dimensional dual-unitary quantum circuits}},
  year = {2022},
  journal = {Quantum},
  volume = {6},
  pages = {631},
  doi = {10.22331/q-2022-01-24-631}
}

@article{BertiniClaeysProsen2026SpaceTimeReview,
  author = {Bertini, Bruno and Claeys, Pieter W. and Prosen, Tomaž},
  title = {{Exactly solvable quantum many-body dynamics from space-time duality}},
  year = {2026},
  journal = {Reviews of Modern Physics},
  volume = {98},
  pages = {025001},
  doi = {10.1103/yx73-dk86},
  url = {https://arxiv.org/abs/2505.11489}
}

@article{ClaeysLamacraft2022Biunitarity,
  author = {Claeys, Pieter W. and Lamacraft, Austen},
  title = {{Emergent quantum state designs and biunitarity in dual-unitary circuit dynamics}},
  year = {2022},
  journal = {Quantum},
  volume = {6},
  pages = {738},
  doi = {10.22331/q-2022-06-15-738},
  url = {https://arxiv.org/abs/2202.12306}
}

@article{ChanDeLuca2024Projected,
  author = {Chan, Amos and De Luca, Andrea},
  title = {Projected State Ensemble of a Generic Model of Many-Body Quantum Chaos},
  journal = {Journal of Physics A: Mathematical and Theoretical},
  volume = {57},
  pages = {405001},
  year = {2024},
  doi = {10.1088/1751-8121/ad7211}
}

@article{Choi1975CompletelyPositive,
  author = {Choi, Man-Duen},
  title = {Completely Positive Linear Maps on Complex Matrices},
  journal = {Linear Algebra and its Applications},
  volume = {10},
  pages = {285--290},
  year = {1975},
  doi = {10.1016/0024-3795(75)90075-0}
}

@article{Stinespring1955Positive,
  author = {Stinespring, W. Forrest},
  title = {Positive functions on $C^*$-algebras},
  journal = {Proceedings of the American Mathematical Society},
  volume = {6},
  pages = {211--216},
  year = {1955},
  doi = {10.1090/S0002-9939-1955-0069403-4},
  url = {https://doi.org/10.1090/S0002-9939-1955-0069403-4}
}

@book{MeckesHaarBookPDF,
 author={Meckes, Elizabeth S.},
 title={The Random Matrix Theory of the Classical Compact Groups},
 publisher={Cambridge University Press},
 year={2019},
 doi={10.1017/9781108303453},
 url={https://doi.org/10.1017/9781108303453}
}

@article{ZhouNahum2019StatisticalMechanics,
  author = {Zhou, Tianci and Nahum, Adam},
  title = {Emergent statistical mechanics of entanglement in random unitary circuits},
  journal = {Physical Review B},
  volume = {99},
  pages = {174205},
  year = {2019},
  doi = {10.1103/PhysRevB.99.174205},
  url = {https://doi.org/10.1103/PhysRevB.99.174205}
}

@article{CollinsSniady2006Haar,
  author = {Collins, Beno{\^i}t and {\'S}niady, Piotr},
  title = {Integration with Respect to the {Haar} Measure on Unitary, Orthogonal and Symplectic Group},
  journal = {Communications in Mathematical Physics},
  volume = {264},
  pages = {773--795},
  year = {2006},
  doi = {10.1007/s00220-006-1554-3},
  url = {https://doi.org/10.1007/s00220-006-1554-3}
}

@article{AubrunLancien2015,
  author = {Aubrun, Guillaume and Lancien, C{\'e}cilia},
  title = {Locally restricted measurements on a multipartite quantum system: data hiding is generic},
  journal = {Quantum Information and Computation},
  volume = {15},
  number = {5--6},
  pages = {513--540},
  year = {2015},
  doi = {10.26421/QIC15.5-6-9},
  url = {https://arxiv.org/abs/1406.1959}
}

@article{Bansal2025,
  author = {Bansal, Nikhil and Mok, Wai-Keong and Bharti, Kishor and Koh, Dax Enshan and Haug, Tobias},
  title = {Pseudorandom Density Matrices},
  journal = {PRX Quantum},
  volume = {6},
  pages = {020322},
  year = {2025},
  doi = {10.1103/PRXQuantum.6.020322},
}

@article{JozsaRobbWootters1994Accessible,
  author = {Jozsa, Richard and Robb, Daniel and Wootters, William K.},
  title = {{Lower bound for accessible information in quantum mechanics}},
  journal = {Physical Review A},
  volume = {49},
  pages = {668--677},
  year = {1994},
  doi = {10.1103/PhysRevA.49.668}
}

@misc{LiuZhang2026SYKScrooge,
  author = {Liu, Zeyu and Zhang, Pengfei},
  year = {2026},
  eprint = {2607.04864},
  archivePrefix = {arXiv},
  url = {https://arxiv.org/abs/2607.04864}
}

@misc{RussottoKlobasCalabreseBertini2026GaussianScrooge,
  author = {Russotto, Angelo and Klobas, Katja and Calabrese, Pasquale and Bertini, Bruno},
  year = {2026},
  eprint = {2610.10279},
  archivePrefix = {arXiv},
  url = {https://arxiv.org/abs/2610.10279}
}

@misc{SarmaHaugPreskillMok2026EquilibriumMagic,
  author = {Sarma, Soumyadeep and Haug, Tobias and Preskill, John and Mok, Wai-Keong},
  year = {2026},
  eprint = {2608.22939},
  archivePrefix = {arXiv},
  url = {https://arxiv.org/abs/2608.22939}
}

@misc{PawlikSierantZakrzewski2026RestrictedTypicality,
  author = {Pawlik, Konrad and Sierant, Piotr and Zakrzewski, Jakub},
  year = {2026},
  eprint = {2609.07832},
  archivePrefix = {arXiv},
  url = {https://arxiv.org/abs/2609.07832}
}

@article{GoldsteinLebowitzTumulkaZanghi2006GAP,
  author = {Goldstein, Sheldon and Lebowitz, Joel L. and Tumulka, Roderich and Zangh{\`\i}, Nino},
  title = {{On the Distribution of the Wave Function for Systems in Thermal Equilibrium}},
  journal = {Journal of Statistical Physics},
  volume = {125},
  pages = {1193--1221},
  year = {2006},
  doi = {10.1007/s10955-006-9210-z},
}

@article{WillsonHeelwegWillard2026MaxEnt,
  author = {Willson, Jacob T. and Heelweg, Henrik J. and Willard, Adam P.},
  title = {{Maximum entropy distributions of wave functions at thermal equilibrium}},
  journal = {Physical Review E},
  volume = {114},
  number = {1},
  pages = {014145},
  year = {2026},
  doi = {10.1103/yk5x-cnzv},
  url = {https://doi.org/10.1103/yk5x-cnzv}
}

\end{document}